\documentclass[twocolumn,tighten,times]{aastex631}
\usepackage{amsmath}
\usepackage{graphicx}
\usepackage{makecell}
\usepackage{overpic}
\hypersetup{linkcolor=blue,citecolor=blue,filecolor=blue,urlcolor=magenta}
\let\oldput\put
\def\put(#1,#2)#3{%
  \oldput(#1,#2){\sffamily #3}%
}
\definecolor{navyblue}{RGB}{0,50,250}
\definecolor{purple}{RGB}{155,0,200}

\newcommand{\um}{$\mu$m}
\newcommand{\Mohawc}{\texttt{Mohawc}}
\newcommand{\mPsi}{\ensuremath{\overline{\Psi_{B}}}}
\newcommand{\mFIRPsi}{\ensuremath{{\Psi_{B}^{\rm{FIR}}}}}
\newcommand{\mRadioPsi}{\ensuremath{{\Psi_{B}^{\rm{Radio}}}}}

\newcommand{\PAint}{\ensuremath{\langle PA^{\rm{int}}_{B} \rangle}}
\newcommand{\PAhist}{\ensuremath{\langle PA^{\rm{hist}}_{B} \rangle}}

\DeclareMathOperator{\atantwo}{atan2}

\shorttitle{Extragalactic magnetism with SOFIA: First results on the $B$-field orientations}
\shortauthors{Borlaff et al.}

\begin{document}

\title{Extragalactic magnetism with SOFIA (SALSA Legacy Program) - V: \\
First results on the magnetic field orientation of galaxies\footnote{The SOFIA Legacy Program for Magnetic Fields in Galaxies provides a software repository at \url{https://github.com/galmagfields/hawc}, and publicly available data at \url{http://galmagfields.com/}}}

\correspondingauthor{Alejandro S. Borlaff}
\email{a.s.borlaff@nasa.gov}

\author[0000-0003-3249-4431]{Alejandro S. Borlaff}
\affil{NASA Ames Research Center, Moffett Field, CA 94035, USA}
\affil{Bay Area Environmental Research Institute, Moffett Field, California 94035, USA}
\affil{Kavli Institute for Particle Astrophysics \& Cosmology (KIPAC), Stanford University, Stanford, CA 94305, USA}

\author[0000-0001-5357-6538]{Enrique Lopez-Rodriguez}
\affil{Kavli Institute for Particle Astrophysics \& Cosmology (KIPAC), Stanford University, Stanford, CA 94305, USA}

\author{Rainer Beck}
\affil{Max-Planck-Institut f\"ur Radioastronomie, Auf dem H\"ugel 69, 53121 Bonn, Germany}

\author[0000-0002-7633-3376]{Susan E. Clark}
\affiliation{Department of Physics, Stanford University, Stanford, California 94305, USA}
\affiliation{Kavli Institute for Particle Astrophysics \& Cosmology (KIPAC), Stanford University, Stanford, CA 94305, USA}

\author[0000-0002-4324-0034]{Evangelia Ntormousi}
\affil{Scuola Normale Superiore di Pisa, Piazza dei Cavalieri 7, Pisa, Italy}
\affil{Institute of Astrophysics, Foundation for Research and Technology-Hellas, Vasilika Vouton, GR-70013 Heraklion, Greece}

\author[0000-0002-8831-2038]{Konstantinos Tassis}
\affil{Department of Physics \& ITCP, University of Crete, GR-70013, Heraklion, Greece} 
\affil{Institute of Astrophysics, Foundation for Research and Technology-Hellas, Vasilika Vouton, GR-70013 Heraklion, Greece}

\author[0000-0002-4059-9850]{Sergio Martin-Alvarez}
\affil{Kavli Institute for Particle Astrophysics \& Cosmology (KIPAC), Stanford University, Stanford, CA 94305, USA}

\author[0000-0001-5357-6538]{Mehrnoosh Tahani}
\affil{Kavli Institute for Particle Astrophysics \& Cosmology (KIPAC), Stanford University, Stanford, CA 94305, USA}

\author[0000-0002-5782-9093]{Daniel~A.~Dale}
\affil{Department of Physics and Astronomy, University of Wyoming, Laramie, WY 82071, USA}

\author[0000-0001-8931-1152]{Ignacio del Moral-Castro}
\affil{Kapteyn Astronomical Institute, University of Groningen, PO Box 800, 9700 AV Groningen, The Netherlands}
\affiliation{Facultad de F\'{i}sica, Universidad de La Laguna, Avda. Astrof\'{i}sico Fco. S\'{a}nchez s/n, 38200, La Laguna, Tenerife, Spain}

\author[0000-0001-6326-7069]{Julia Roman-Duval}
\affil{Space Telescope Science Institute, 3700 San Martin Drive, Baltimore, MD 21218}

\author{Pamela M. Marcum}
\affil{NASA Ames Research Center, Moffett Field, CA 94035, USA}

\author[0000-0001-5747-7086]{John E. Beckman}
\affiliation{Instituto de Astrof\'{i}sica de Canarias, C/ V\'{i}a L\'actea, E-38200 La Laguna, Tenerife, Spain}
\affiliation{Facultad de F\'{i}sica, Universidad de La Laguna, Avda. Astrof\'{i}sico Fco. S\'{a}nchez s/n, 38200, La Laguna, Tenerife, Spain}

\author{Kandaswamy Subramanian}
\affiliation{IUCAA, Post Bag 4, Ganeshkhind, Pune 411007, India}
\affiliation{Department of Physics, Ashoka University, Rajiv Gandhi Education City, Rai, Sonipat 131029, Haryana, India}

\author[0000-0002-8281-8388]{Sarah Eftekharzadeh}
\affiliation{NASA Goddard Space Flight Center, 8800 Greenbelt Road, Greenbelt, MD 20771, USA}

\author{Leslie Proudfit}
\affil{SOFIA Science Center, NASA Ames Research Center, Moffett Field, CA 94035, USA}


\begin{abstract}
We present the analysis of the magnetic field ($B$-field) structure of galaxies measured with far-infrared (FIR) and radio (3 and 6 cm) polarimetric observations. We use the first data release of the Survey of extragALactic magnetiSm with SOFIA (SALSA) of 14 nearby ($<20$ Mpc) galaxies with resolved ($5\arcsec-18\arcsec$; $90$ pc--$1$ kpc) imaging polarimetric observations using SOFIA/HAWC+ from $53$ to $214$ \um. We compute the magnetic pitch angle ($\Psi_{B}$) profiles as a function of the galactocentric radius. We introduce a new magnetic alignment parameter ($\zeta$) to estimate the disordered-to-ordered ratio of spiral $B$-fields. We find FIR and radio wavelengths to not generally trace the same $B$-field morphology in galaxies. The $\Psi_{B}$ profiles tend to be more ordered across all galactocentric radii in radio ($\zeta_{\rm{6cm}} = 0.93\pm0.03$) than in FIR ($\zeta_{\rm{154\mu m}} = 0.84\pm0.14$). For spiral galaxies, FIR $B$-fields are $2-75$\% more turbulent than the radio $B$-fields. For starburst galaxies, we find that FIR polarization is a better tracer of the $B$-fields along the galactic outflows than radio polarization. Our results suggest that the $B$-fields associated with dense, dusty, turbulent star-forming regions, those traced at FIR, are less ordered than warmer, less-dense regions, those traced at radio, of the interstellar medium. The FIR $B$-fields seem to be more sensitive to the activity of the star-forming regions and molecular clouds within a vertical height of few hundred pc in the disk of spiral galaxies than the radio $B$-fields.
\end{abstract}


\section{Introduction} \label{sec:INT}


\begin{figure*}[ht!]
\begin{center}
   \includegraphics[width=\textwidth, trim={0 29 0 85}, clip]{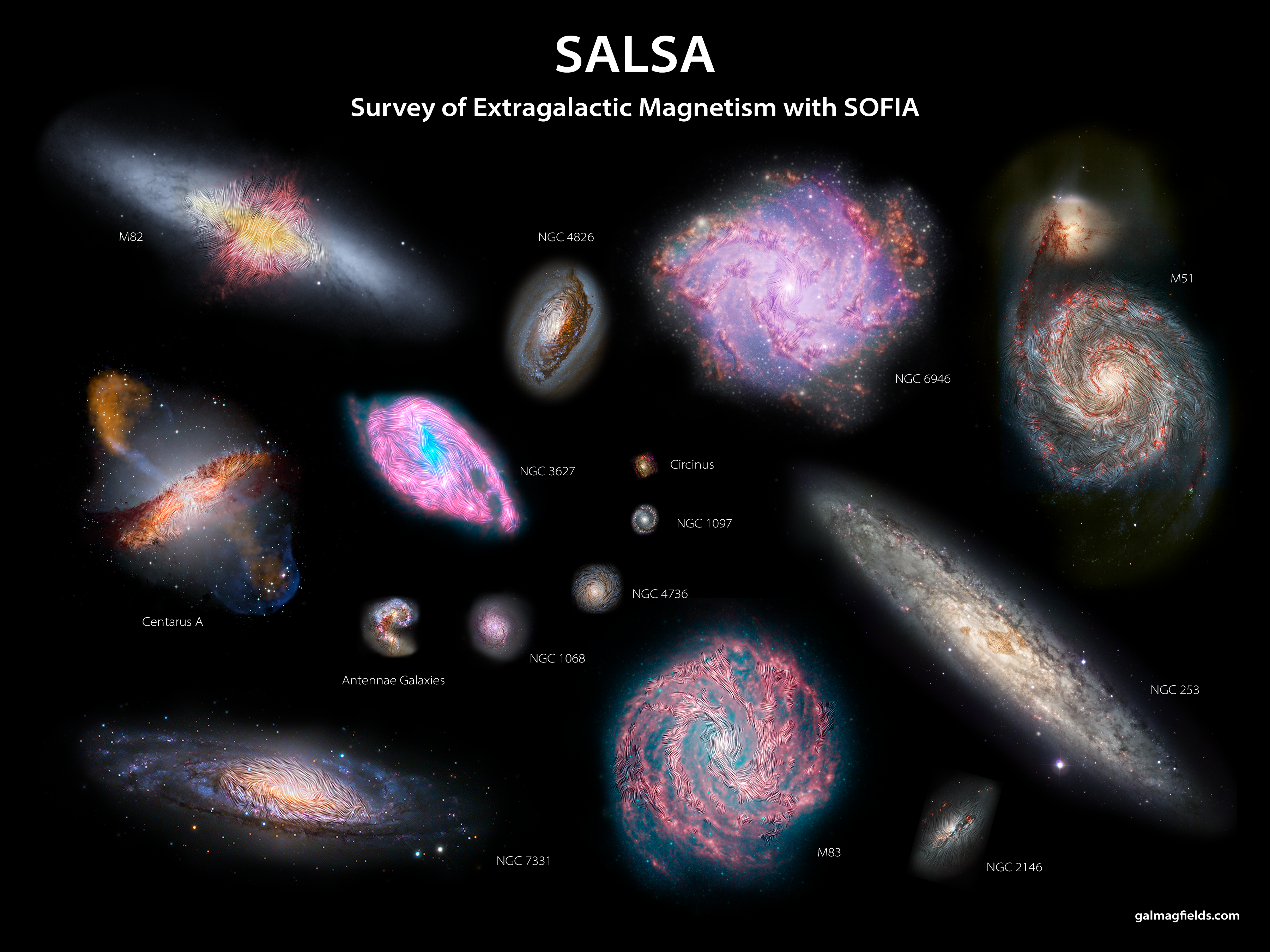}
\end{center}
\vspace{-0.5cm}
\caption{Far-infrared magnetic field orientations in the plane of the sky represented over the galaxies in the SALSA sample \citep[including the Antennae galaxies from][SALSA~VI]{SALSAVI} using the Line Integral Convolution \citep[LIC,][]{LIC} technique over archival background observations. References to the specific RGB background image of each object are detailed in Appendix \ref{App:RGBComponents}. Objects are shown in relative angular sizes based on SOFIA/HAWC+ limit detection radius. Note that this figure is only for illustrative purposes, and no analysis was derived from it.}
\label{fig:SALSA_poster}
\end{figure*}

Polarimetric observations allow the tracing of magnetic fields ($B$-field) in galaxies \citep[e.g.,][]{beck+2013inbook_641}. Synchrotron emission is expected to emerge largely from the warm and diffuse ISM, due to a combination of a large volume fraction for the ubiquitous cosmic ray electron population and thermal gas emission. While this emission is considerably sensitive to Faraday rotation, it should be relatively less affected by depolarization from disordered $B$-field structures due to larger coherence scales for the field and a prevalence of the mean magnetic field \citep{Evirgen2017}. The FIR polarization arises from thermal emission of magnetically aligned dust grains tracing a weighted density medium along the line-of-sight (LOS). Based on SOFIA/HAWC+ observations of extra-galactic objects, we estimate that this polarized FIR emission is associated to a relatively dense ($\log_{10}(N_{\rm~HI+H_{2}}~[\rm{cm}^{-2}])=[20,23]$) and cold ($T_{\rm~d}=[20,50]$ K) component of the ISM \citep[][SALSA~IV]{SALSAIV}. The addition of FIR polarimetric observations by the Survey of extragALactic magnetiSm with SOFIA (SALSA; PI: Lopez-Rodriguez, E. \& Mao, S. A.) Legacy Program \citep[][SALSA~IV]{SALSAIV} enables for the first time to explore the differences with radio polarimetric observations allowing us to investigate extragalactic $B$-fields in different ISM phases (Fig. \ref{fig:SALSA_poster}).

In fact, while the large-scale ordered structure of the $B$-field \footnote{\emph{Ordered fields} are defined to be what polarized FIR emission or synchrotron emission at short radio wavelengths (where Faraday depolarization is small) measures within the telescope beam, projected to the plane perpendicular to the line of sight. Contributions for ordered fields come from \emph{anisotropic random fields}, whose average over the beam vanishes, and \emph{regular fields}, whose average over the beam is finite. To distinguish between these contributions, additional Faraday rotation data in the radio 
range are needed \citep{beck+2013inbook_641}.} measured in radio and FIR are compatible in the less-dense interarm regions, significant divergence has been found in the spiral arms of M\,51 \citep[SALSA I:][]{Borlaff2021,Surgent2023arXiv230207278S}. This result may be due to the different thermodynamical properties between the warm diffuse ISM and the cold dust filaments associated with star-forming regions and molecular clouds. In the closest merger of two spiral galaxies (the Antennae galaxies), the FIR $B$-field closely follows the relic spiral arm, while the radio $B$-field is mostly radial \citep[][SALSA~VI]{SALSAVI}. This result suggests that the FIR $B$-field may be cospatial with the molecular gas in the midplane of the relic spiral arm, while the radio $B$-field seems to be more aligned with the dynamics of the HI gas on an extraplanar layer above the disk affected by the merger activity. In addition, the 1 kpc-scale starburst ring of NGC\,1097 has a compressed and constant FIR $B$-field at the intersection with the galactic bar, while the radio $B$-field has a spiral morphology tightly cospatial with the diffuse neutral gas \citep[][SALSA~II]{SALSAII}. To observe the $B$-field within the cold and dense ISM affected by the star formation activity and molecular clouds, FIR polarimetric observations are necessary.



In this paper, we explore the structure of the $B$-field in a quantitative approach probed by the FIR and radio for a sample of 14 galaxies with diverse morphologies (grand-design spirals, post-merger, starbursts). In addition, we will compare the results with the $B$-field structure measured using radio polarimetric observations for cases in which ancillary data are available (see Sec.\,\ref{subsec:ObsRadio}). Sec.\,\ref{subsec:BmapsFIR} is dedicated to the analysis of the $B$-field orientation maps. In Sec.\,\ref{subsec:PitchB} we quantitatively analyze the structure of the $B$-field pitch angle for spiral galaxies in the sample. Finally, Sec.\,\ref{subsec:Dtheta} introduces the structural parameter, $\zeta$, dedicated to quantify the ordering ratio of the spiral $B$-field. The discussion and conclusions are presented in Secs.\,\ref{sec:DIS} and \,\ref{sec:CON} respectively.

\section{Data}\label{subsec:ObsRed}

\subsection{Far-infrared polarimetric observations}\label{subsec:ObsFIR}

We analyze the $B$-field orientations of the  galaxies using the same FIR polarimetric observations as presented in SALSA~III and IV \citep{SALSAIV, SALSAIII}. The associated FIR polarimetry data are from the SALSA Legacy Program obtained with the High-Angular Wideband Camera Plus \citep[HAWC+;][]{Vaillancourt2007,Dowell2010,Harper2018} on board the 2.7-m Stratospheric Observatory For Infrared Astronomy (SOFIA). SOFIA/HAWC+ has an optimal combination of spatial resolution, field of view, and sensitivity for observational tests of the theoretical connections between $B$-field, turbulence, and star formation. FIR polarimetric observations ---practically impossible using ground-based observatories due to atmospheric absorption--- allow us to efficiently detect the thermal polarized emission of magnetically aligned dust grains. 

In summary, the FIR polarimetric observations were performed using four bands, $53$, $89$, $154$, and $214$~\um, with SOFIA/HAWC+. Table~\ref{tab:HAWC} shows the HAWC+ configuration for every band. The observations were performed using the on-the-fly-mapping (OTFMAP) polarimetric mode described in SALSA~III \citep{SALSAIII}. The exception is M\,51: this galaxy was observed using the chopping and nodding technique and reduced as described by \citet{Borlaff2021}. Table~\ref{tab:FIRRadioObs} shows the galaxy sample and the associated bands. The pixel scale of the final data product is equal to the detector pixel scale (i.e. Nyquist sampling) in any given band (Table \ref{tab:HAWC}). For a detailed description of the analysis of the polarization fraction see SALSA~IV \citep{SALSAIV}. Table \ref{tab:GalaxySample} shows the properties of the galaxy sample.

\begin{deluxetable}{cccccc}
\tablecaption{HAWC+ configuration. \emph{Columns, from left to right:} 
(a) Band name. 
(b) Central wavelength of the band in \um. 
(c) FWHM bandwidth in \um. 
(d) Pixel scale of the band in \arcsec. 
(e) FWHM in \arcsec. 
(f) FOV for polarimetric observations in \arcmin.
\label{tab:HAWC} 
}
\tablecolumns{6}
\tablewidth{0pt}
\tablehead{\colhead{Band}	&	\colhead{$\lambda_{c}$} &	\colhead{$\Delta \lambda$}	&	\colhead{Pixel scale}	&	\colhead{Beam size}	& \colhead{Polarimetry FOV}  \\ 
 	& \colhead{(\um)}	& \colhead{(\um)} &  \colhead{ (\arcsec)} & \colhead{(\arcsec)}  & \colhead{(\arcmin)} \\
\colhead{(a)} & \colhead{(b)} & \colhead{(c)} & \colhead{(d)} & \colhead{(e)} & \colhead{(f)} 
}
\startdata
A	& $53$	&	$8.7$&	$2.55$	&	$4.85$	&	$1.4\times1.7$ \\
C	& $89$	&	$17$	&	$4.02$	&	$7.8$	&	$2.1\times2.7$	\\
D	& $154$	&	$34$	&	$6.90$	&	$13.6$	&	$3.7\times4.6$	\\
E	& $214$	&	$44$	&	$9.37$ 	&	$18.2$	&	$4.2\times6.2$  \\	
\enddata
\end{deluxetable}

\subsection{Radio polarimetric observations}\label{subsec:ObsRadio}


Ancillary radio polarimetric observations at $3$ and $6$~cm are used for the analysis. Table \ref{tab:FIRRadioObs} shows the radio polarimetric data that were publicly available and kindly shared by the authors of the associated manuscripts. These datasets were obtained using observations of the Effelsberg 100-m single-dish radio-telescope, in combination with datasets from the Karl G. Jansky Very Large Array (VLA) when available. The beam sizes for the radio polarization vary from target to target between 6\arcsec\ to 15\arcsec. The physical beam sizes (in pc) of the original available  datasets are described in Table \ref{tab:FIRRadioObs}. We refer to the original manuscripts shown in Table \ref{tab:FIRRadioObs} for a complete description of the observations and analysis of the datasets used in our work. Longer wavelength ($18$, $20$ cm) observations can be strongly affected by Faraday rotation \citep{Beck2019}, and are thus not considered in this work. 

For each galaxy and radio wavelength, the Stokes $IQU$ were convolved with a Gaussian kernel to match the beam size (full-width-half-maximum; FWHM) of HAWC+ at the bands shown in Table \ref{tab:FIRRadioObs}. Then, we reprojected each dataset to the largest beam size per galaxy and per band, matching the same pixel scale and field-of-view (FOV). Finally, the fractional degree ($P$) and angle of polarization ($PA$), and polarized flux ($PI$) were computed, accounting for the level of polarization bias. To avoid biased results due to the number of measurements across the galaxy, we only use radio polarization measurements that are spatially coincident with the HAWC+ observations. The radio polarization maps are used to spatially correlate the polarization arising from synchrotron emission with that arising from thermal emission by means of magnetically aligned dust grains observed with HAWC+. Throughout this manuscript, the results at $3$ and $6$~cm are very similar, within their uncertainties.

\begin{deluxetable*}{lccccr}
\tablecaption{FIR and radio polarimetric data. \emph{Columns, from left to right:} 
(a) Galaxy name.
(b) Central wavelength of the HAWC+ bands. 
(c) Physical scale of the beam at each HAWC+ band,
(d) Central wavelength of the band at radio wavelengths.
(e) Physical scale of the beam at each radio band,
(f) References of the radio polarimetric observations.}
\label{tab:FIRRadioObs} 
\tablecolumns{6}
\tablewidth{0pt}
\tablehead{\colhead{Galaxy} & 	\colhead{FIR}  & \colhead{$\theta_{\rm{FIR,beam,pc}}$} & \colhead{Radio} & \colhead{$\theta_{\rm{radio,beam,pc}}$} &  \colhead{References} \\ 
 					   &  \colhead{(\um)} &  \colhead{(pc)}	& \colhead{(cm)}	& \colhead{(pc)} \\
\colhead{(a)} & \colhead{(b)} & \colhead{(c)} & \colhead{(d)} & \colhead{(e)} & \colhead{(f)}} 
\startdata
Centaurus A 	&	89				& 128          &  -  &   -	&	-\\
Circinus 		&	53, 89, 214		& 98, 157, 367     & -  &	-	&	-\\
M~51 			&	154				& 566           &	3, 6	& 333 & \citet{Fletcher2011}	\\
M~82 			&	53, 89, 154, 214	& 91, 146, 254, 340 &	3 + 6 	& 112	& \citet{Adebahr2013,Adebahr2017}	\\
M~83 			&	154				& 307 &	6 	& 226  & 	\citet{Frick2016}	\\
NGC~253 	&	89, 154			& 132, 231 &	3, 6	& 119 &	\citet{Heesen2011}	\\
NGC~1068 	&	53, 89			& 335, 539 &	- 	&	- &  - \\
NGC~1097 	&	89, 154			& 722, 1259 &	3, 6	& 556	& \citet{Beck2005}	\\
NGC~2146 	&	53, 89, 154, 214	& 400, 644, 1123, 1504 &	- 	&	- &  - \\
NGC~3627 	&	154				& 587 &	3, 6 & 561	& \citet{Soida2001}	\\
NGC~4736 	&	154				& 349 &	3, 6 &	206, 385 & \citet{Chyzy2008}	\\
NGC~4826 	&	89				& 210 &	-	&	- &  - \\
NGC~6946 	&	154 				& 448 &	6 	& 495	& \citet{Beck1991,Beck2007}\\
NGC~7331 	&	154				& 1025 &	- 	&	- & - \\	
\enddata
\end{deluxetable*}

\begin{deluxetable*}{lcccccl}
\tablecaption{Galaxy Sample. \emph{Columns, from left to right:} 
(a) Galaxy classification and name. $\circledcirc$: face-on spiral galaxies with a spiral $B$-field pattern, $\divideontimes$: starburst and/or edge-on galaxies without a clear spiral pattern. 
(b) Galaxy distance in Mpc. 
(c) Physical scale in pc per arcsec. 
(d) Galaxy type. 
(e) Inclination of the galaxy in degrees (face-on $i=0^{\circ}$, edge-on $i=90^{\circ}$. 
(f) Position angle of the large axis of the galaxy in the plane of the sky. 
(g) References associated to the distance, inclination and tilt angles. 
\label{tab:GalaxySample} 
}
\tablecolumns{6}
\tablewidth{0pt}
\tablehead{\colhead{Galaxy} & 	\colhead{Distance$^{1}$}  & \colhead{Scale} & \colhead{Type$^{\star}$} & 
\colhead{Inclination (i)$^{2}$} &	\colhead{Tilt ($\theta$)$^{2}$} &  \colhead{References} \\ 
 	&  \colhead{(Mpc)}	& \colhead{(pc/\arcsec)}	&
\colhead{($^{\circ}$)} & \colhead{($^{\circ}$)} & \colhead{($^{\circ}$)} \\
\colhead{(a)} & \colhead{(b)} & \colhead{(c)} & \colhead{(d)} & \colhead{(e)} & \colhead{(f)} & \colhead{(g)}} 
\startdata
$\divideontimes$ Centaurus A 	&	$3.42$	&	$16.42$	&	S0pec/Sy2/RG	&	$83\pm6$		&	$114\pm4$	&	
$^{1}$\citet{Ferrarese2007}; $^{2}$\citet{Quillen2010}	\\
$\divideontimes$ Circinus 		&	$4.20$	&	$20.17$	&	SA(s)b/Sy2	&	$40\pm10$	&	$205\pm10$	&	
$^{1}$\citet{Tully2009}; $^{2}$\citet{Jones1999}			\\
$\circledcirc$ M~51 		&	$8.58$	&	$41.21$	&	Sa			&	$22.5\pm5$	&	$-7\pm3$		&
$^{1}$\citet{McQuinn2017}; $^{2}$\citet{Colombo2014}	\\
$\divideontimes$ M~82 		&	$3.85$	&	$18.49$	&	I0/Sbrst		&	$76\pm1$		&	$64\pm1$		&
$^{1}$\citet{Vacca2015}; $^{2}$\citet{Mayya2005}		\\
$\circledcirc$ M~83 		&	$4.66$	&	$22.38$	&	SAB(s)c		&	$25\pm5$		&	$226\pm5$	&
$^{1}$\citet{Tully2013}; $^{2}$\citet{Crosthwaite2002}	\\
$\divideontimes$ NGC~253 	&	$3.50$	&	$16.81$	&	SAB(s)c/Sbrst	&	$78.3\pm1.0$	&	$52\pm1$		&
$^{1}$\citet{RS2011}; $^{2}$\citet{Lucero2015}		\\
$\circledcirc$ NGC~1068 	&	$14.40$	&	$69.16$	&	(R)SA(rs)b/Sy2	&	$40\pm3$		&	$286\pm5$	&
$^{1}$\citet{BH1997}; $^{b}$\citet{Brinks1997}			\\
$\circledcirc$ NGC~1097 	&	$19.10$	&	$92.21$	&	SB(s)b/Sy1	&	$41.7\pm0.6$	&	$133.0\pm0.1$	&
$^{1}$\citet{Willick1997}; $^{2}$\citet{Hsieh2011} 		\\
$\divideontimes$ NGC~2146 	&	$17.20$	&	$82.61$	&	SB(s)ab/Sbrst	&	$63\pm2$		&	$140\pm2$	&
$^{1}$\citet{Tully1988}; $^{2}$\citet{Tarchi2004} 		\\
$\circledcirc$ NGC~3627 	&	$8.90$	&	$42.75$	&	SAB(s)b		&	$52\pm1$		&	$176\pm1$	&
$^{1}$\citet{Kennicutt2003}; $^{2}$\citet{Kuno2007} 		\\
$\circledcirc$ NGC~4736 	&	$5.3$	&	$25.46$	&	SA(r)ab		&	$36\pm7$		&	$292\pm2$	&	
$^{1}$\citet{Kennicutt2003}; $^{2}$\citet{Dicaire2008} 		\\
$\circledcirc$ NGC~4826 	&	$5.60$	&	$26.89$	&	(R)SA(rs)ab	&	$65\pm5$		&	$125\pm5$	&
$^{1}$\citet{Kennicutt2003}; $^{2}$\citet{Braun1994} 		\\
$\circledcirc$ NGC~6946 	&	$6.80$	&	$32.66$	&	Sc			&	$38.4\pm3.0$	&	$239\pm1$	&
$^{1}$\citet{Karachentsev2000}; $^{2}$\citet{Daigle2006} \\
$\circledcirc$ NGC~7331 	&	$15.7$	&	$75.40$	&	SA(s)b		&	$78.1\pm2.7$	&	$165\pm1.2$	&
$^{1}$\citet{Kennicutt2003}; $^{2}$\citet{Daigle2006}		\\	
\enddata
\tablenotetext{{\star}}{Galaxy type from NASA/IPAC Extragalactic Database (NED; \url{https://ned.ipac.caltech.edu/})}
\end{deluxetable*}


\section{Magnetic field orientation maps}\label{subsec:BmapsFIR}

\subsection{Methodology}
\label{subsec:BmapsFIR_methods}

\begin{figure}[t!]
\begin{center}
\includegraphics[width=0.49\textwidth, trim=0 0 0 0]{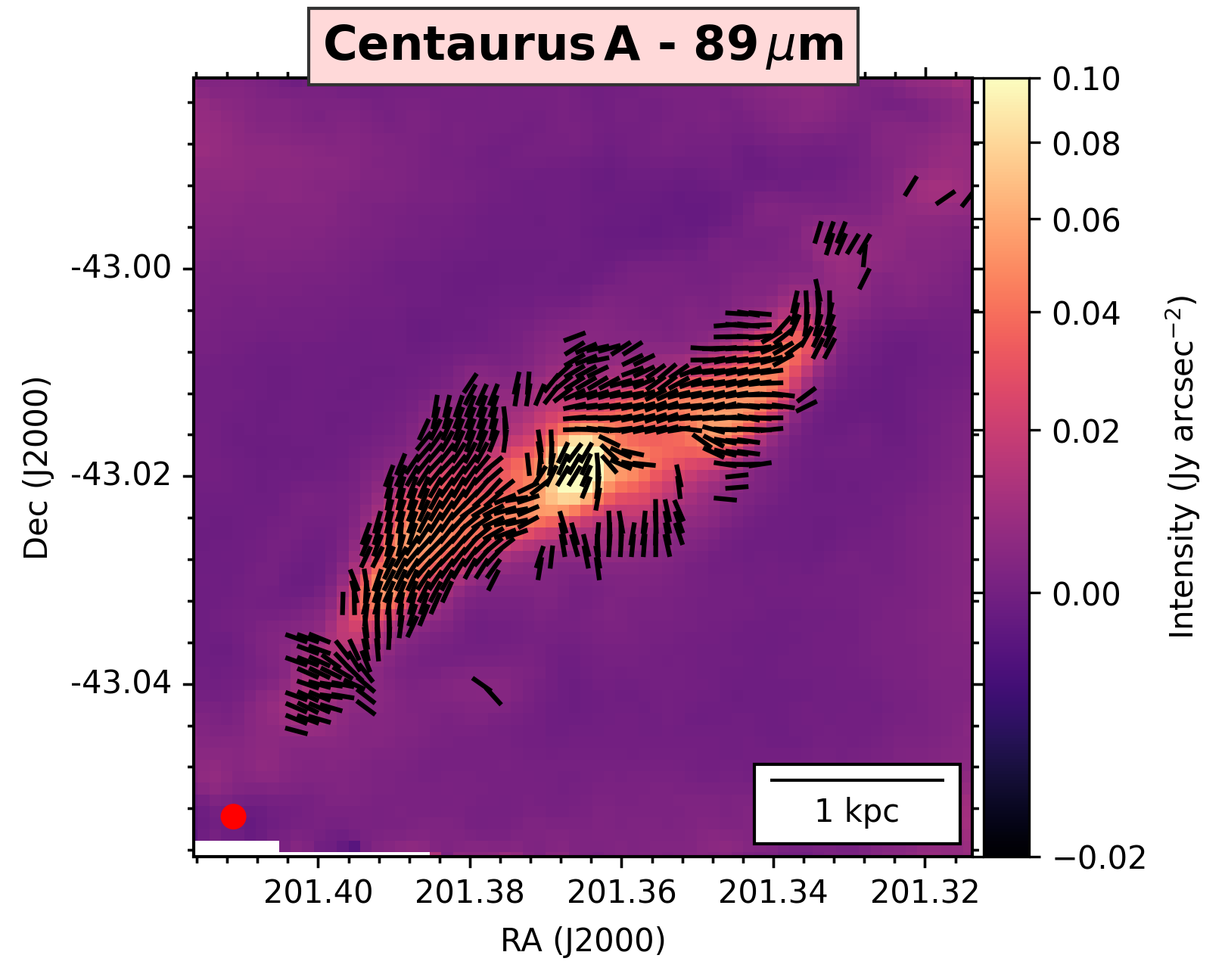}
\end{center}

\caption{FIR $B$-field orientation maps of Centaurus A. The colorscale shows the surface brightness in FIR (89 \um). Polarization measurements have been set to have a constant length to show the inferred $B$-field orientation (black dashes). Polarization measurements with $PI/\sigma_{PI} \ge 2.5$, $P\le20$\%, and $I/\sigma_{I} \ge 10$ were selected.
\label{fig:FIR_morphology_CenA}}
\end{figure}

\begin{figure*}[ht!]
\begin{center}
   \includegraphics[width=0.324\textwidth, trim=0 0 0 0]{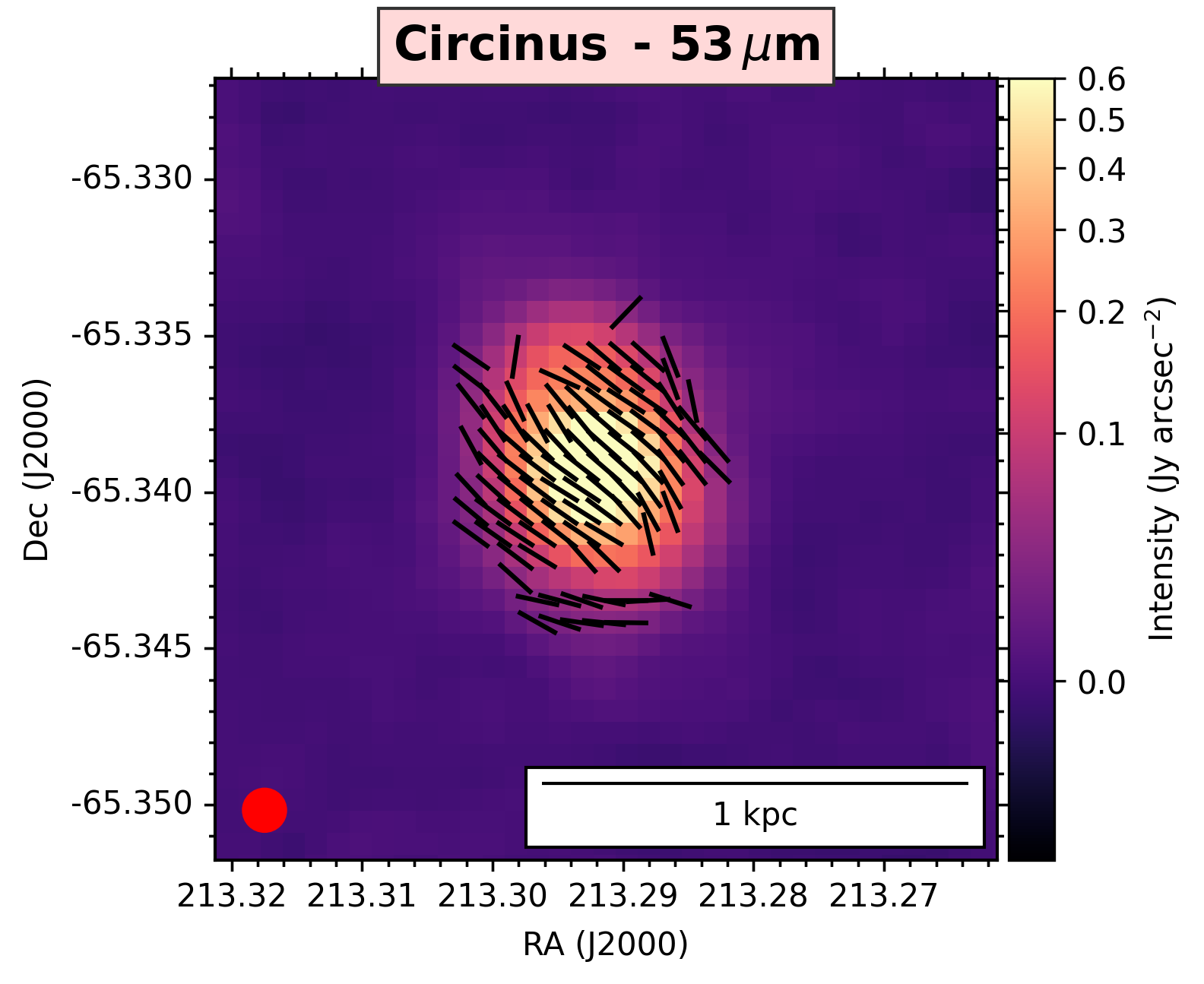}
   \includegraphics[width=0.324\textwidth, trim=0 0 0 0]{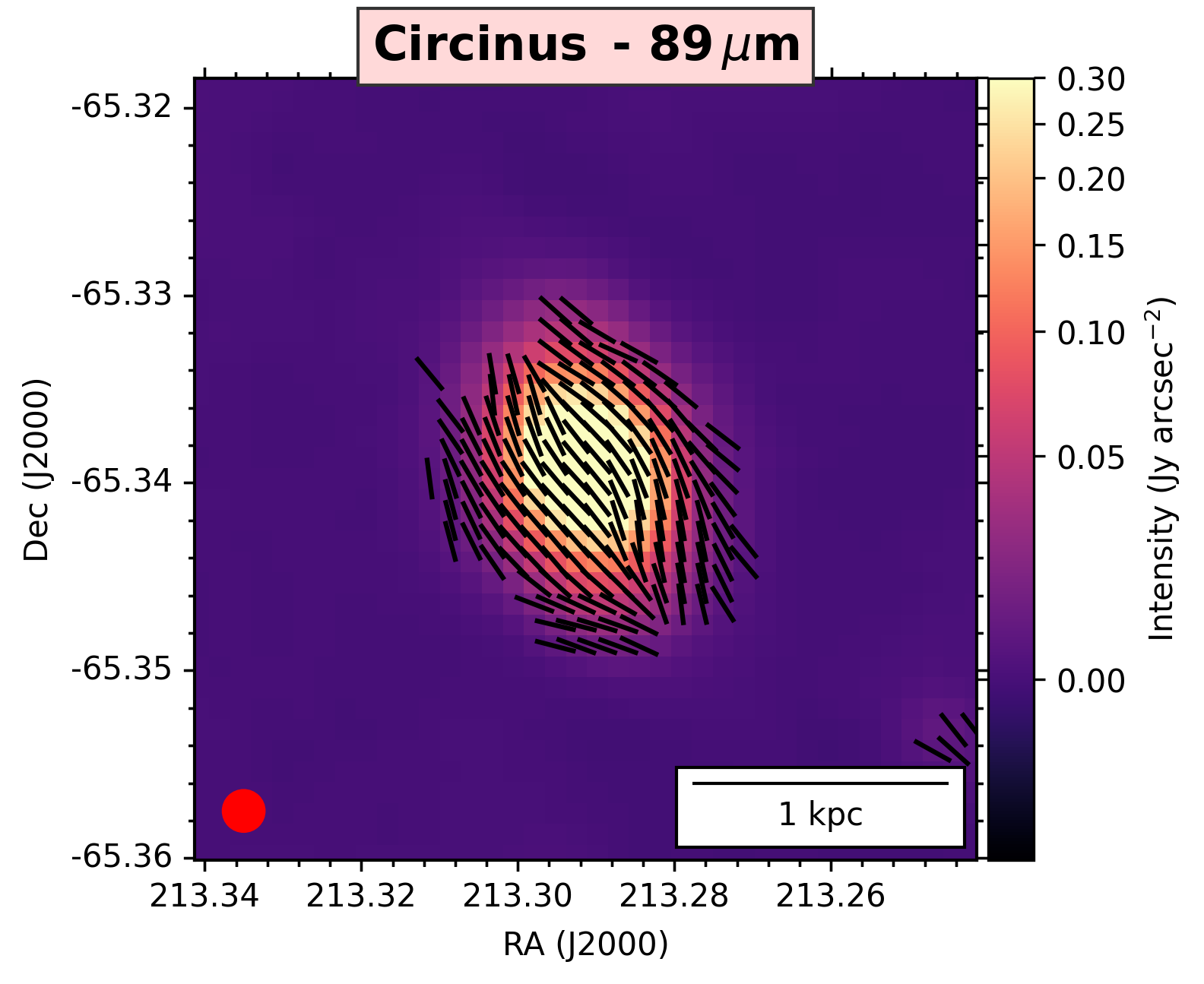}
   \includegraphics[width=0.337\textwidth, trim=0 0 0 0]{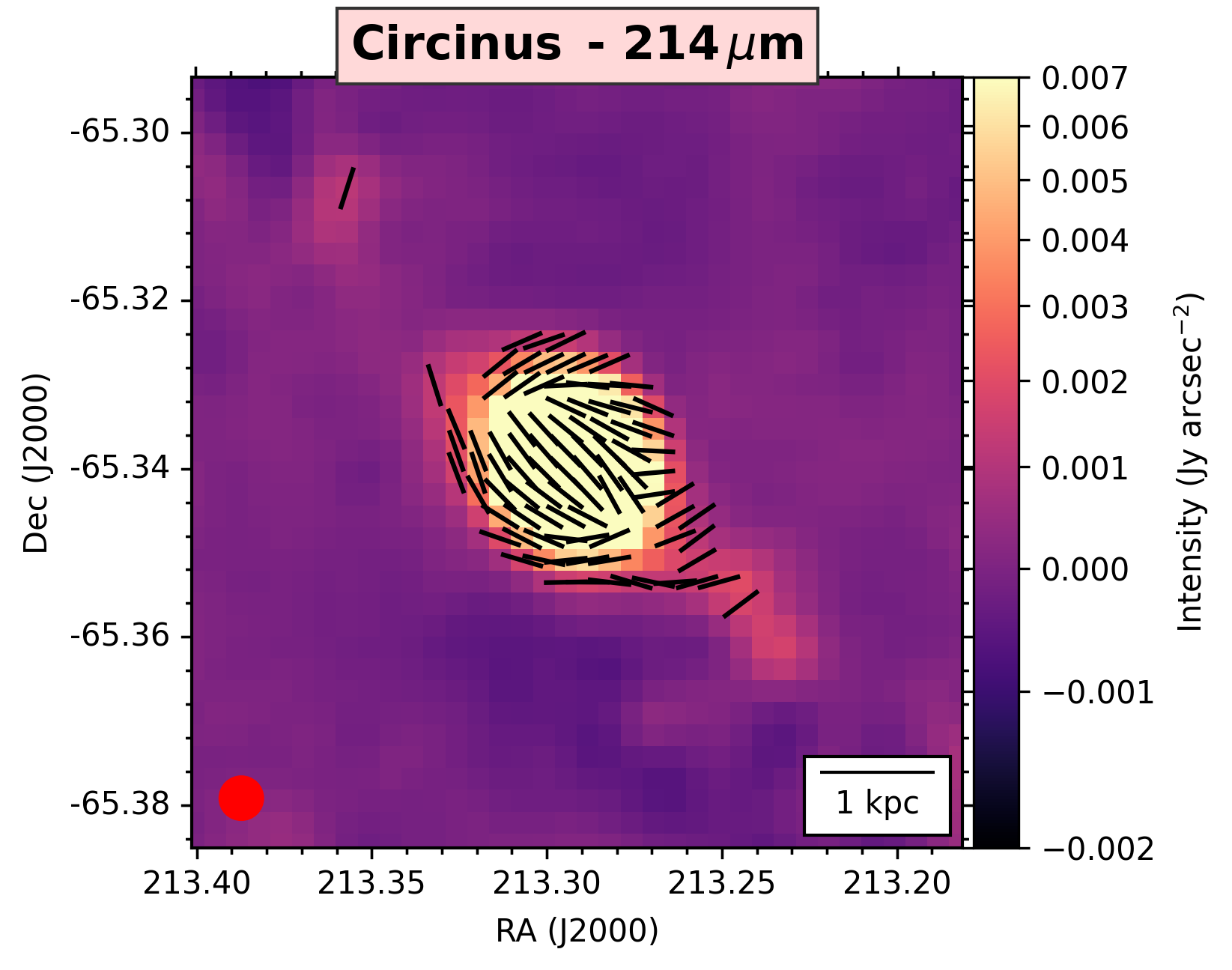}
\end{center}
\caption{FIR $B$-field orientation maps of Circinus (53, 89, and 214 \um, from left to right). The colorscale shows the surface brightness in each band. Polarization measurements have been set to have a constant length to show the inferred $B$-field orientation (black lines). We refer to Table~\ref{tab:PPA} for the quality cuts in $PI/\sigma_{PI}$, $P$, and $I/\sigma_{I}$ for each band. Note that the FOV is different for each panel, in order to improve the visualization of the individual polarization measurements.
\label{fig:FIR_morphology_Circinus}}
\end{figure*}

\begin{figure*}[ht!]
\begin{center}
   \includegraphics[width=0.50\textwidth, trim=0 0 0 0]{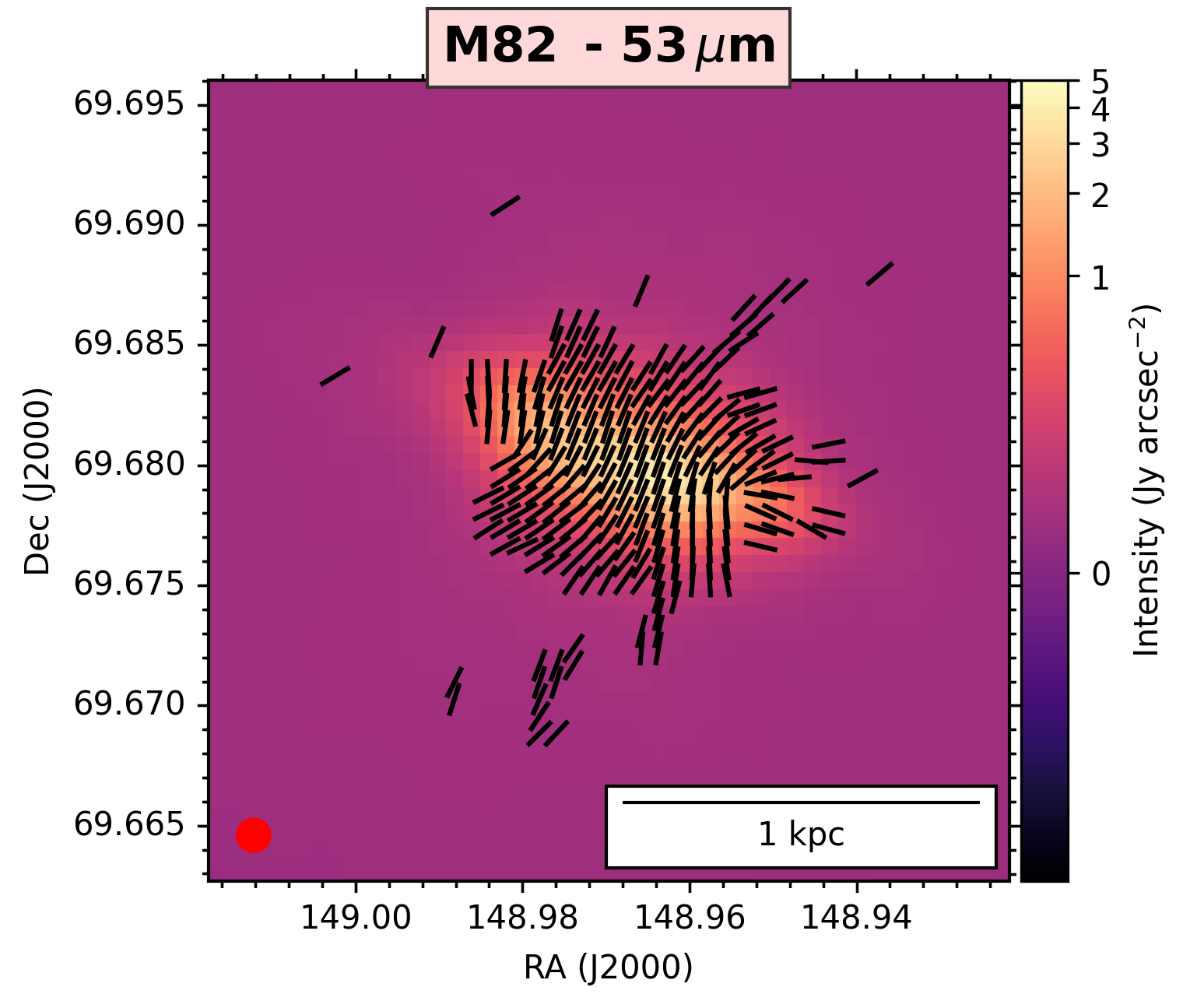}
   \includegraphics[width=0.49\textwidth, trim=0 0 0 0]{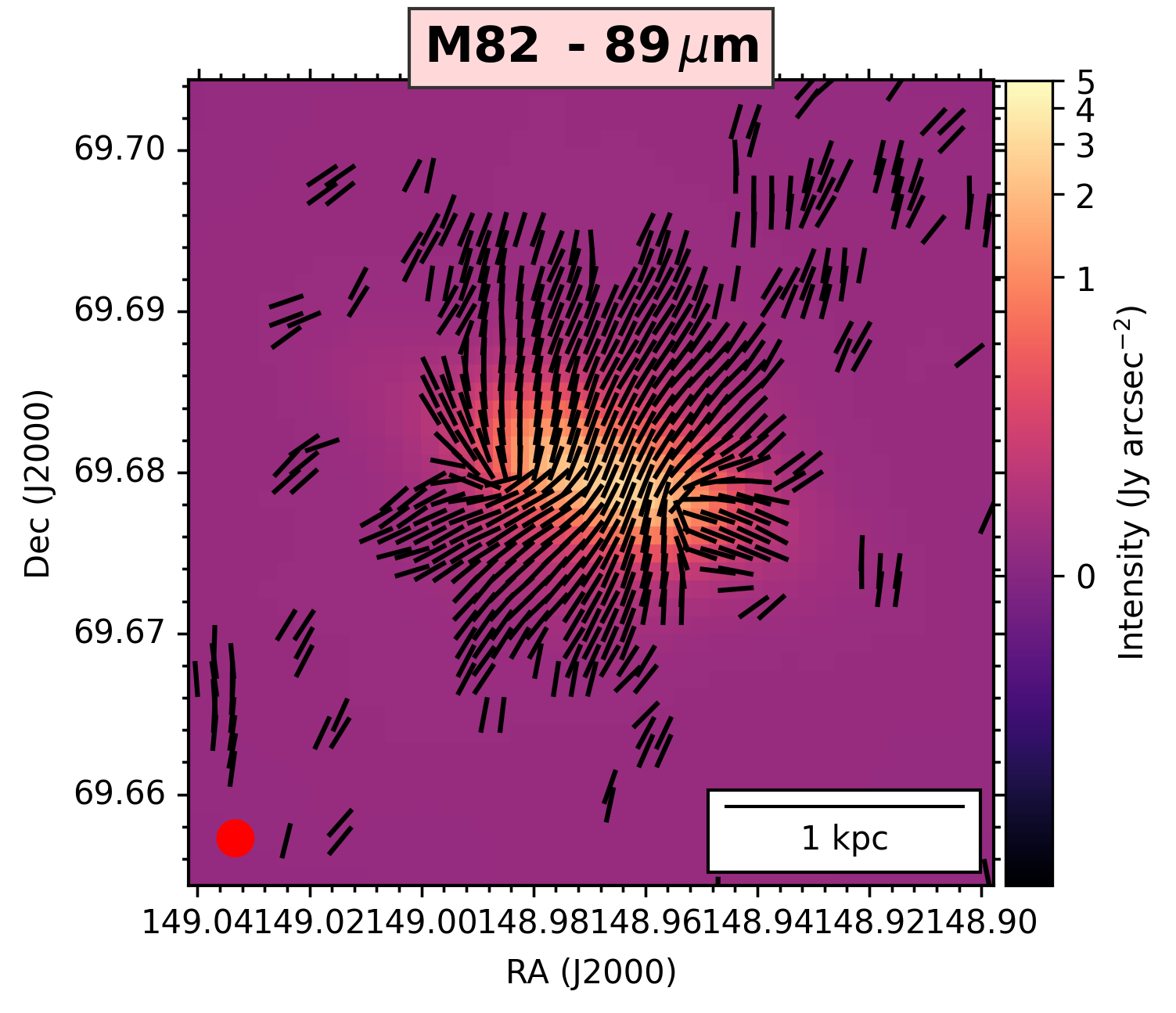}
\end{center}
   
  \begin{center}
   \includegraphics[width=0.485\textwidth, trim=0 0 0 0]{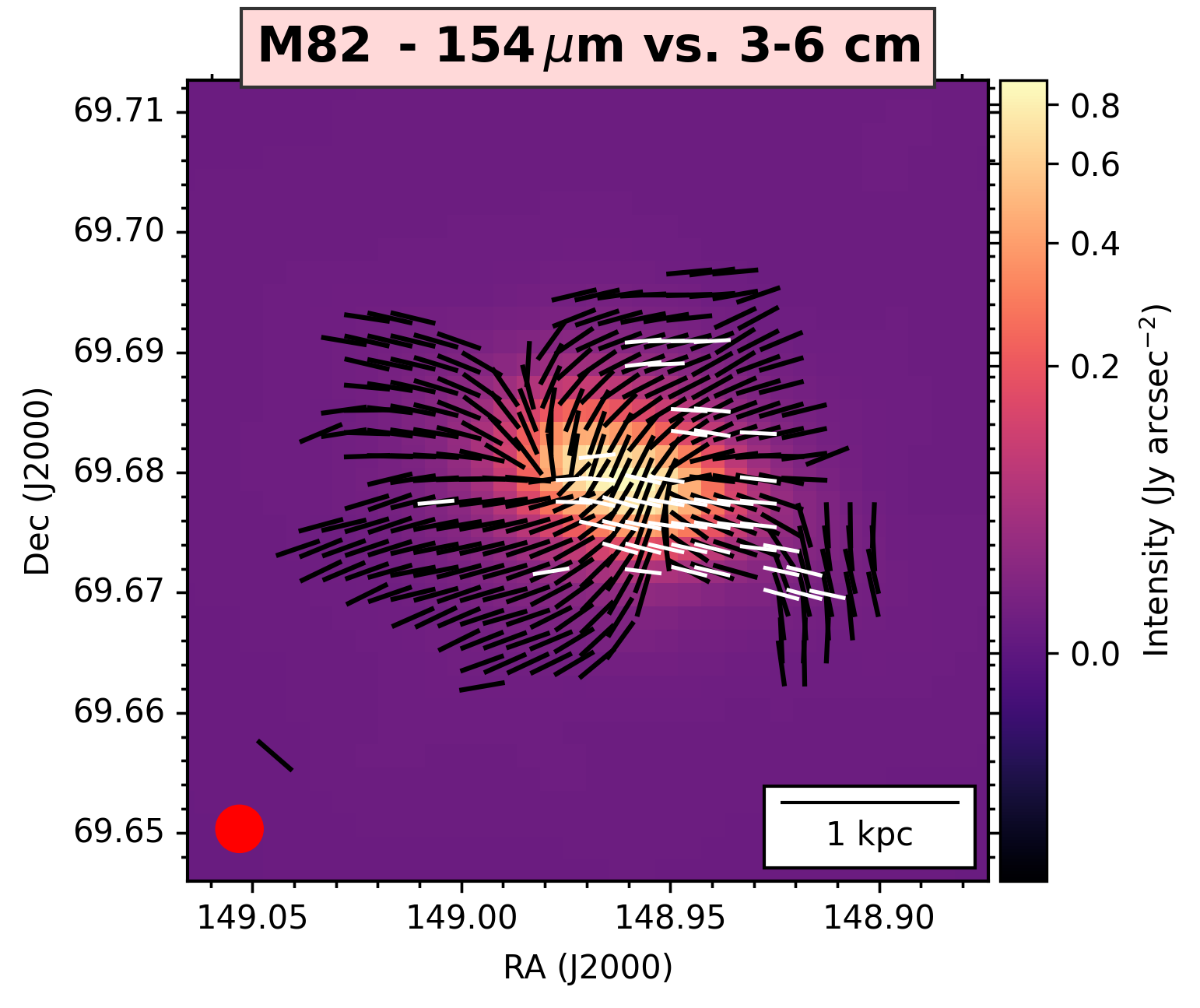}
   \includegraphics[width=0.505\textwidth, trim=0 0 0 0]{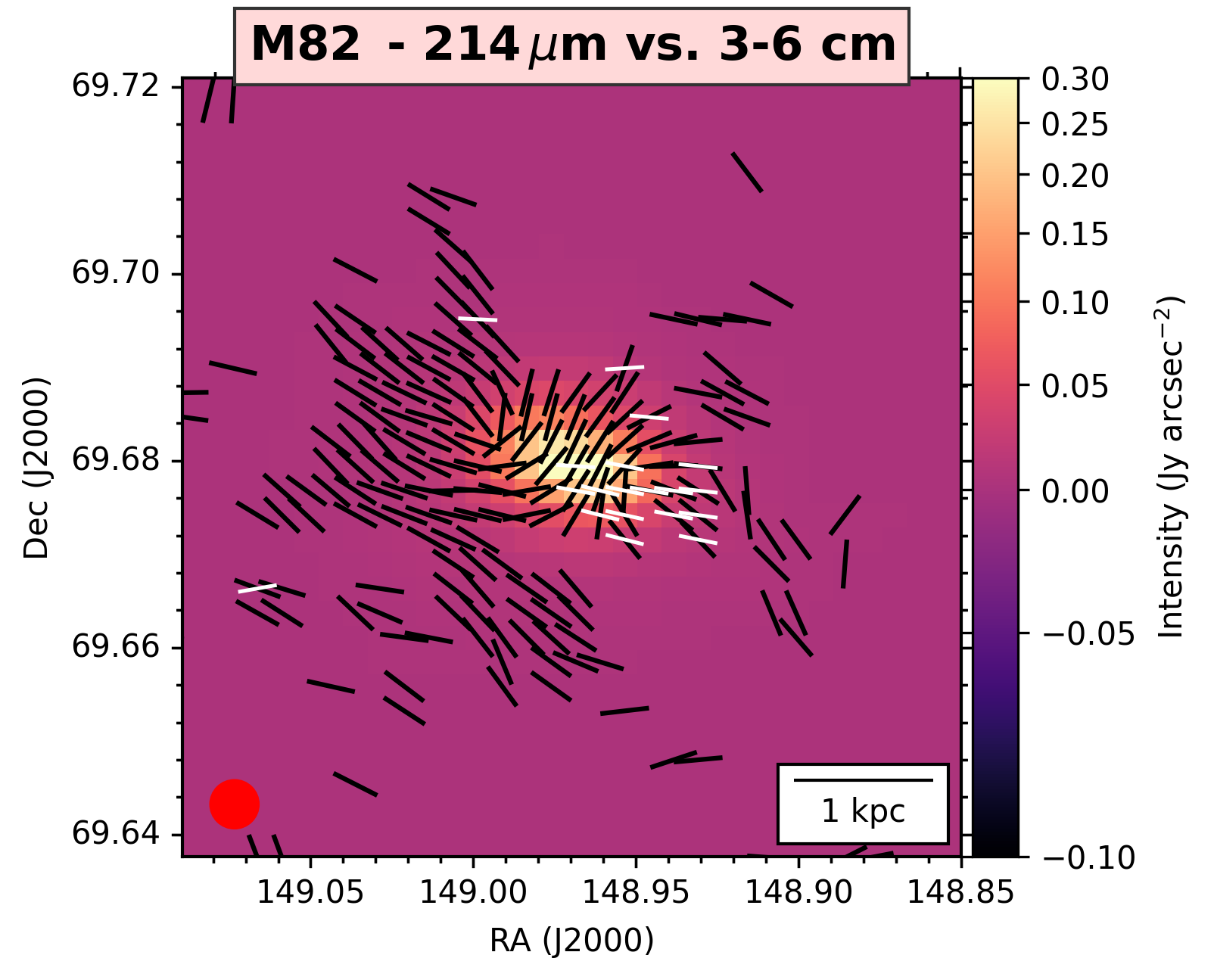}
\end{center}

\caption{FIR $B$-field orientation maps of M~82 (53, 89, 154, and 214 \um, from left to right, top to bottom). The maps in 154\,\um~and 214\,\um~show the radio polarimetric observations from \citet{Adebahr2013,Adebahr2017} in 3 and 6 cm (averaged). The colorscale shows the surface brightness in each band. Polarization measurements have been set to have a constant length to show the inferred $B$-field orientation (black lines), while the radio polarization observations are represented with white lines. We refer to Table~\ref{tab:PPA} for the quality cuts in $PI/\sigma_{PI}$, $P$, and $I/\sigma_{I}$ for each band. Note that the FOV is different for each panel, in order to improve the visualization of the individual polarization measurements.
\label{fig:FIR_morphology_M82}}
\end{figure*}

\begin{figure*}[ht!]
\begin{center}
   \includegraphics[width=0.49\textwidth, trim=0 0 0 0]{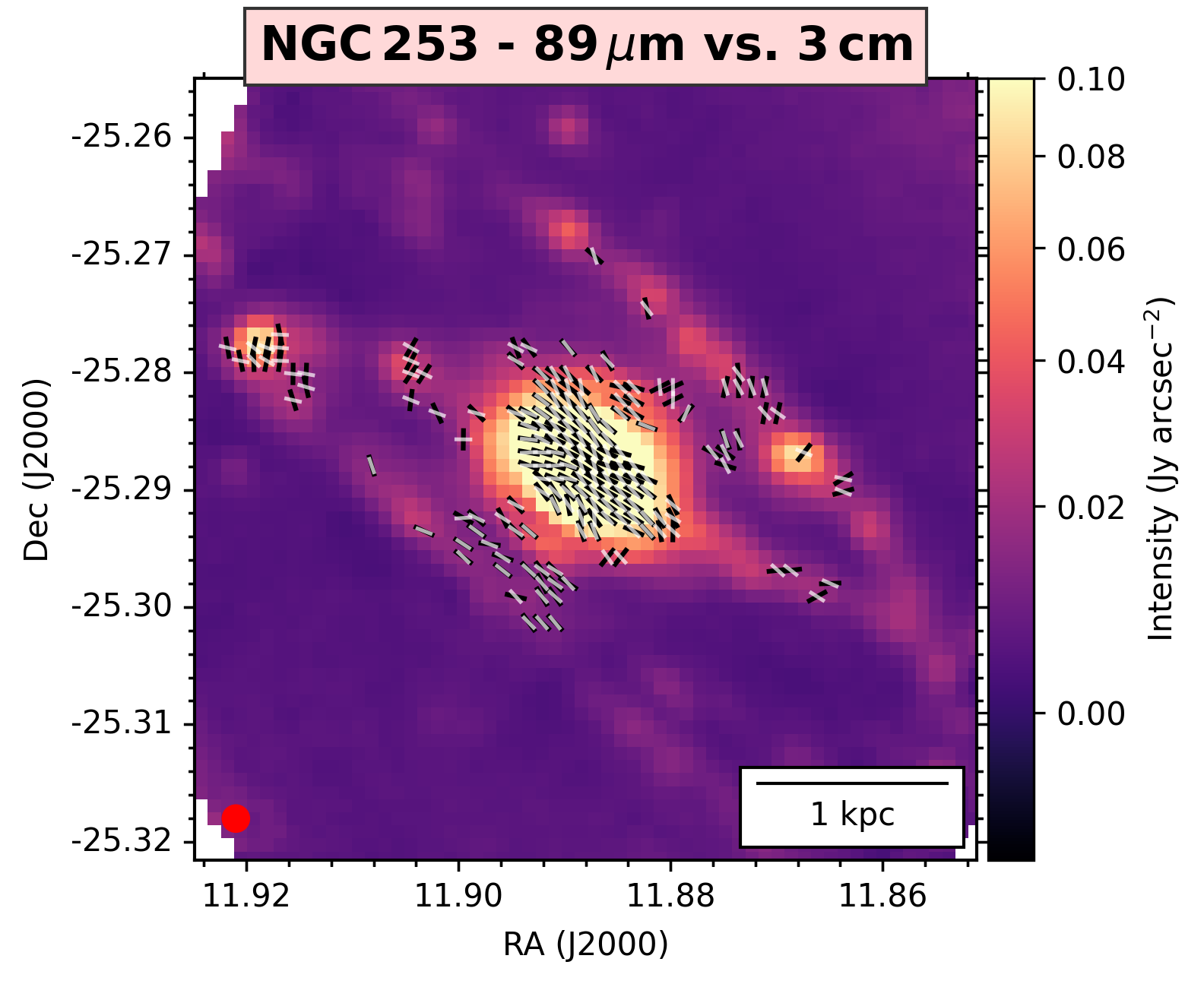}
\includegraphics[width=0.49\textwidth, trim=0 0 0 0]{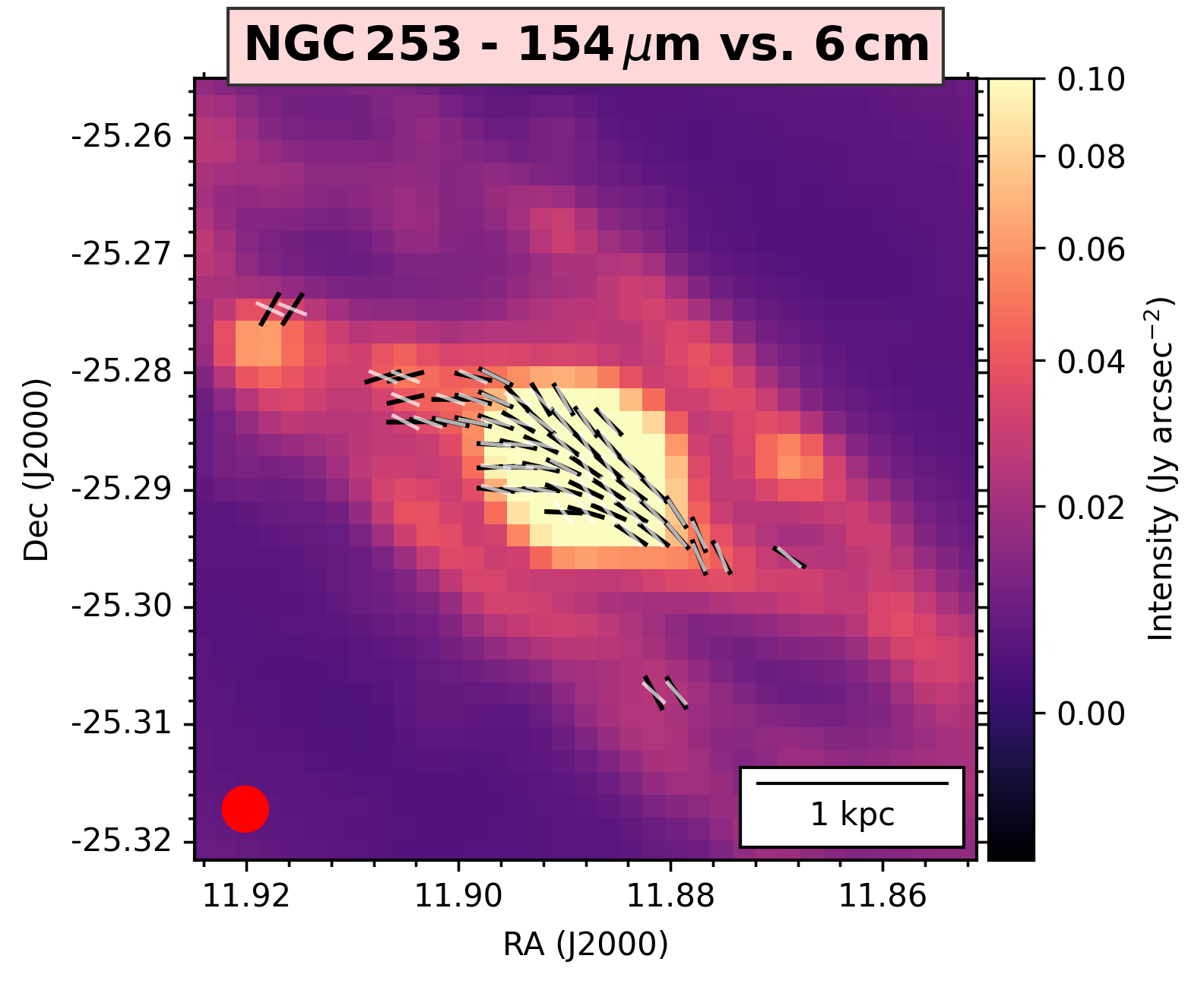}
\end{center}

\caption{FIR and radio $B$-field orientation maps of NGC\,253 (89 \um\ vs. 3 cm and 154 \um\ vs. 6 cm, from left to right). The FIR $B$-field is represented with black lines, while the radio polarization observations \citep{Heesen2011} are represented with white lines. The colorscale shows the surface brightness in each band. The polarization vectors have been set to a constant length to show the inferred $B$-field orientation. We refer to Table~\ref{tab:PPA} for the quality cuts in $PI/\sigma_{PI}$, $P$, and $I/\sigma_{I}$ for each band.
\label{fig:FIR_morphology_NGC253}}
\end{figure*}

\begin{figure*}[ht!]
\begin{center}
\includegraphics[width=0.49\textwidth, trim=0 0 0 0]{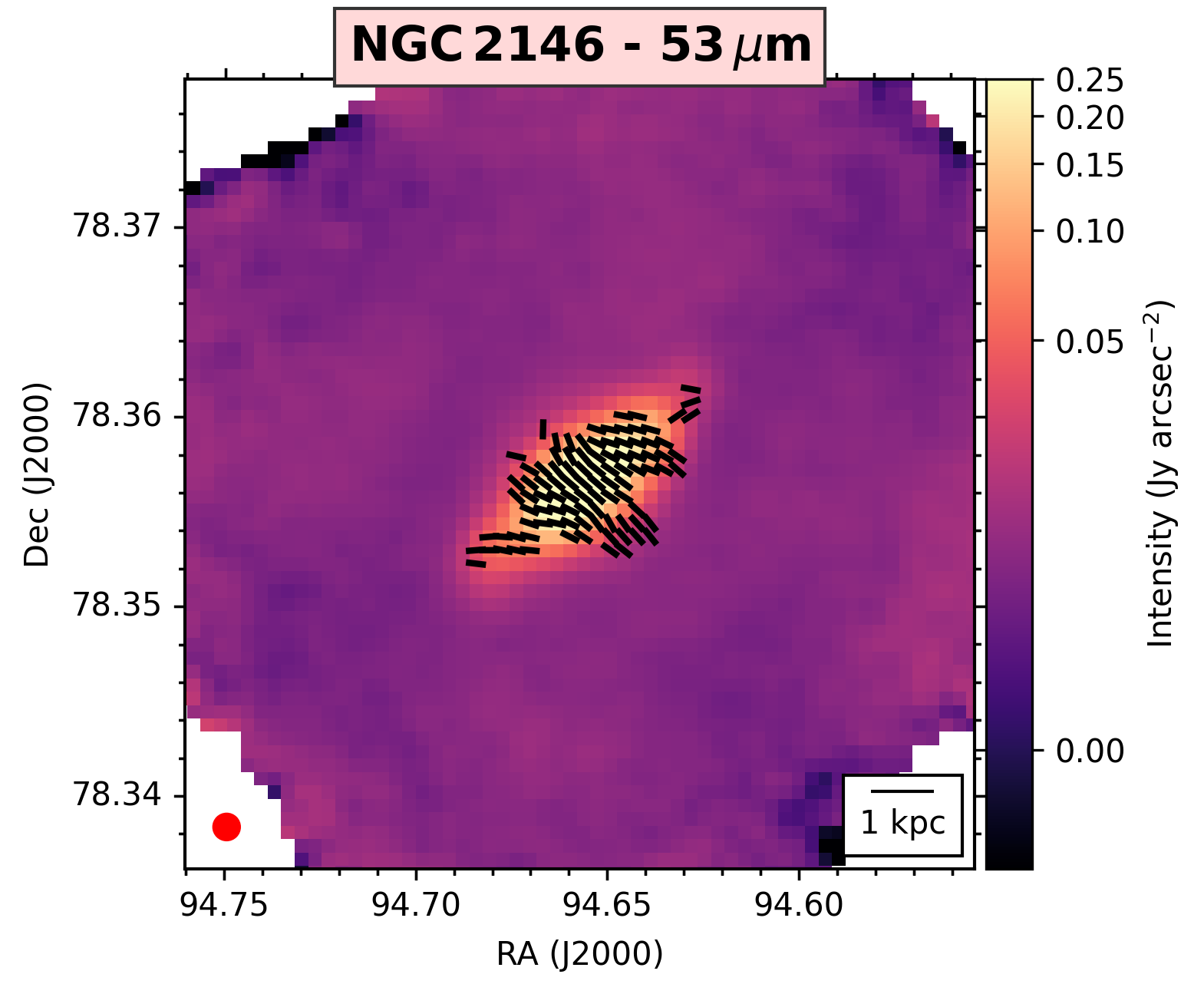}
\includegraphics[width=0.49\textwidth, trim=0 0 0 0]{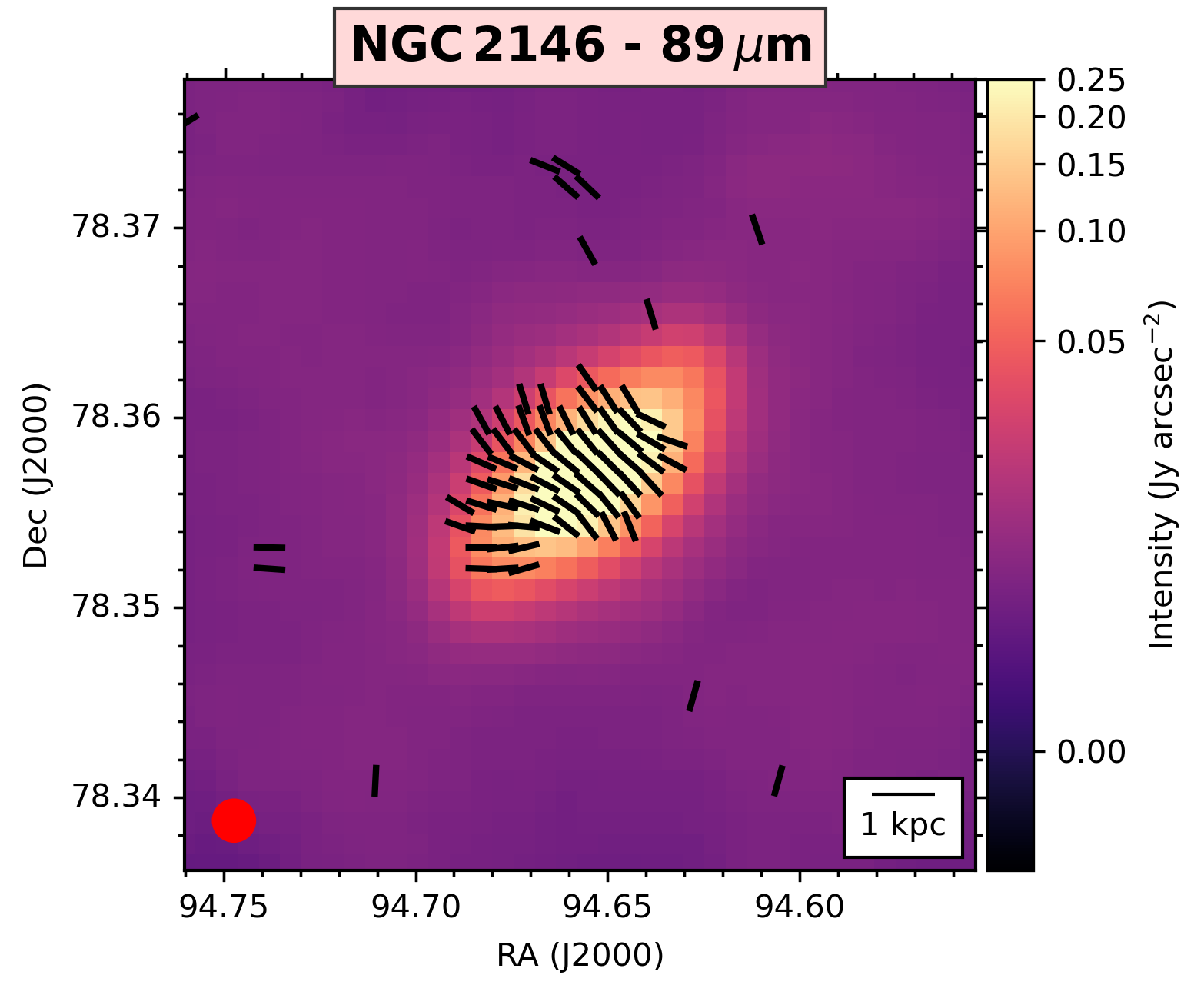}
\end{center}

\begin{center}
\includegraphics[width=0.49\textwidth, trim=0 0 0 0]{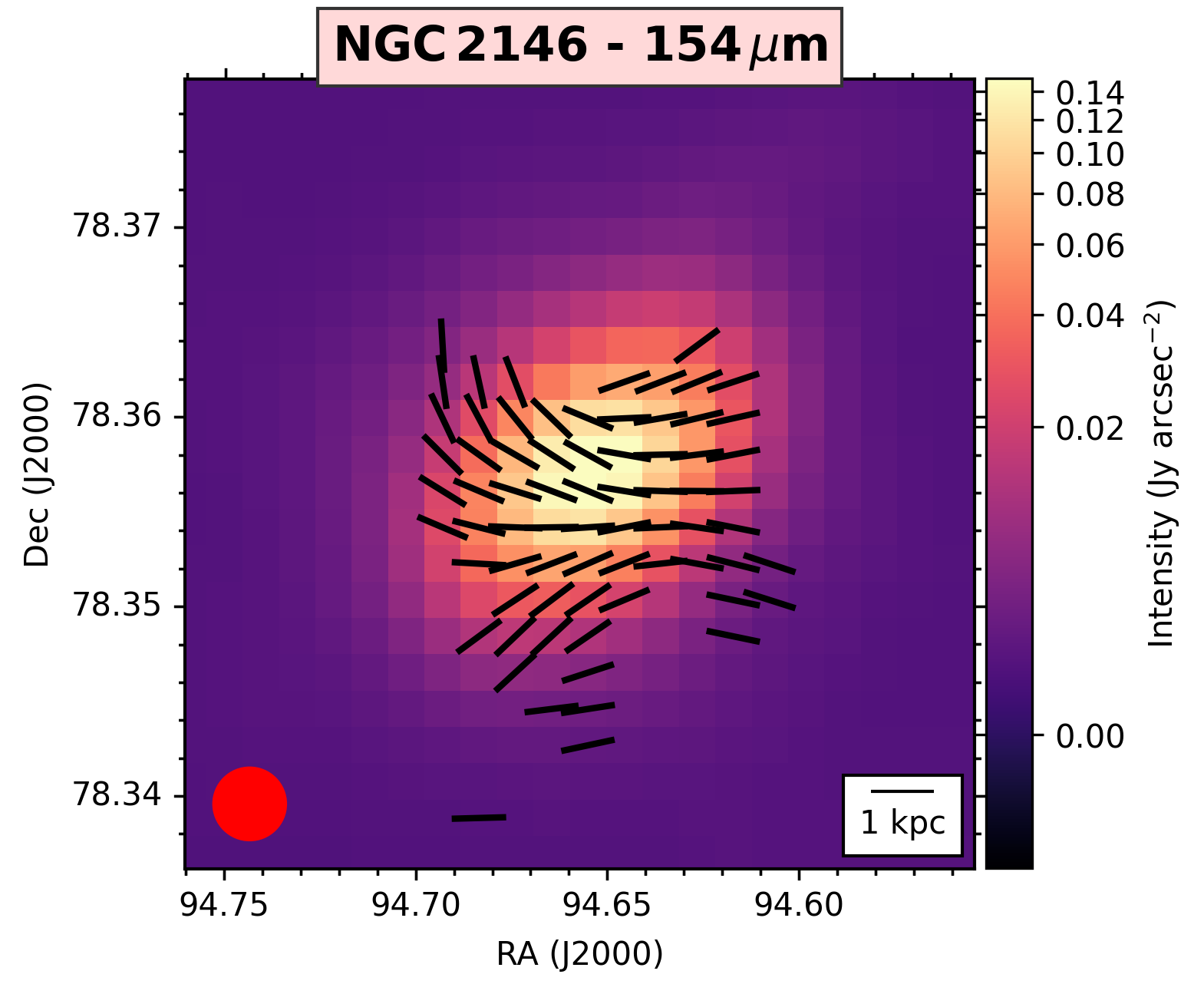}
\includegraphics[width=0.49\textwidth, trim=0 0 0 0]{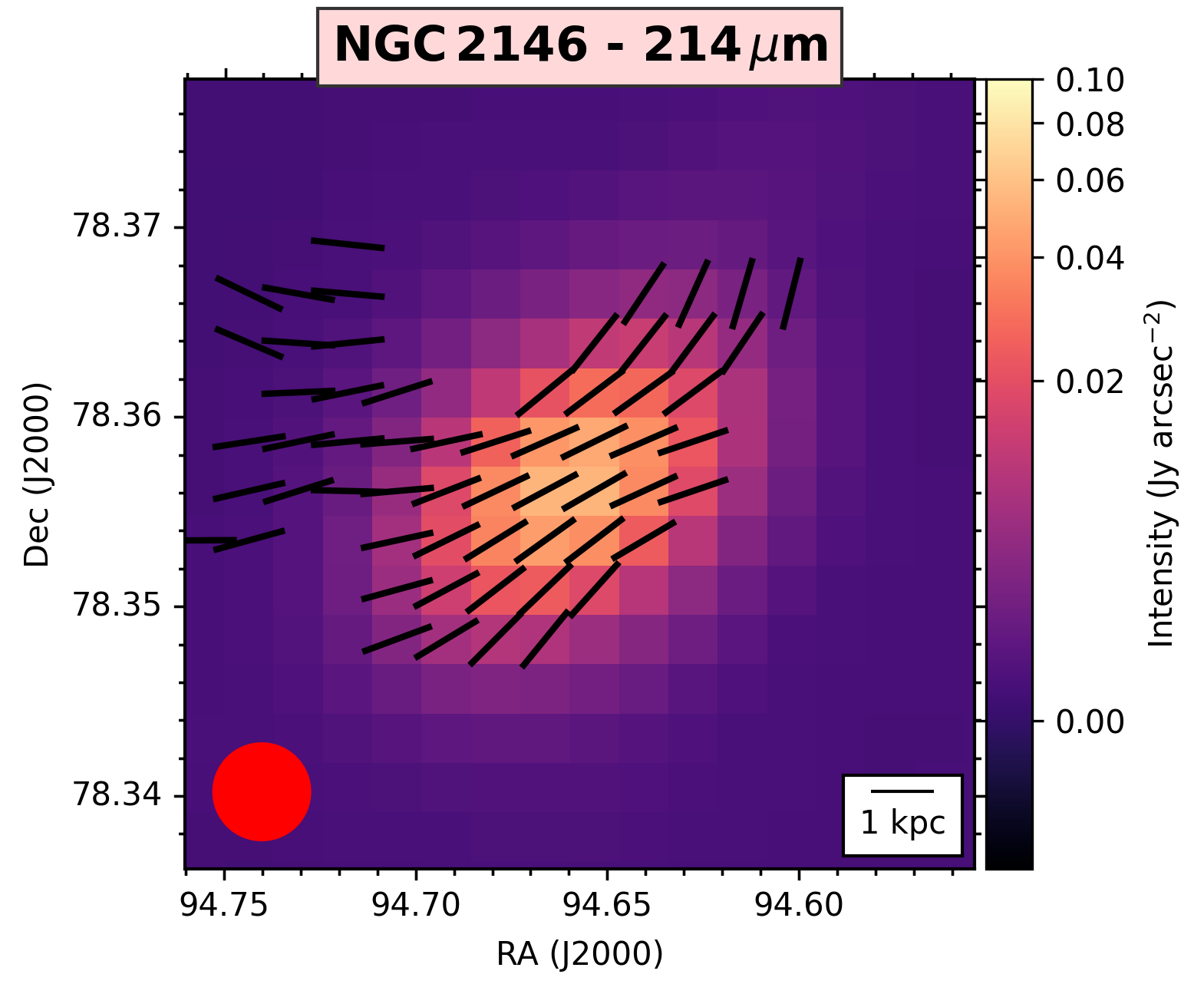}
\end{center}
\caption{FIR and radio $B$-field orientation maps of NGC\,2146 (53 \um, 89 \um, 154 \um, 214 \um, from left to right). FIR $B$-field is represented with black lines. The colorscale shows the surface brightness in each band. Polarization measurements have been set to have a constant length to show the inferred $B$-field orientation. Measurements with $PI/\sigma_{PI} \ge 3.0$, $P\le20$\%, and $I/\sigma_{I} \ge 20$ were selected.
\label{fig:FIR_morphology_NGC2146}}
\end{figure*}

\begin{figure*}[ht!]
\begin{center}
\includegraphics[width=0.32\textwidth, trim=0 0 0 0]{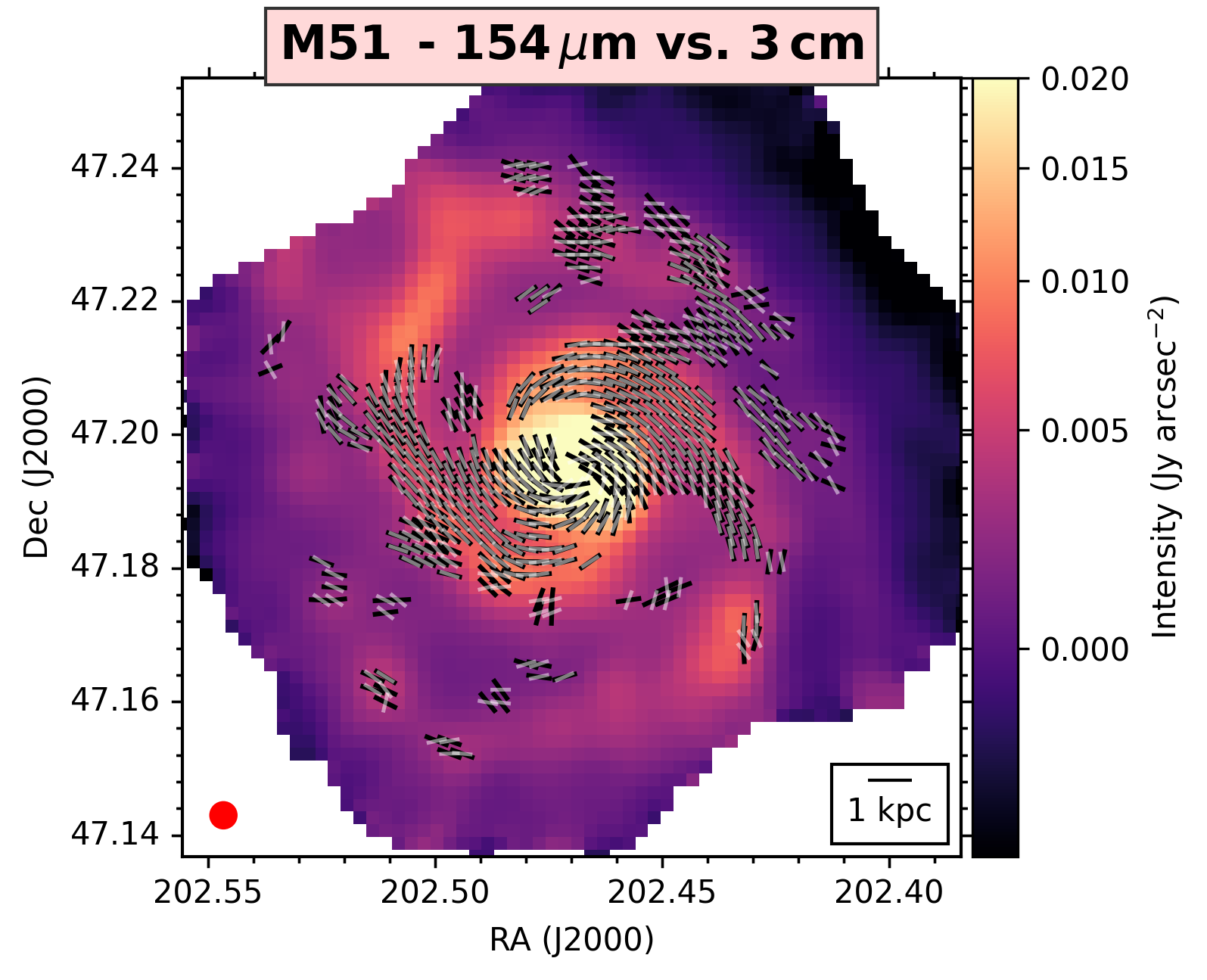}
\includegraphics[width=0.32\textwidth, trim=0 0 0 0]{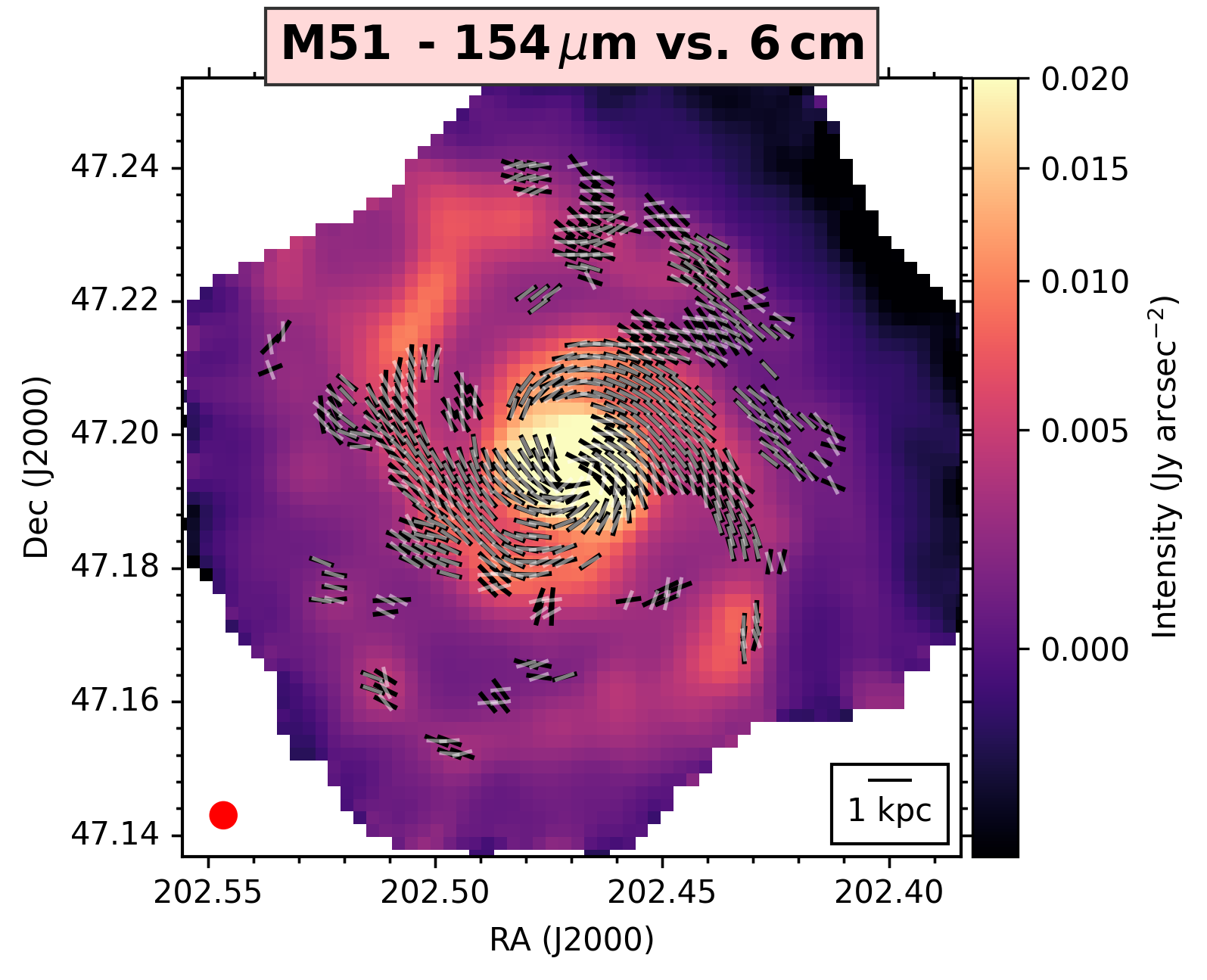}
\includegraphics[width=0.33\textwidth, trim=0 0 0 0]{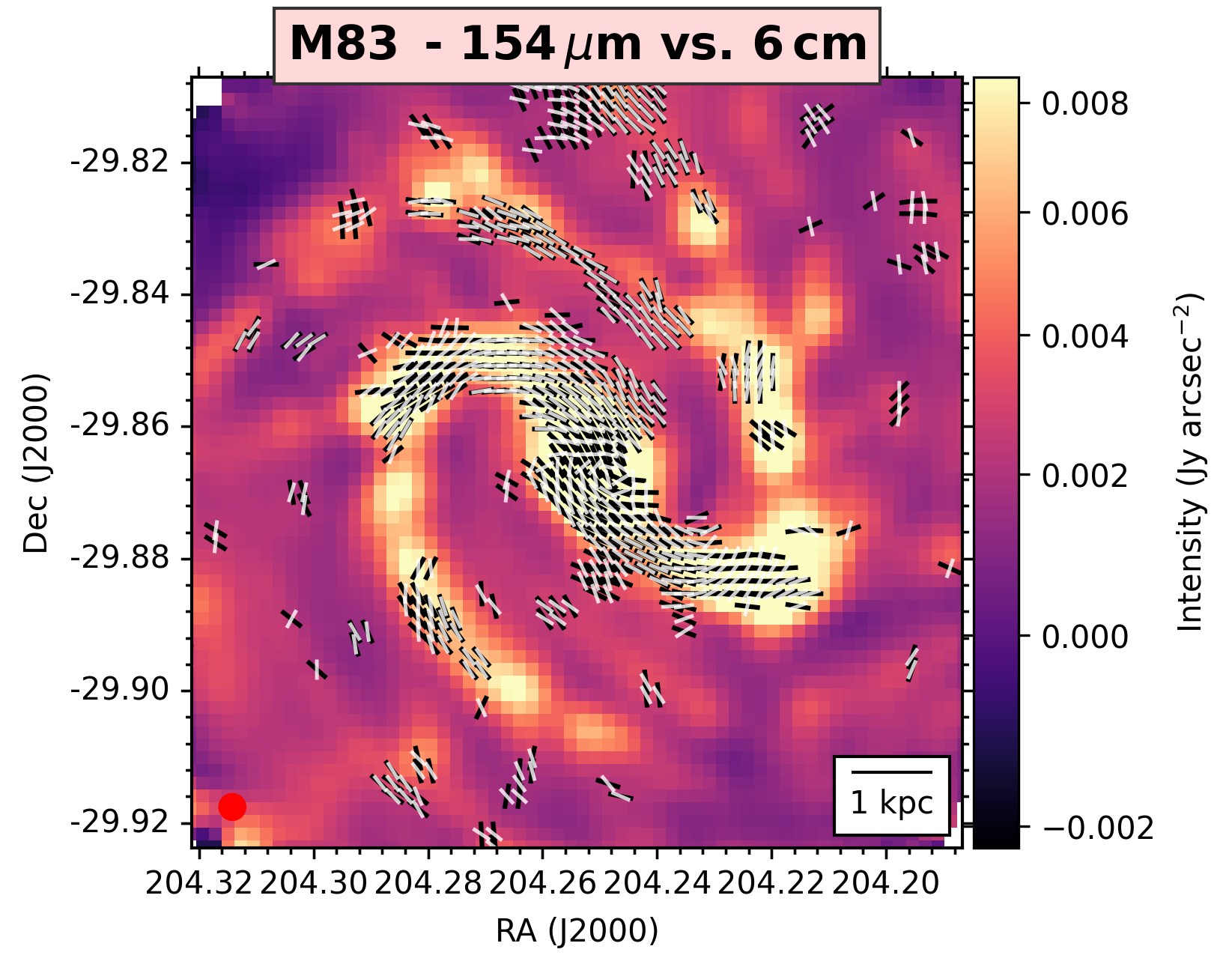}

\includegraphics[width=0.33\textwidth, trim=0 0 0 0]{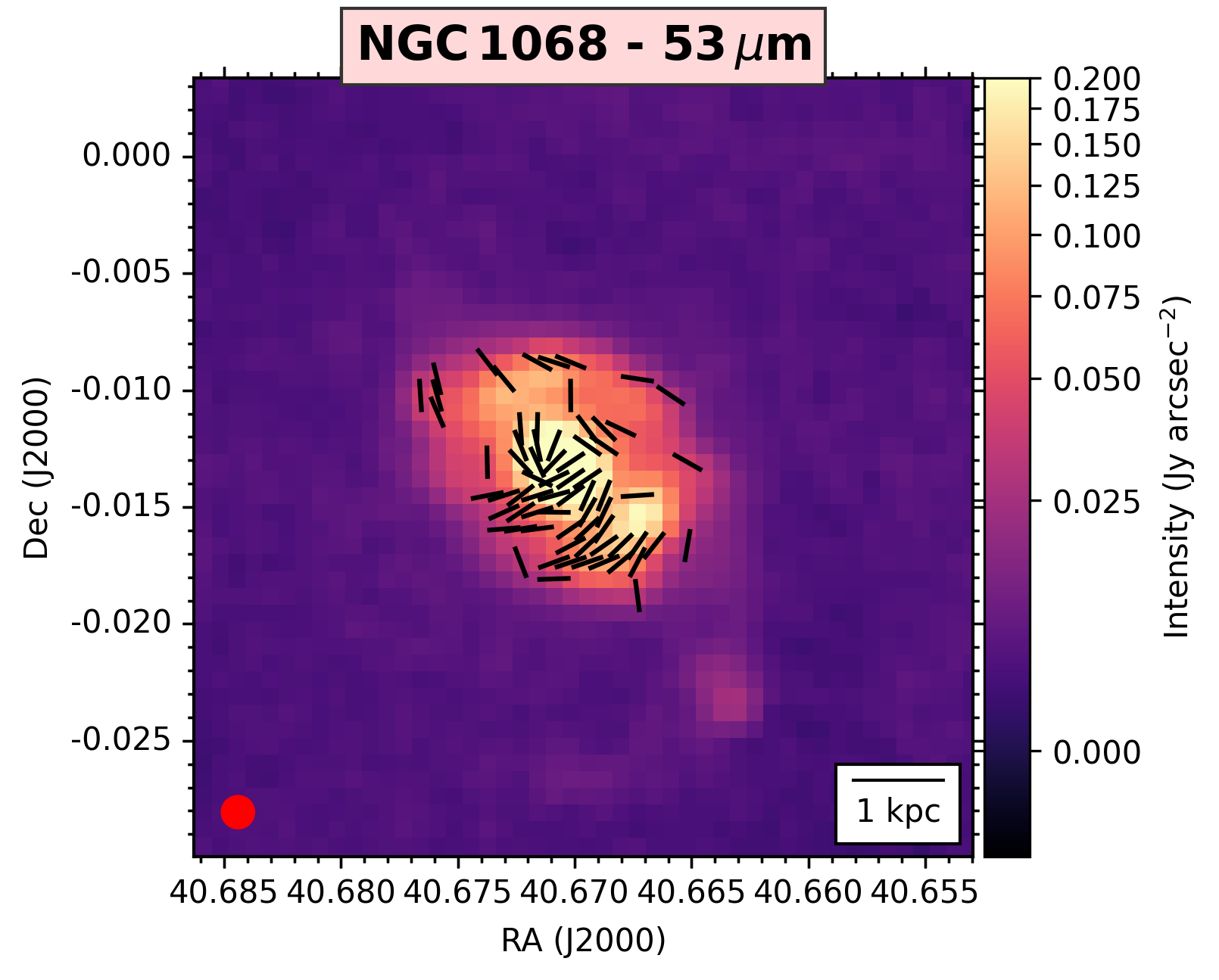}
\includegraphics[width=0.33\textwidth, trim=0 0 0 0]{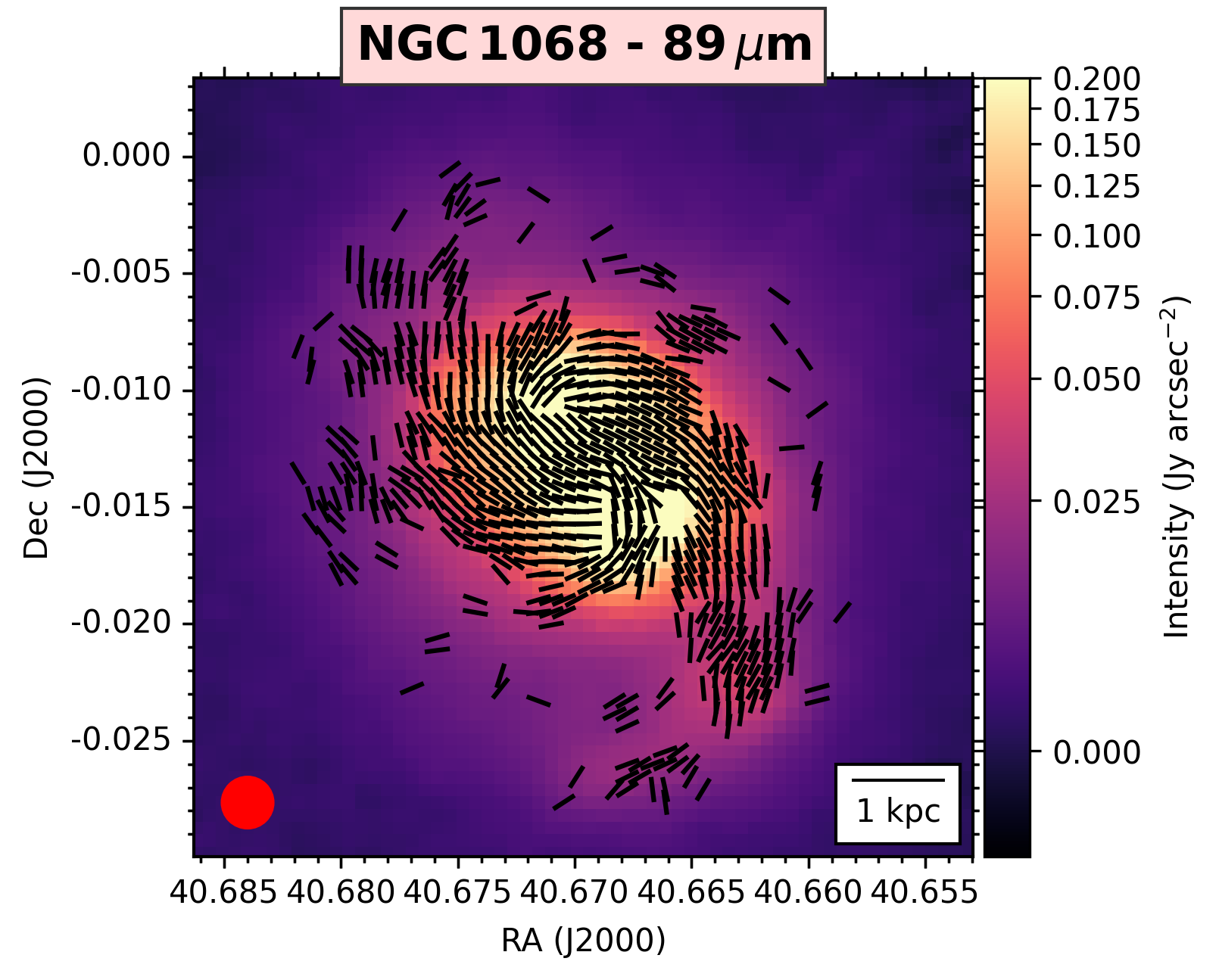}
\includegraphics[width=0.32\textwidth, trim=0 0 0 0]{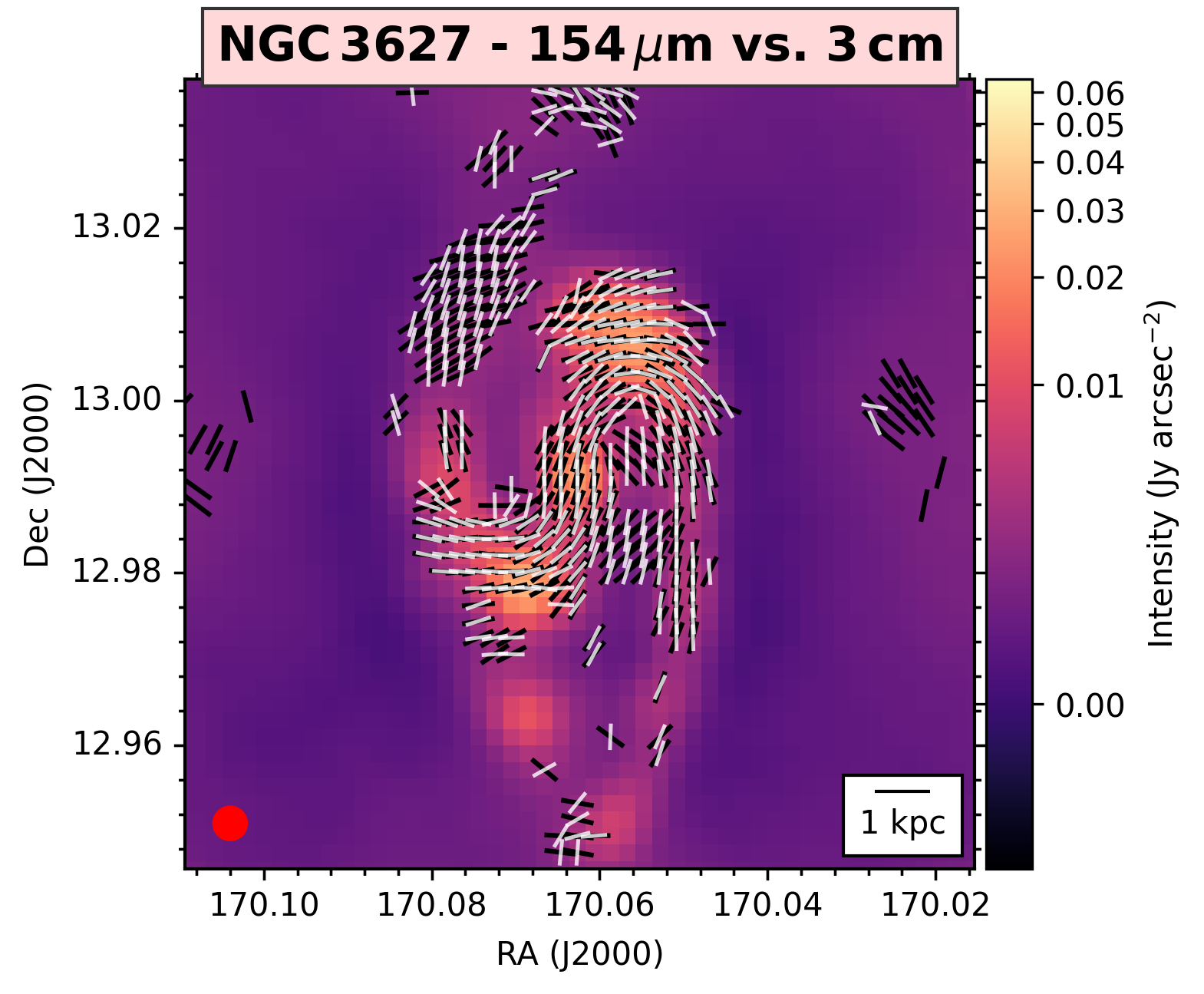}

\includegraphics[width=0.33\textwidth, trim=0 0 0 0]{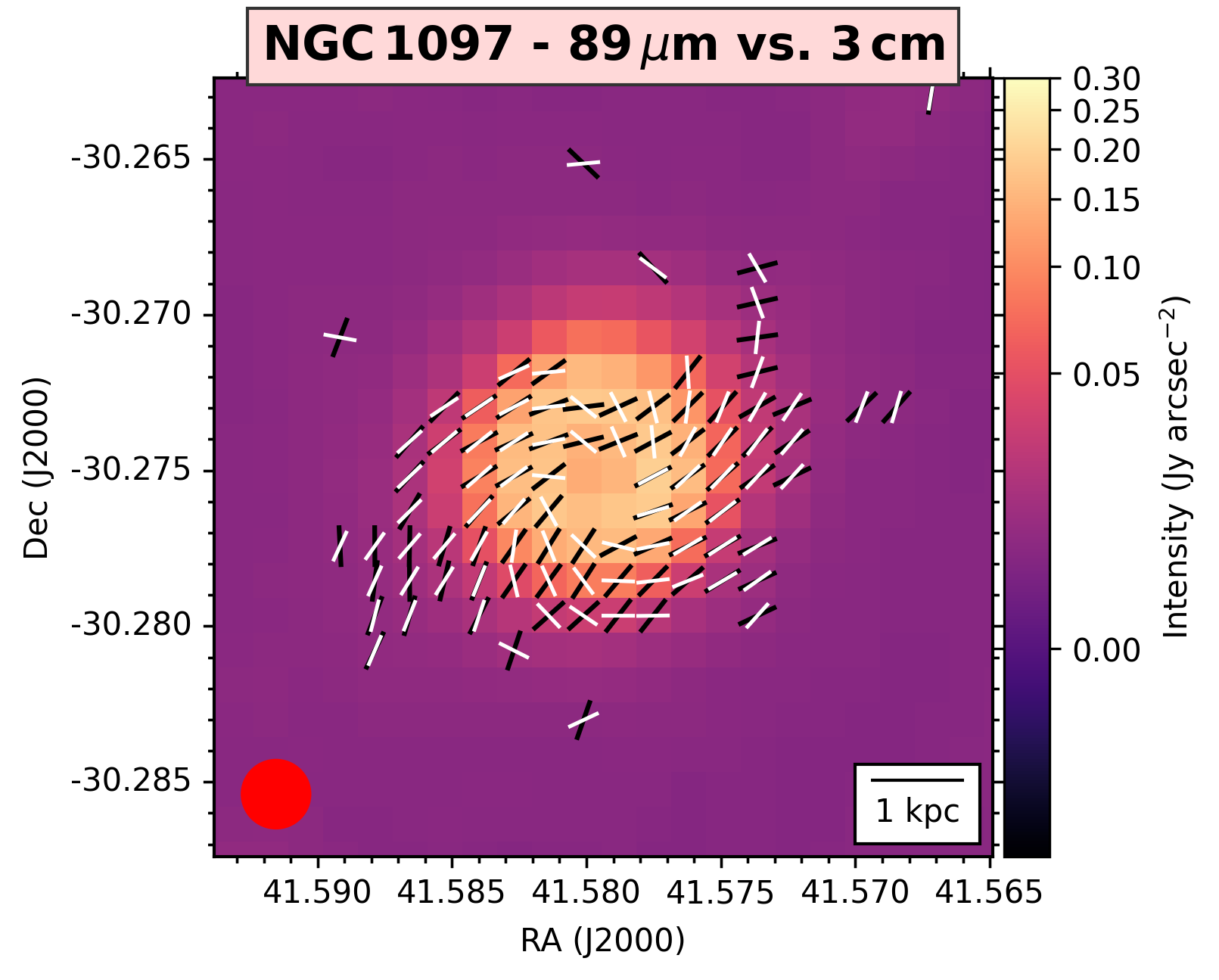}
\includegraphics[width=0.33\textwidth, trim=0 0 0 0]{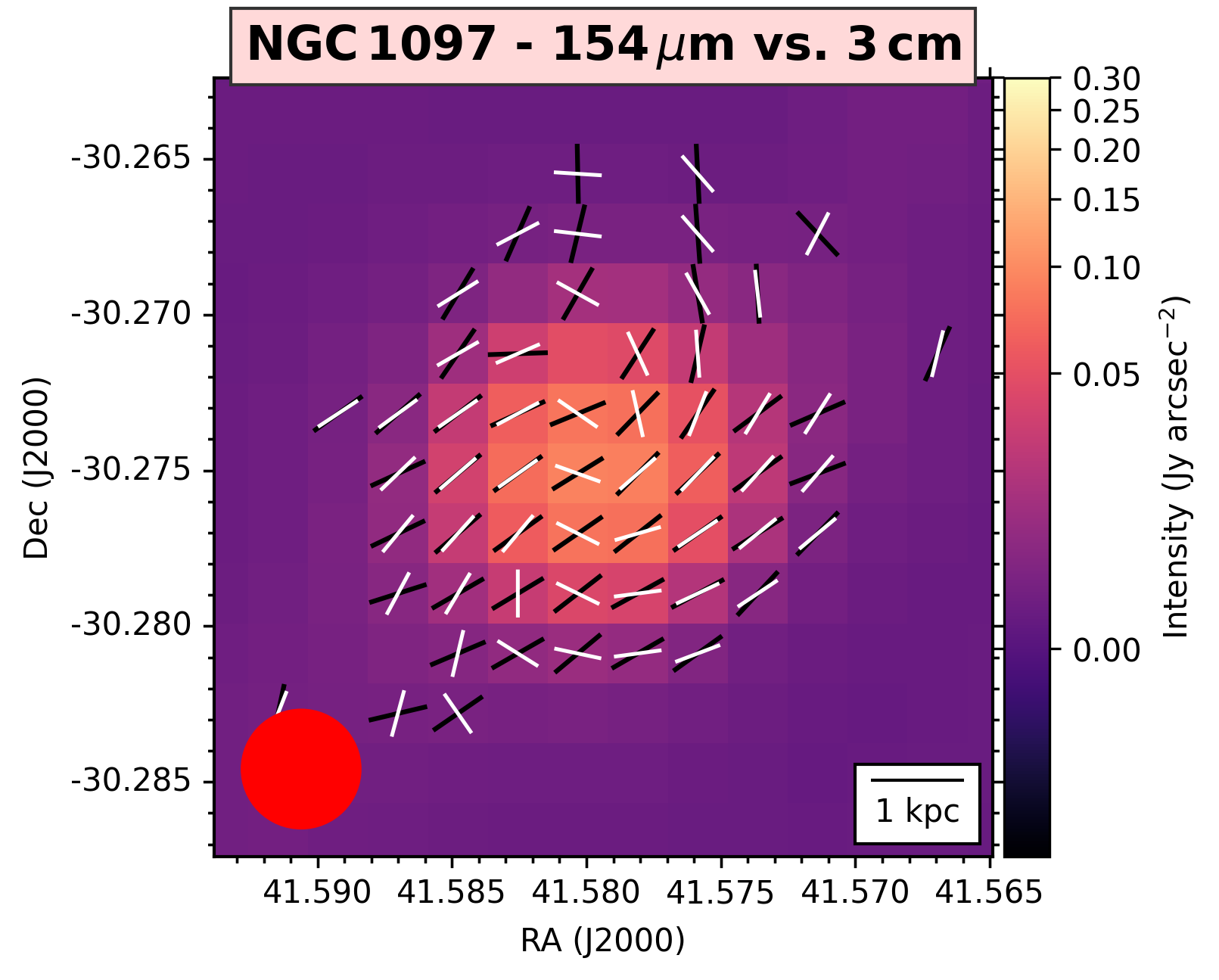}
\includegraphics[width=0.32\textwidth, trim=0 0 0 0]{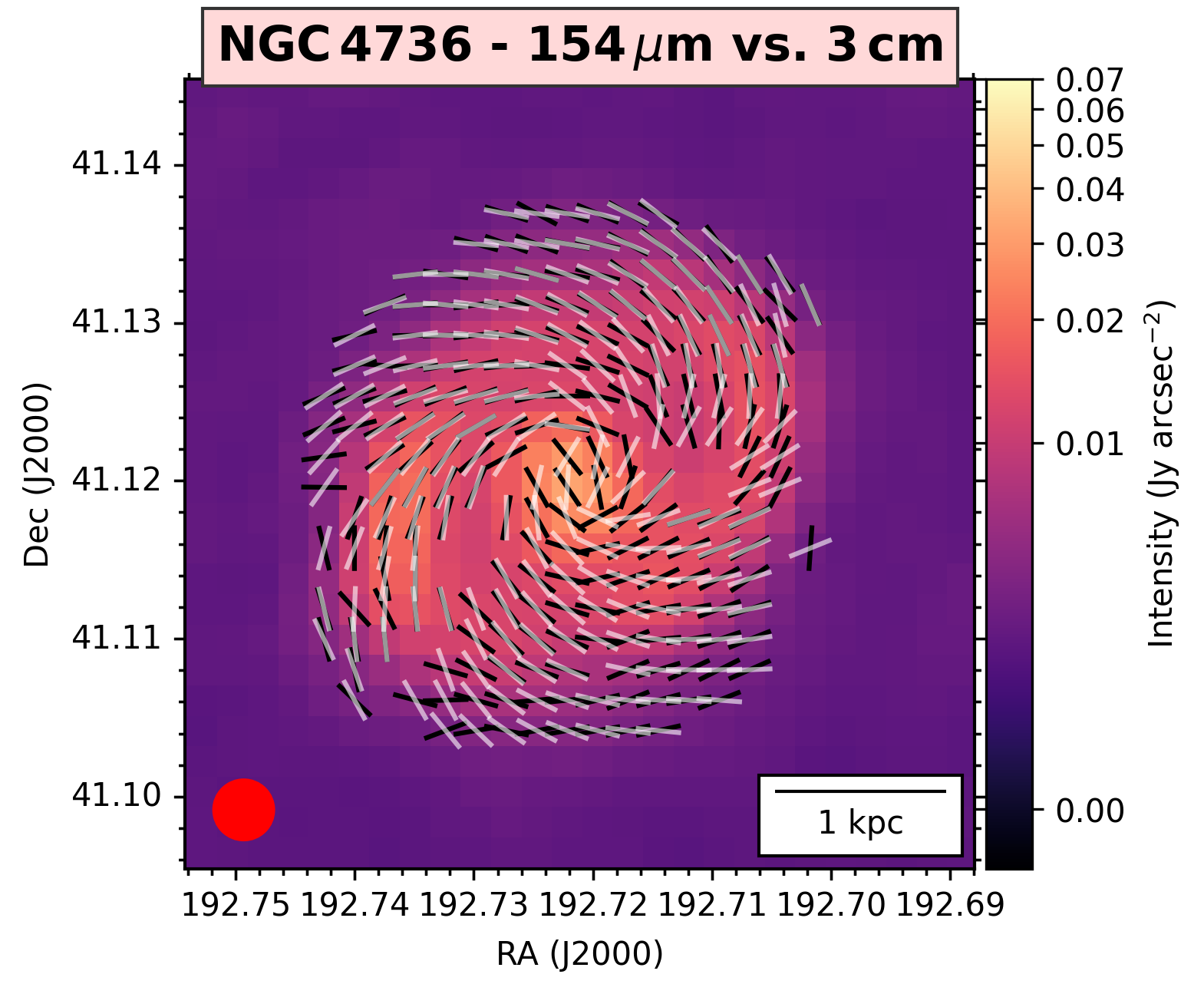}

\includegraphics[width=0.32\textwidth, trim=0 0 0 0]{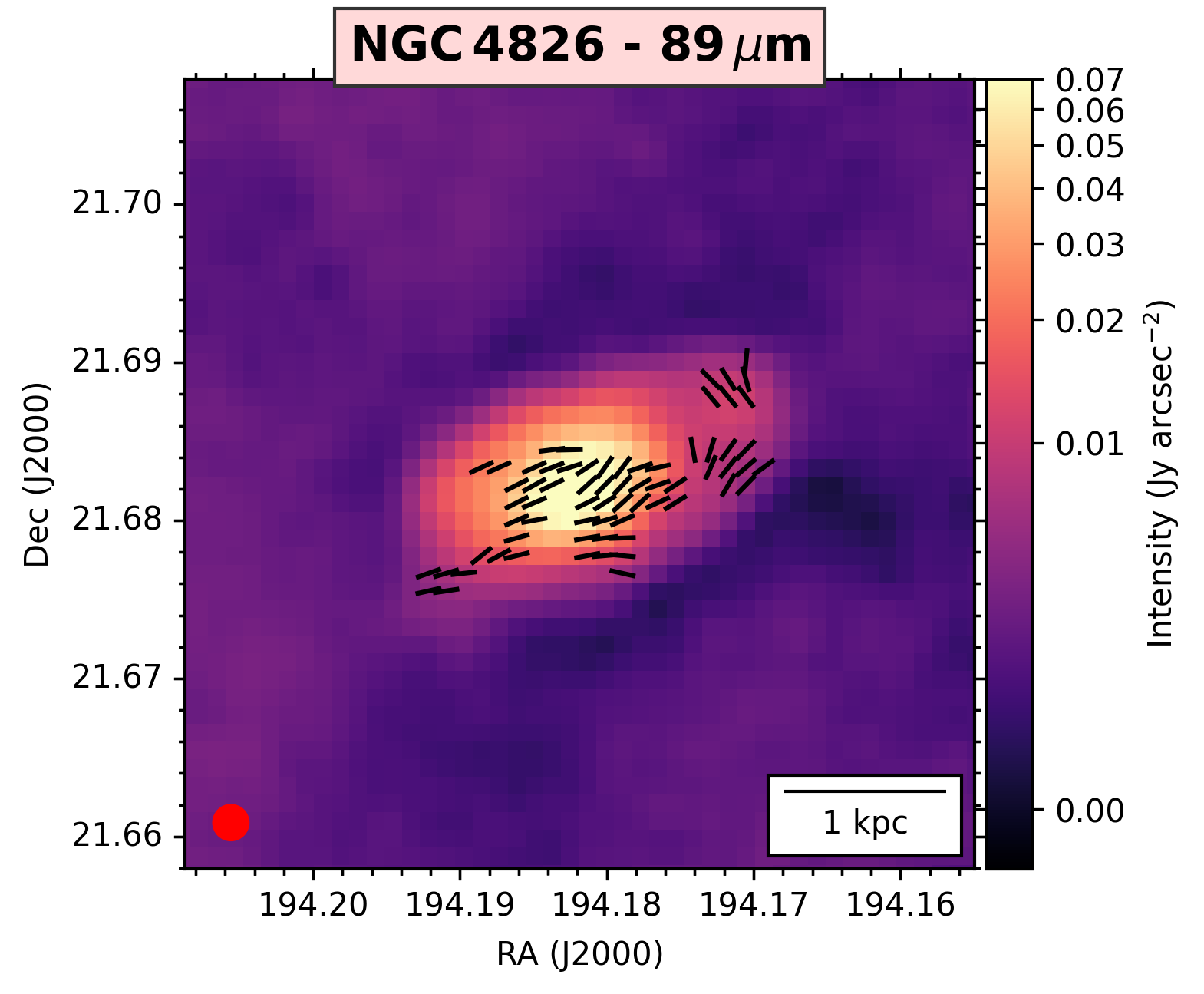}
\includegraphics[width=0.33\textwidth, trim=0 0 0 0]{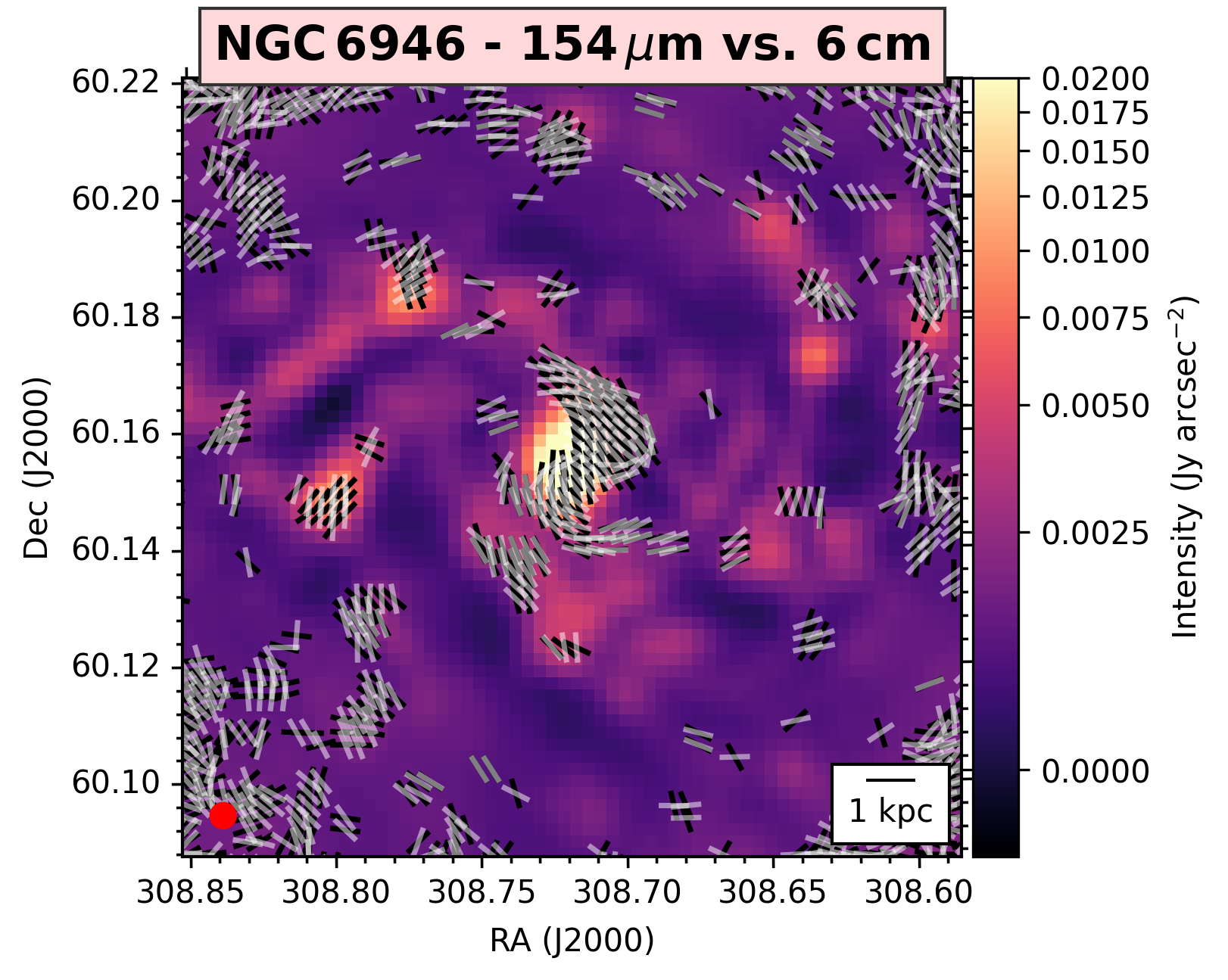}
\includegraphics[width=0.33\textwidth, trim=0 0 0 0]{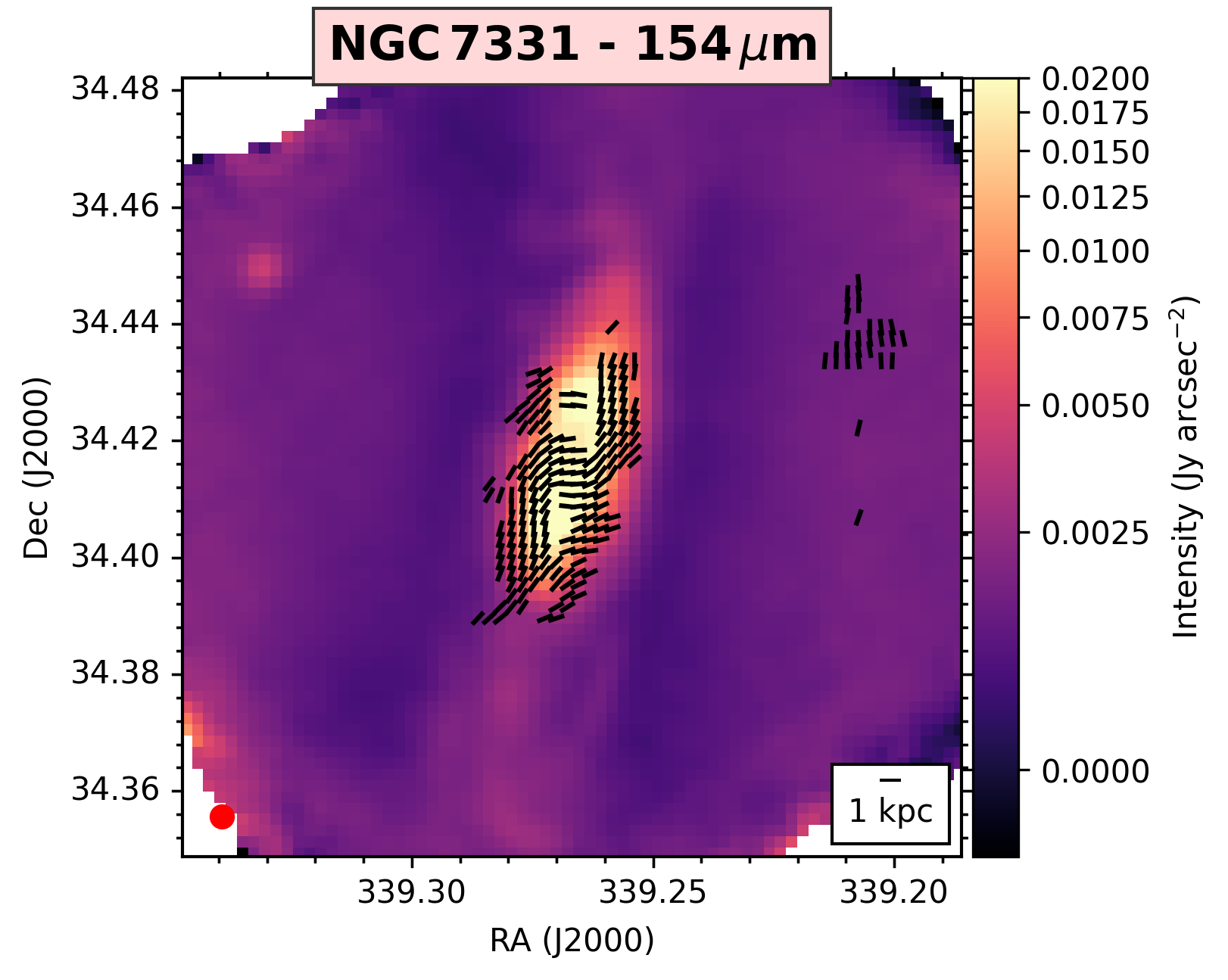}

\end{center}
\caption{FIR and radio $B$-field orientation maps of the spiral galaxies. From left to right, top to bottom: M\,51, M\,83, NGC\,1068, NGC\,3627, NGC\,1097, NGC\,4736, NGC\,4826, NGC\,6946, and NGC\,7331 (wavelengths indicated in the titles). FIR $B$-field (53, 89, 154, and 214 \um) is represented with black lines, while radio polarization observations (3, 6 cm) are represented with white lines, if available (see Table \ref{tab:FIRRadioObs}). The colorscale shows the surface brightness in each band. Polarization measurements have been set to have a constant length to show the inferred $B$-field orientation. Measurements with $PI/\sigma_{PI} \ge 2.0$, $P\le20$\%, and $I/\sigma_{I} \ge 20$ were selected.
\label{fig:FIR_morphology_spirals}}
\end{figure*}

Figures \ref{fig:FIR_morphology_CenA} -- \ref{fig:FIR_morphology_spirals} show the $B$-field orientation maps of the galaxies associated with the first data release of SALSA \citep[SALSA IV, ][]{SALSAIV} at FIR wavelengths and radio wavelengths (when available, see Table \ref{tab:FIRRadioObs}). The FIR polarization measurements with at least $PI/\sigma_{PI} \ge 2.0$ and $P\le 20$\% were selected for all galaxies, with customized $I/\sigma_{I}$ to avoid contamination from the background, as shown in the Appendix (Table \ref{tab:PPA}), where $\sigma_{PI}$ and $\sigma_{I}$ are the uncertainties per pixel of the polarized flux density and Stokes $I$, respectively.The polarization fraction limit of $P = 20\%$ is given
by the maximum polarization fraction ($p_{\rm max} = 19.8\%$) of interstellar dust measured by \textit{Planck} observations \citep{Planck2015}. Note that more stringent quality cuts were applied to some specific galaxies, to avoid systematic effects. We refer to \citet{LopezRodriguez2021ApJ...914...24L} for specific information about the quality cuts chosen for each galaxy and its relation to the SOFIA/HAWC+ OTFMAP polarimetric observation mode. We have rotated the measured polarization angles by $90^{\circ}$ to show the inferred $B$-field orientations in the plane of the sky. For radio polarization measurements, all pixels with $PI/\sigma_{PI} \le 2.0$ and $I/\sigma_{I} \le 2$ were excluded, as well as those pixels that contain no significative FIR polarization emission. For visualization purposes, all figures in this manuscript have a constant polarization length to show the $B$-field orientation. 

Measuring the average $B$-field orientation allows us to quantify the effects of the different spatial resolutions in the observations. We provide two different estimations of the average $B$-field position angle: \PAhist\ and \PAint. \PAhist\ is defined as the average of the $B$-field position angles of the individual polarization measurements after quality cuts. \PAhist\ is estimated using the circular mean as

\begin{equation}\label{eq:Phist}
\langle PA^{\rm{hist}}_{\rm{B}} \rangle = \frac{1}{2}\atantwo \left(\sum_{i=1}^{N} \sin (2PA_{\rm{B},i}),{\sum_{i=1}^{N} \cos (2PA_{\rm{B},i}})\right)
\end{equation}
\noindent
where $PA_{\rm{B},i}$ is the $B$-field orientations of $N$ individual polarization measurements.

\PAint\ is the integrated $B$-field orientation of the galaxy disk, estimated as

\begin{equation}\label{eq:PABint}
\PAint = \frac{1}{2}\arctan \left( \frac{\langle U \rangle}{\langle Q \rangle} \right) + \frac{\pi}{2} \\
\end{equation}
\noindent
where $\langle U \rangle$ and $\langle Q \rangle$ are the average of the Stokes $QU$ for pixels with $I/\sigma_{I} \ge 20$ in the FIR observations. 

The difference between \PAhist\ and \PAint\ is that the latter is the average orientation of the $B-$field in the case of unresolved observations (assigning higher weights to pixels with larger $PI$). The uncertainty is a measurement of the accuracy of the polarization angle of an unresolved galaxy. In contrast, \PAhist\ is an estimation of the circular mean $B-$field orientation on all the analyzed independent beams across the galaxy disk, not weighted by $PI$, allowing to contrast with \PAint\ how the average position angle of the polarization maps is modified with sufficient spatial resolution. The uncertainty is a measurement of the angular dispersion of the distribution of $B$-field orientations within the galaxy disk. The uncertainties of \PAint\ and \PAhist\ are measured using the definition of circular standard deviation from \citep{MARDIA1972ibc1}, implemented in \texttt{SciPy}\footnote{scipy.stats.circstd: \url{https://docs.scipy.org/doc/scipy/reference/generated/scipy.stats.circstd.html}}.

\begin{figure*}[ht!]
\centering
\includegraphics[angle=0,width=\textwidth]{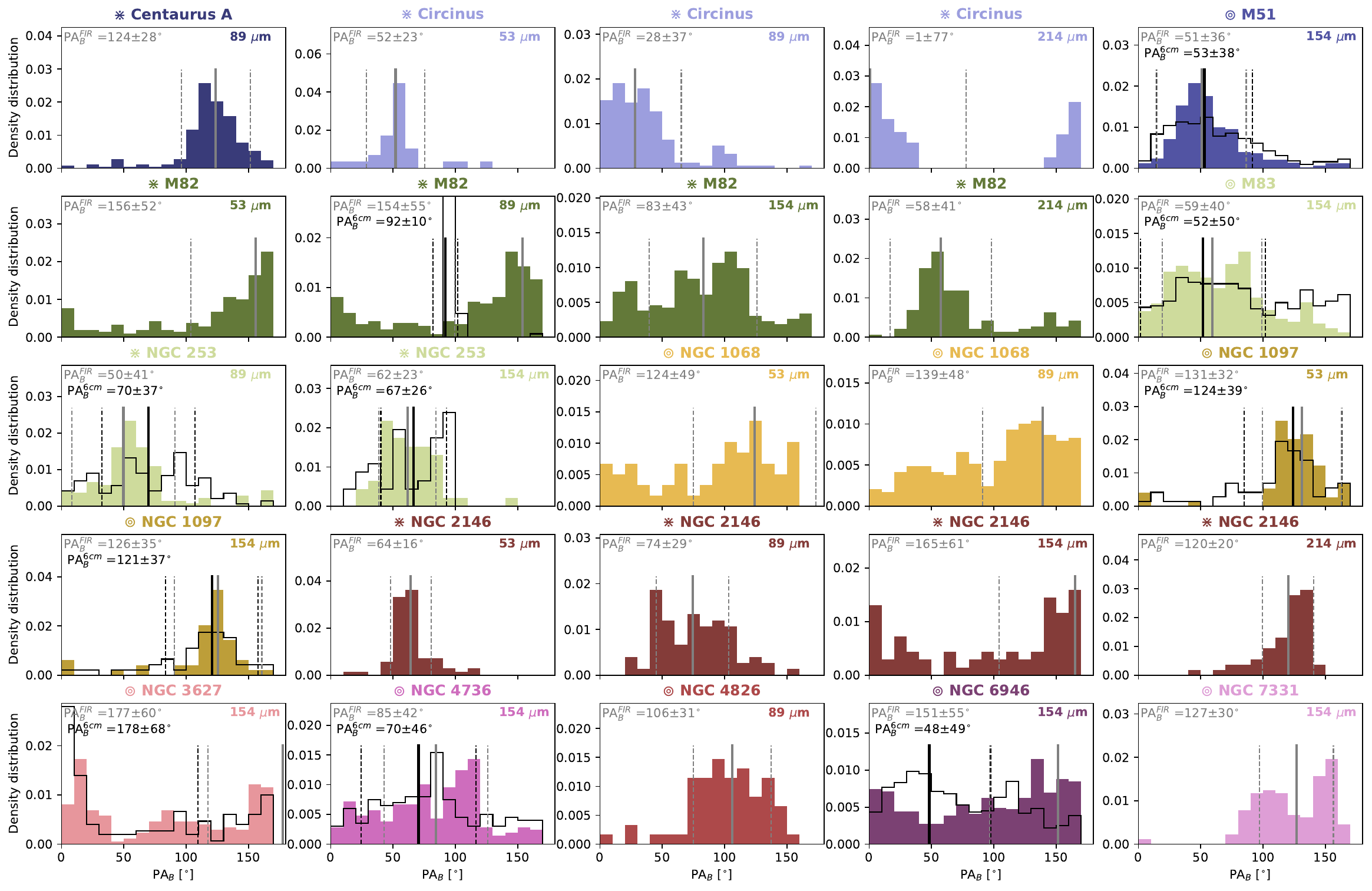}
\caption{Histograms of the $B$-field orientations. The histograms are set to $10^{\circ}$ bins, each galaxy is shown in a unique color with the symbols as in Table \ref{tab:GalaxySample}, and the mean (grey solid line, Eq. \ref{eq:Phist}) and $1\sigma$ uncertainty (grey dashed line) for the FIR observations are shown. For  the galaxies with radio polarimetric observations, the histograms of the $B$-field orientations at $6$ cm (black step histogram) with the mean (black solid line, Eq. \ref{eq:Phist}) and $1\sigma$ uncertainty (black dashed line) are shown. Note that the radio observations were smoothed to match the HAWC+ observations are each band.
\label{fig:PAhist}}
\epsscale{2.}
\end{figure*}

Figure \ref{fig:PAhist} shows the histograms of the $B$-field orientations in bins of $10^{\circ}$ for all the individual polarization measurements per galaxy and per band at FIR (filled histograms) and $6$ cm (black solid lines) wavelengths. We show the estimations of  \PAhist~per galaxy and per band. The $1\sigma$ uncertainty is estimated as the standard deviation of the histogram. As mentioned in Sec.\,\ref{subsec:ObsRadio}, we only use radio polarization measurements that are spatially coincident with the HAWC+ observations. The comparison between \PAhist\ and \PAint\ for the FIR and radio observations are shown in Fig.\,\ref{fig:fig4}. The tabulated data can be found in Appendix \ref{App:PAradio} (Table \ref{tab:PPA}).

\begin{figure}[ht!]
\includegraphics[angle=0,width=\columnwidth]{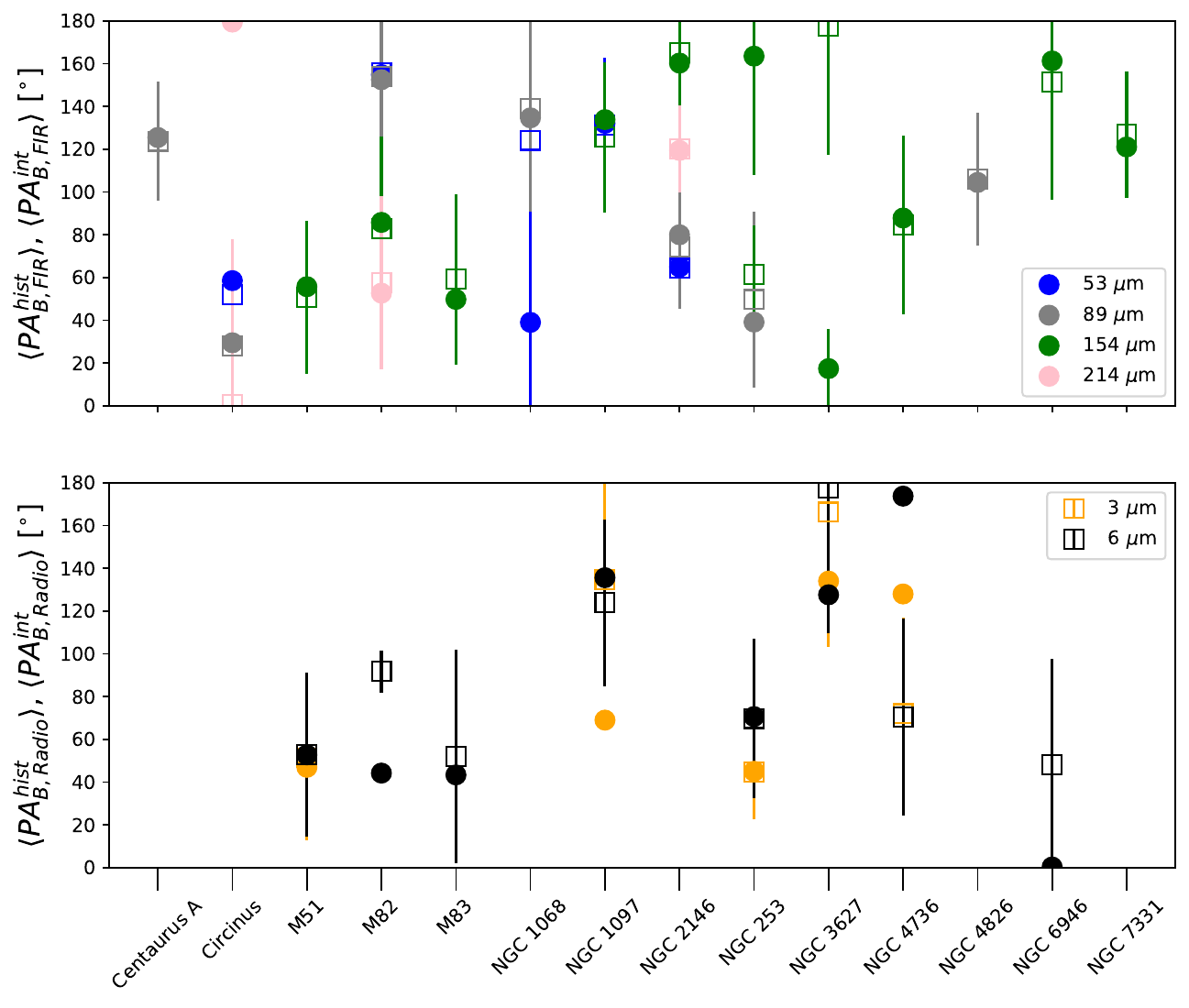}
\caption{Mean and integrated $B$-field orientation of galaxies. $B$-field orientation at FIR (top) and radio (bottom) from individual measurements, $\langle $PA$^{\rm{hist}}_{B} \rangle$ (open square, Eq.\,\ref{eq:Phist}), and from the integrated Stokes $QU$, $\langle $PA$^{\rm{int}}_{B} \rangle$ (filled circle, Eq.\,\ref{eq:PABint}) are shown. Tabulated values are shown in Table \ref{tab:PPA}.
\label{fig:fig4}}
\epsscale{2.}
\end{figure}

\subsection{Results}
\label{subsec:BmapsFIR_results}

We present the results from the analysis of the morphology of the $B$-field estimated using FIR and radio observations. We divide our sample (Table \ref{tab:GalaxySample}) into starburst and/or high inclination galaxies ($\divideontimes$; without a clear spiral pattern, Sec.\,\ref{subsec:results_Bmaps_edgeon}) and low inclination spiral galaxies ($\circledcirc$; with a spiral $B$-field pattern, Sec.\,\ref{subsec:results_Bmaps_faceon}). In general, some spiral galaxies with lower inclination tend to present well-defined spiral magnetic arms. We use these to measure their magnetic pitch angles. Starburst and edge-on galaxies show  multiple B-field components associated with the galactic outflows, if present, and the B-fields parallel to the disk. We note that this classification addresses strictly the projected B-field structure of the objects, which can be different from their respective morphological classification.

\subsubsection{Starbursts and/or high inclination galaxies}
\label{subsec:results_Bmaps_edgeon}

Centaurus A (Fig.\,\ref{fig:FIR_morphology_CenA}) shows a $B$-field orientation tightly following the $3$ kpc warped molecular disk as previously reported by \citet{ELR2021}. The \PAhist\ and \PAint\ are measured to be $124\pm28^{\circ}$ and $126\pm9^{\circ}$, respectively. The large angular dispersion of $\pm28^{\circ}$ is attributed to turbulence at spatial scales smaller than the beam of the observations driven by the merger activity and star-forming regions across the warped molecular disk.

We measure an almost constant $B$-field orientation within the central $1$ kpc of the Circinus galaxy (Fig.\,\ref{fig:FIR_morphology_Circinus}). Note that the $B$-field orientation bends around the core at all wavelengths, which may indicate that the measured $B$-field is arising from the central starburst ring \citep{Elmouttie1998,Zschaechner2016}. The \PAhist\ changes from $52\pm23^{\circ}$ at $53$ \um~to $28\pm37^{\circ}$ at $89$ \um, with a similar, within the uncertainties, \PAint. At $214$ \um, the \PAint\ is dominated by the highly polarized regions in the bar of the galaxy with a large-scale $B$-field at $179\pm2^{\circ}$. The $B$-field orientation in the central $1$ kpc at $214$ \um~is consistent with the $B$-field at $53$ and $89$ \um. Radio polarimetric observations at $6$ cm with a beam size of $9\arcsec$ show two radio lobes cospatial with the ionization cones and a spiral $B$-field at scales of $1\arcmin$ north and south of the unpolarized core \citep{Elmouttie1998b}.

We present a new analysis of the $154$ \um~$B$-field orientation measurements of NGC~1097. The galaxy shows a constant FIR $B$-field orientation at $154$ \um\ within the central $1$ kpc starburst ring in agreement with the previously B-field orientations reported at $89$ \um~by \citet[][SALSA II]{SALSAII}.The \PAhist\ and \PAint\ are measured to be $126\pm35^{\circ}$ and $134\pm16^{\circ}$, respectively (Table \ref{tab:PPA}). The $154$ \um~$B$-field is parallel to the galaxy's bar (which goes from South-East to North-West in the image). The radio polarimetric observations show a spiral $B$-field pattern within the central 1 kpc of the starburst ring of NGC~1097 \citep{Beck2005}. The differences in the FIR and radio B-fields were suggested to arise from a) the diffuse gas dragging matter towards the core at radio, and b) the dense gas located in the galactic shocks at the intersections of the bar and starburst ring \citep{SALSAII}. The former generates a spiral B-field and the latter generates a constant B-field in the shocks.

M\,82's $B$-field orientation at FIR and radio polarization maps show large morphological differences (Fig.\,\ref{fig:FIR_morphology_M82}). The $53-214$ \um\ show a $B$-field orientation perpendicular to the galaxy's disk indicating the presence of a galactic outflow with a diameter of $1$ kpc at the center of the galaxy extending $\sim2-3$ kpc above and below the galaxy's disk. The $B$-field orientation is parallel to the disk of the galaxy from a radius of $1$ to $3$ kpc. The M\,82 results are compatible with the previous $53$ and $154$ \um~observations reported by \citet{Jones2019} and  \citet{ELR2021a}. Here, we present deeper observations covering a wider wavelength range, i.e. $53-214$~\um, and larger extension ($> 1$ kpc). At $3$ and $6$ cm, the $B$-field orientation in M\,82 is mainly parallel to the galactic disk \citep{Adebahr2017}. Depolarization due to Faraday rotation or the lifetime of cosmic ray electrons along the galactic outflows may cause unpolarized regions (Sec.\,\ref{subsec:DIS_SB}). Interestingly, the radio $B$-field to the north and south of the galaxy disk is coincident with the $B$-field orientation at $154$~\um\ (Fig. \ref{fig:FIR_morphology_M82}).

For NGC\,253 (Fig.\,\ref{fig:FIR_morphology_NGC253}), the $B$-field orientation is predominately parallel to the disk of the galaxy at both $89$ and $154$ \um. A $B$-field parallel to the disk of the galaxy at $89$ \um\ was also measured using HAWC+ by \citet{Jones2019}. At $89$ \um, both \PAhist\ and \PAint\ are in agreement, within the uncertainties, with a $B$-field parallel to the disk. However, a change of $\Delta PA_{B} = |\PAint\ - \PAhist| = 102\pm59^{\circ}$ is detected at $154$ \um. Our hypothesis is that this angular change is mainly due to the highly polarized, $\sim3-5$\%, regions with a $B$-field twisting within the central 1 kpc of the galaxy with an hourglass shape. This twisted shape may be the signature of the galactic outflow associated with its central starburst having hotter dust temperatures than the dust located in the disk. This $B$-field structure has also been observed at radio wavelengths \citep{Heesen2011}. Deeper and higher angular resolution observations are required to statistically detect this trend at FIR wavelengths. Note that our released $89$ and $154$ \um~observations only account for the $33$\% and $18$\% of the total requested time (SALSA III and IV), but these data are deeper and complementary observations to those presented by \citet{Jones2019}. 

For the starburst galaxy NGC\,2146 (Fig.\,\ref{fig:FIR_morphology_NGC2146}), the $B$-field orientation is parallel to the galactic outflow within the $53-154$ \um~wavelength range. At $53$ and $89$ \um, the $B$-field orientation is aligned with the galactic outflow with a diameter of $\sim2$ kpc at the center of the galaxy. At $214$ \um, the $B$-field orientation is dominated by a component parallel to the disk of the galaxy. The $B$-field orientation at $154$ \um~is more complex showing the transition between a $B$-field dominated by the galactic outflow at $53-89$ \um~to a $B$-field dominated by the galactic disk at $214$ \um. 

For both M\,82 and NGC\,2146, the histograms show a change in \PAhist\ and \PAint\ from $53$ to $214$ \um~(Fig.\,\ref{fig:fig4} and Table \ref{tab:PPA}). Specifically, \PAhist\ and \PAint\ changes from $\sim156^{\circ}$ and $\sim64^{\circ}$ (parallel to the outflows) to $\sim58^{\circ}$ and $\sim120^{\circ}$ (parallel to the disk) in M\,82 and NGC\,2146, respectively. These results show that the resolved and unresolved FIR polarimetric observations are sensitive to the $B$-fields in the warm, $\sim30-50$ K at $50-100$ \um, dust of galactic outflows and cold, $\sim10-30$ K at $100-220$ \um, dust in the galactic disk.

In summary:
\begin{itemize}

\item The edge-on galaxy Centaurus A has a highly inclined $B$-field mostly parallel to the $\sim3$ kpc warped disk of the galaxy. The large-scale $B$-field shows large angular dispersion across the galactic disk driven by the merger and star-forming activity \citep{ELR2021a}. 

\item The $B$-field orientation in Circinus is mostly constant and parallel to the inner-bar and the starburst ring in the central 1 kpc.   

\item The starburst galaxies M\,82, NGC\,253, and NGC~2146 show signatures of a $B$-field orientation potentially associated and parallel to galactic outflows up to $\sim3$ kpc scales above and below the disk. For M\,82, the $B$-field is parallel to the galactic outflow at all wavelengths \citep{Jones2019,ELR2021}. The median orientation of the FIR $B$-field of NGC\,2146 changes with wavelength from parallel (53 - 89\,\um) to perpendicular (154 - 214\,\um) with respect to the galactic outflow driven by the central starburst. For NGC\,253, the $B$-field is mostly parallel to the galaxy disk with a signature of an hour-glass shape driven by the starburst activity in the central 1 kpc.

\item For M\,82, the radio $B$-field orientation is parallel to the galaxy disk, although some hint of the galactic outflow is also measured and coincident with the $154$~\um\,$B$-field.

\end{itemize}

\subsubsection{Spiral galaxies}
\label{subsec:results_Bmaps_faceon}

The FIR $B$-field orientation maps of the spiral galaxies are shown in Fig.\,\ref{fig:FIR_morphology_spirals} (including NGC\,1097). The histograms of $B$-field orientations for the polarization measurements are presented in Fig.\,\ref{fig:PAhist}. In general, we find that low inclination objects (i.e., NGC~1068, NGC~4736, NGC~6946) present relatively flatter PA$_{B}$ distributions at FIR and radio wavelengths than galaxies with higher inclinations (i.e., M\,82, NGC\,253, NGC\,2146). This result is because the spiral B-field is more centrosymetric at lower inclinations, while the spiral B-field is more asymmetric due to the cancelation of B-field along the LOS as inclination increases \citep[see section 5.3 by][]{SALSAIV}.

All galaxies show a large-scale spiral $B$-field, except for NGC\,6946, which shows a highly disordered FIR $B$-field spatially located with the star-forming regions in the spiral arms. Fig.\, \ref{fig:FIR_morphology_spirals} shows the NGC\,6946 $B$-field pattern traced by both FIR and radio polarimetric observations, revealing clear large structural differences. At radio wavelengths, a large-scale well-ordered spiral $B$-field has been measured \citep{Beck1991,Beck2007} mostly aligned with the interarm regions (`magnetic arms'), instead of the morphological spiral arms. However, we detect signs of a spiral pattern in radio and FIR also in the morphological spiral arms (see Sec.\,\ref{subsec:PitchB}). At FIR wavelengths, we detect a spiral $B$-field in the central $\sim3$ kpc diameter. At larger galactocentric radius ($R>3$ kpc), the $B$-field orientation is highly disordered in the spiral arms. The histogram of $B$-field orientations (Fig.\,\ref{fig:PAhist}) shows an almost flat distribution with a $\langle $PA$^{\rm{hist,FIR}}_{B} \rangle= 151\pm55^{\circ}$ at $154$ \um~and $\langle $PA$^{\rm{hist,Radio}}_{B} \rangle= 48\pm49^{\circ}$ at $6$ cm.

Another remarkable result is the well-ordered $\sim1$ kpc-scale spiral $B$-field over the starburst ring of NGC\,4736 at both FIR and radio wavelengths (Fig.\,\ref{fig:FIR_morphology_spirals}). For both tracers, the measured spiral $B$-field seems to be uncorrelated with the underlying morphological structure. We measure a similar \PAhist\ of $85\pm42^{\circ}$ and $70\pm46^{\circ}$ at $154$ \um~and $6$ cm, respectively. We note that the southern region of the galaxy has slightly different $B$-field orientations at radio and FIR wavelengths, while in the northern region both FIR and radio wavelengths have similar $B$-field orientations. Secs.\,\ref{subsec:PitchB} and \ref{subsec:Dtheta} present a quantification of these differences.

NGC\,3627 presents a well-ordered spiral $B$-field that follows the inner morphological spiral pattern at 154 \um, 3~cm, and 6~cm. Interestingly, beyond $90\arcsec$ from the galactic center, the FIR magnetic orientation deviates from the morphological arms (see North-East region), as opposed to the observed structure in 3 and 6 cm. In Sec.\,\ref{subsec:PitchB} we will analyze in detail how this change in orientation translates into a break in the magnetic pitch angle profile very similar to the results obtained in M\,51 \citep{Borlaff2021}.

We detect a spiral $B$-field morphology in both NGC\,4826 and NGC\,7331. For each galaxy, the \PAhist\ and \PAint\ are in agreement, within their uncertainties (Table \ref{tab:PPA}), which indicates that the $B$-field is aligned with the major axis of the galaxy. This result is due to the high inclination $>65^{\circ}$ of these galaxies. NGC\,4826 presents a non-regular spiral $B$-field, with the majority of the polarization measurements located in the southern half of the galaxy. Interestingly, only the southern section was found to be polarized also in 18 and 22 cm by \citet{Heald2009A&A...503..409H}, a potential consequence of Faraday depolarization. Sec.\,\ref{subsec:PitchB} shows the associated radial profiles of the magnetic pitch angles of these galaxies. Radio polarimetric observations covering the several kpc-scale of the host galaxy of NGC\,7331 are unavailable. 

For M\,51 and M\,83, the overall spiral $B$-field morphology is the same in both FIR and radio polarimetric observations, with some significative differences at large radii. The $B$-field orientation map of M\,51 is the same as presented by \citet{Borlaff2021}, where the pitch angles of the spiral arms at the outskirts of M\,51 are different at FIR and radio wavelengths. Sec.\,\ref{subsec:PitchB} shows the associated radial profiles of the magnetic pitch angles of these galaxies. A similar result is observed in the outskirts of M\,83 when comparing the radio polarization and FIR magnetic pitch angle profiles (Fig.\,\ref{fig:FIR_morphology_spirals}).


The $B$-field orientation in NGC\,1068 is compatible with the previous $89$~\um~observations reported by \citet{ELR2020}. Here, we present deeper $89$ \um~polarimetric observations and new $53$ \um~polarimetric observations. We detect a $6$ kpc-scale spiral $B$-field at $89$~\um, in comparison with the $3$ kpc scale detected by \citet{ELR2020}. The new $53$ \um~polarimetric observations show a change of $\sim90^{\circ}$ at the core of NGC~1068 with respect to the $89$ \um. This angular offset may be due to a transition from dichroic absorption to dichroic emission or that the $53$ \um\ is resolving the beginning of the spiral arm within the central $1$ kpc. 
 
In summary:

\begin{itemize}
\item The spiral galaxies M\,51, M\,83, NGC\,1068, NGC\,3627, and NGC\,4736 have kpc-scale spiral ordered $B$-fields. 

\item NGC\,4736 has a FIR and radio spiral $B$-field without clear spatial correspondence with the underlying morphology in total emission. The total emision is cospatial with the starburst ring at $\sim1$ kpc around the core.

\item NGC\,6946 has a highly disordered FIR $B$-field, in comparison with the well-ordered large-scale spiral $B$-field at radio wavelengths. The disordered FIR $B$-field seems to be potentially connected to star formation activity in the spiral arms.

\item Significant large-scale differences between FIR and radio polarization observations are found in several regions of M\,51, NGC\,1097, and NGC\,3627.

\item For all these galaxies, deviations from the large-scale spiral $B$-fields are found in star-forming regions and at the edges of the nuclear bar.

\end{itemize}


\section{Magnetic pitch angles}\label{subsec:PitchB}

The morphology of the large-scale B-field in the disk of spiral galaxies is typically defined using the pitch angle $\Psi_{B} = \arctan(B_{\rm{r}}/B_{\phi})$, where $B_{\rm{r}}$ is the radial B-field component, and $B_{\phi}$ is the azimuthal B-field component \citep[e.g.,][]{Krasheninnikova1989}. Note that the B-field is three dimensional, so from an observational point of view these two quantities are the B-fields components projected on the plane of the sky. Observationally, $\Psi_{B}$ is defined as the angle between the line tangent to a circle in the galaxy plane and the direction of the $B$-field at a specified radius. Positive values of $\Psi_{B}$ represent a clockwise sense of the spiral structure, while negative values represent a counter-clockwise direction. Values of $\Psi_{B}$ close to zero represent a circular B-field (tightly-wounded), while higher absolute values of $\Psi$ represent a more open (less-wounded) spiral structure. The methodology to estimate the magnetic pitch angle maps for FIR and radio polarization observations (\Mohawc) is described in Appendix \ref{App:PitchBmethods} and previously presented by \citet{Borlaff2021}. This method was implemented in \textsc{python} and is available on the project website\footnote{\Mohawc: \url{https://github.com/Borlaff/MOHAWC}}.

We compute the magnetic pitch angles of 8 spiral galaxies M\,51, M\,83, NGC\,1068, NGC\,3627, NGC\,4736, NGC\,4826, NGC\,6946, and NGC\,7331. The other galaxies have a) vertical $B$-fields along the galactic outflows (i.e. M\,82, NGC\,253, and  NGC\,2146); b) parallel $B$-fields along the galactic disk in edge-on galaxies  (i.e. Centaurus A); or c) constant $B$-fields in the central 1 kpc (i.e. Circinus, NGC\,1097). For these galaxies, we do not perform any analysis based on the magnetic pitch angles with exception of the global $B$-field orientation histograms and integrated $B$-field orientation in the full disk of galaxies as described in Sec.\,\ref{subsec:BmapsFIR}. 

\begin{figure*}[ht!]
\centering
  \begin{overpic}[width=\textwidth]{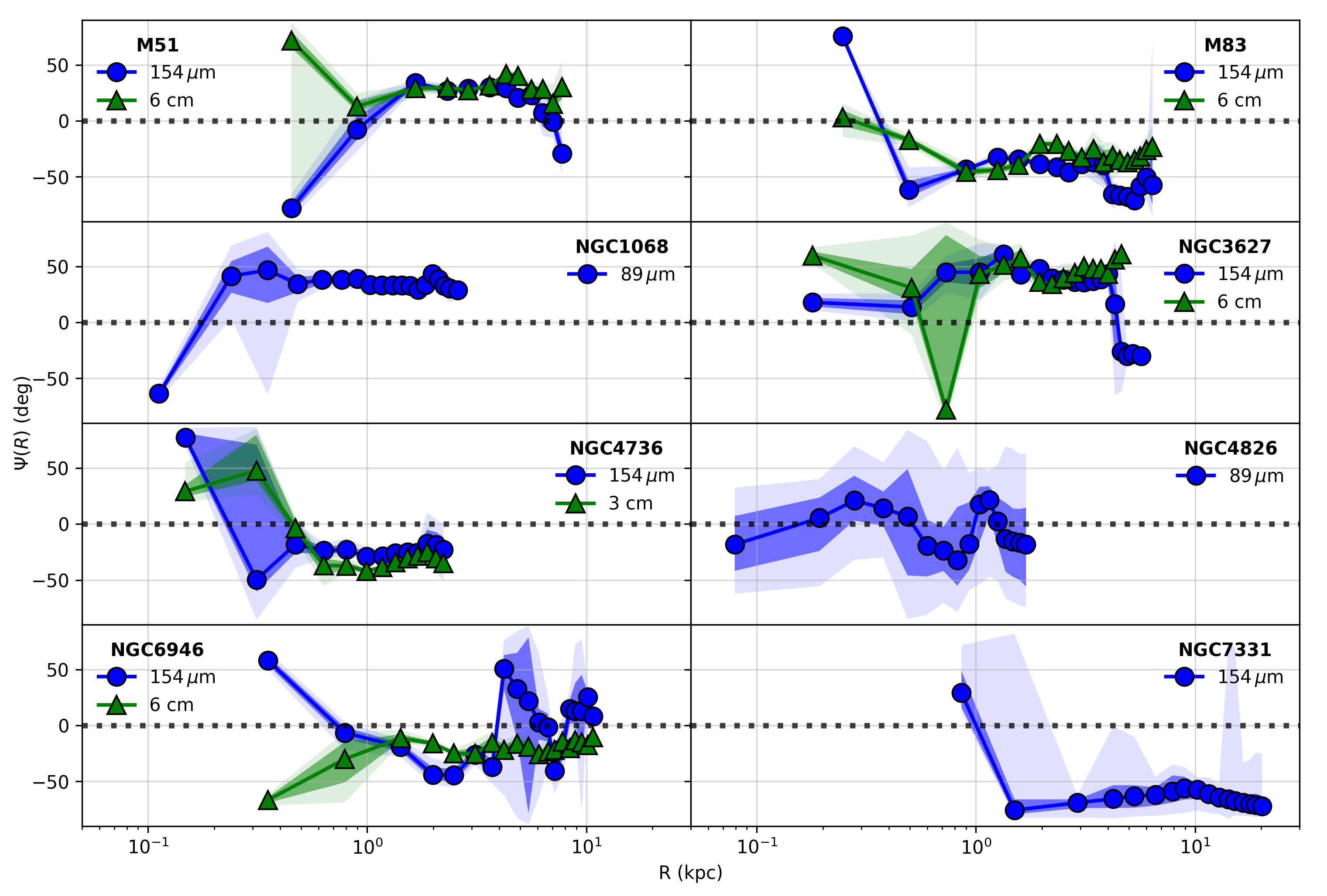}
\put(48.4,58.7){\includegraphics[scale=0.12]{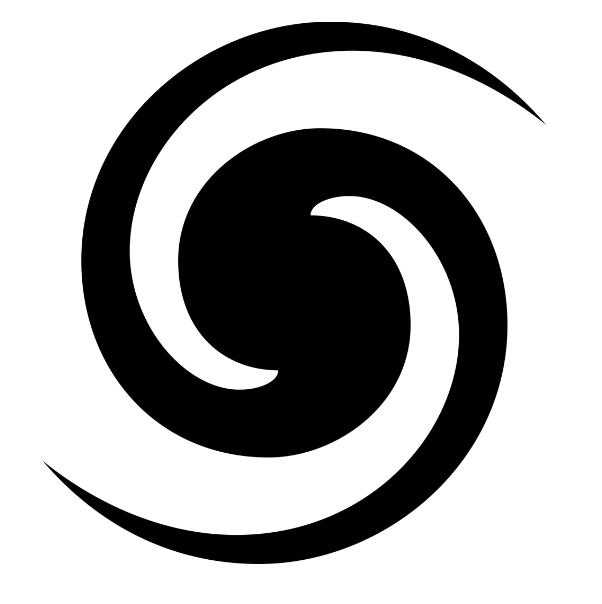}}  
\put(48.4,50.9){\includegraphics[scale=0.12]{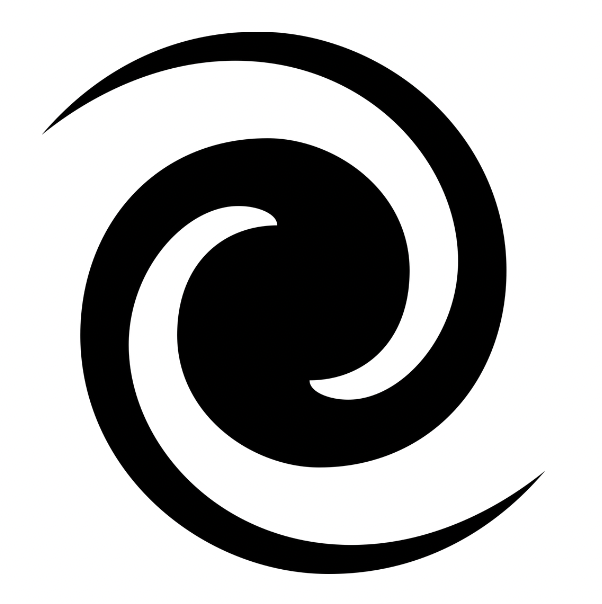}}  
\end{overpic}
\caption{Radial profiles of the magnetic pitch angle ($\Psi_{B}$) of spiral galaxies at FIR (blue) and radio (green) wavelengths. The 68\% ($1\sigma$) and 95\% ($2\sigma$) statistical significance error contours are shown as shadowed areas.}
\label{fig:pitch_angle_profiles}
\end{figure*}

Figure \ref{fig:pitch_angle_profiles} shows the magnetic pitch angles as a function of the galactocentric radius at FIR, \mFIRPsi, of the sample of spiral galaxies. Our FIR observations trace the $B$-fields from $0.1$ kpc to $5$ kpc in radius for most galaxies, and up to $15$ kpc in NGC~7331. We also compute the magnetic pitch angles at 3 and 6 cm, \mRadioPsi, whenever available, using the same LOS measurements as those at FIR wavelengths. Table \ref{tab:pB} shows the median magnetic pitch angles across the full galaxy disk.

\begin{deluxetable*}{lccccccc}
\tablecaption{\textbf{Magnetic pitch angles of spiral galaxies}. \emph{Columns, from left to right:} 
(a) Galaxy name.
(b) FIR wavelength.
(c) FIR magnetic pitch angle. The errors represent the standard deviation of the pitch angle profile, not the uncertainties of the average value. Note: These values are estimated from the pitch angle profile bins (not pixel-to-pixel), and their uncertainties are intended to represent its large scale variability. 
(d) Average $\zeta$ alignment of the FIR spiral magnetic field.
(e) Radio wavelength.
(f) Radio magnetic pitch angle.
(g) Average $\zeta$ alignment of the radio spiral magnetic field.
(h) Difference in $\zeta$ alignment between the radio and FIR spiral magnetic field.
\label{tab:pB} 
}
\tablecolumns{9}
\tablewidth{0pt}
\tablehead{\colhead{Galaxy} & 	\colhead{Band}  & \colhead{$\langle \Psi_{B}^{\rm{FIR}} \rangle$} & 
\colhead{$\zeta^{\rm{FIR}}$} & \colhead{Band}  & \colhead{$\langle \Psi_{B}^{\rm{Radio}} \rangle$}   & \colhead{$\zeta^{\rm{Radio}}$} & \colhead{$\Delta\zeta^{\rm{Radio-FIR}}$}\\ 
 					   &  \colhead{(\um)}	& \colhead{($^{\circ}$)}	& 
 			(\%)		   & \colhead{(cm)}	& \colhead{($^{\circ}$)} & (\%) & (\%)\\
\colhead{(a)} & \colhead{(b)} & \colhead{(c)} & \colhead{(d)} & \colhead{(e)} & \colhead{(f)} & \colhead{(g)} & \colhead{(h)}} 
\startdata
M~51 			&	154				    &	$28.2^{+1.3}_{-40.5}$	&   $88.7^{+1.2}_{-1.2}$ &   3\,cm   & $29.80^{+10.7}_{-5.6}$ &   $95.28^{+0.62}_{-0.96}$   &    $+6.4^{+3.0}_{-2.4}$ \\ 
	&  &	&   &   6\,cm   &  $29.8^{+10.7}_{-5.6}$   & $94.31^{+0.85}_{-0.45}$  &    $+7.8^{+2.9}_{-2.5}$\\
M~83 			&	154				    &	$-38.4^{+2.3}_{-27.0}$	&   $80.71^{+0.69}_{-3.4}$   &   6\,cm   &   $-31.2^{+10.4}_{-6.7}$ & $90.5^{+0.5}_{-2.1}$  &  $+12.7^{+4.2}_{-3.6}$ \\  
NGC~1068   &	89				    &	$33.1^{+5.8}_{-2.7}$	&   $86.4^{+5.1}_{-5.2}$   &	-	&	-   &   - & - \\
NGC~3627 	    &	154				    &	$36.6^{+8.9}_{-63.2}$		&   $80.97^{+0.16}_{-1.57}$   &   6\,cm  &   $47.8^{+9.2}_{-12.8}$ &   $91.93^{+1.16}_{-0.28}$ & $+11.0^{+4.8}_{-3.8}$ \\
NGC~4736 	    &	154				    &	$-24.9^{+6.8}_{-4.0}$	&   $93.6^{+0.1}_{-4.7}$  &   3\,cm   &   $-34.0^{+33.4}_{-3.4}$   &   $95.83^{+0.01}_{-0.35}$  & $+1.7^{+1.5}_{-1.0}$ \\ 
NGC~4826 	    &	89				    &	$-6^{+22}_{-12}$   &	$57.7^{+4.7}_{-4.2}$    &   -	&	-   & - & - \\
NGC~6946 	    &	154 				&	$-6^{+40}_{-31}$	&	$51^{+18}_{-18}$  &   6\,cm	&	$-17.4^{+2.5}_{-7.0}$    &   $88.0^{+1.8}_{-3.9}$&    $+76.0^{+6.2}_{-6.2}$  \\
NGC~7331 	    &	154				    &	$-66.2^{+7.7}_{-4.0}$	&	$86.62^{+0.31}_{-1.53}$   &   -	&	-   &   - & -\\	
\enddata
\end{deluxetable*}

The main result is that FIR and radio wavelengths do not generally trace the same $B$-field morphology in spiral galaxies. This result has also been recently measured by independent analyses, using alternative pitch angle estimation methods \citep{Surgent2023arXiv230207278S}. As reported in \citet{Borlaff2021}, M~51's spiral magnetic field presents a similar structure in both radio and FIR within the central 5 kpc, showing a significant difference in the outskirts of the disk. At radii $R>5$ kpc, a tightening of the FIR spiral magnetic field is observed (\mPsi\ in FIR changes from approximately $25^{\circ}$ to $0^{\circ}$ and negative values). This reduction of the FIR pitch angle is not observed in radio, which displays a pitch angle that opens up with galactocentric radius. Similar discrepancies between FIR and radio $B$-field structure are observed in the outskirts of M\,83. M\,83's FIR polarization structure is more loosely wounded than at radio wavelengths (average $\mPsi = -39.9^{+2.2\circ}_{-3.3}$ at 154\,\um\ and $\mPsi = -29.7^{+3.6\circ}_{-3.9}$ at 6 cm), and there is a trend towards more negative values in FIR not detected in radio. This behavior is opposite to what it is observed in NGC\,3627 ($\mPsi = -38.2^{+1.2\circ}_{-1.3}$ at 154\,\um\ and $\mPsi = -47.71^{+0.13\circ}_{-0.01}$ at 6 cm), and in the outskirts of M\,51 ($R>5$ kpc). 

NGC\,4736 has a similar spiral $B$-field orientation in both FIR and radio wavelengths ($\mPsi = -25.33^{+1.51\circ}_{-0.66}$ at 154\,\um~ and $\mPsi = -35.0^{+4.2\circ}_{-2.3}$ at 3 cm) within the central 2 kpc. As shown in Fig.\,\ref{fig:FIR_morphology_spirals}, the northern spiral arm of NGC~4736 has similar pitch angles in both FIR and radio wavelengths. However, the southern arm has a more tightly wounded magnetic pitch angle at FIR than at radio wavelengths. 

The lack of a well-ordered spiral magnetic structure in NGC\,6946 at FIR wavelengths is clear in the magnetic pitch angle profile. NGC\,6946 has a regular $B$-field at radio wavelengths ($\Psi^{\rm{6\,cm}}=-17.6^{+1.7\circ}_{-4.1}$) and mostly random at FIR wavelengths, although signs of a certain spiral structure can be detected with high variability ($\Psi^{\rm{FIR}}=-6.5^{+5.0\circ}_{-12.5}$). The $B$-field morphology at FIR wavelengths is very patchy and mostly coincident with compact regions of high FIR emission, potentially associated with star-forming regions across the disk \citep[SALSA IV,][]{SALSAIV}. This result is in contrast to the regular spiral $B$-field (`magnetic arms') observed at radio wavelengths \citep{Beck1991,Beck2007}. The difference between the radio polarization pattern and the FIR observations is clearly visible in Fig.\,\ref{fig:FIR_morphology_spirals}. 

For those galaxies without radio polarimetric observations, we found that a) the highly inclined galaxies NGC~4826 ($\mPsi = -5^{+12\circ}_{-11}$ at 89\,\um) and NGC~7331 ($\mPsi = -66.0^{+1.9\circ}_{-2.8}$ at 154\,\um) have an ordered spiral $B$-field morphology at FIR wavelengths, and b) the spiral galaxy NGC 1068 ($\mPsi = 33.23^{+0.43\circ}_{-0.13}$ at 89\,\um) has a magnetic pitch angle that is more tightly wounded with increasing galactocentric radius ($\rho=-0.618$, $p=0.006$, Spearman correlation test).

In summary:

\begin{itemize}
\item FIR and radio wavelengths do not generally trace the same $B$-field morphology in spiral galaxies.

\item Magnetic pitch angle profiles traced with radio polarization observations present a more constant distribution across the disk than those obtained with FIR (e.g., the outskirts of M~51, M~83, or the entire disk of NGC\,6946).

\item NGC\,6946 has a large-scale well-ordered spiral $B$-field at radio wavelengths, while the $B$-field is highly disordered at FIR wavelengths. The FIR polarization is mainly cospatial with the star-forming regions, while the radio polarization is cospatial with the morphological interarm regions.

\end{itemize}




\section{Angular dispersion}\label{subsec:Dtheta}


\subsection{Methodology}\label{subsubsec:Dthetamethods}

Another interesting quantity is the dispersion of the local $B$-field orientation and the average magnetic pitch angle at each galactrocentric radius. In addition to the magnetic pitch angle, \Mohawc\ calculates the residuals between the magnetic pitch angle model and the measured $B$-field orientation maps on a pixel-by-pixel basis. These residuals can be used to provide a quantitative measurement of the goodness-of-fit for the axisymmetric spiral $B$-field model, and by extension, to compare how `ordered' are the magnetic fields as a function of the tracer. By doing so, we can determine which regions are dominated by large-scale ordered spiral magnetic fields (low residuals) or by `disordered' $B$-fields (high residuals)\footnote{Note that we use the term `disordered' rather than turbulent $B$-fields, because `turbulent' is typically defined as the coherence length of the $B$-field in the ISM of $50-100$ pc \citep{Ruzmaikin1988,brandenburg+2005physrep417_1,Haverkorn2008}. These scales are well-below the physical resolution, $\theta_{\rm{beam,pc}}=[210,1025]$ pc (Table \ref{tab:GalaxySample}), of the FIR and radio observations of spiral galaxies, and hence cannot contribute to polarized emission.}. We define the magnetic alignment parameter, $\zeta$, as:

\begin{equation}
\label{eq:BrBo_2}
\zeta = \cos{(2\delta\theta)}
\end{equation}
\noindent
where $\delta\theta$ is defined in Eq.\,15 from \citet{Clark2019ApJ887136C} as the angle difference from two angular measurements, $\theta_1$ and $\theta_2$,

\begin{equation}
\label{eq:BrBo_1}
\delta\theta = \frac{1}{2} \arctan{\Bigg[\frac{\sin(2\theta_1)\cos(2\theta_2) - \cos(2\theta_1)\sin(2\theta_2)}{\cos(2\theta_1)\cos(2\theta_2) + \sin(2\theta_1)\sin(2\theta_2)}\Bigg]}
\end{equation}

$\zeta$ can range from 1 (perfect alignment) to -1 (perpendicular fields)\footnote{Note that this definition is different to the histogram of relative orientations (HRO) used to quantify the relative orientation between the morphological structure of molecular clouds and the B-field orientation \citep[e.g.,][]{Planck_IR_XXXII,Dylan2018}.}. $\zeta = 1$ thus represents a perfect alignment of the median axisymmetric spiral $B$-field with the local field (large scale $B$-field dominates). A random polarization field (disordered dominated) will yield an average value of $\zeta = 0$. Since the axisymmetric model is measured from the polarization maps, $\zeta = -1$ values are not expected unless extreme small-scale local changes in the $B$-field direction are present.

The $\zeta$ radial profiles can be constructed using the same method as the magnetic pitch angle profiles (see Appendix \ref{App:PitchBmethods}). We use the pitch angle probability distribution of each pixel to calculate $\zeta$ as a function of the galactocentric radius and the associated uncertainties (68\%, 95\%, equivalent to the 1$\sigma$, 2$\sigma$). For all the analyses, we consider a critical level of at least p = 0.05 (95\%) to declare statistical significance. This method is implemented in \textsc{python} and is available on the project website. Most of the large $\zeta$ regions are usually at the cores as expected given the limited statistics within the central 2-3 beams. Low alignment for the inner beams can be caused both by statistical limitations (low number of valid pixels at the core), observational limitations (beam depolarization), and intrinsic physical properties of the core (higher turbulence). Due to this, we remove the central 2-3 beams for the analysis.

\subsection{Results}\label{subsubsec:Dthetaresults}

\begin{figure*}[ht!]
\begin{center}
\includegraphics[width=0.24\textwidth, trim=0 0 0 0]{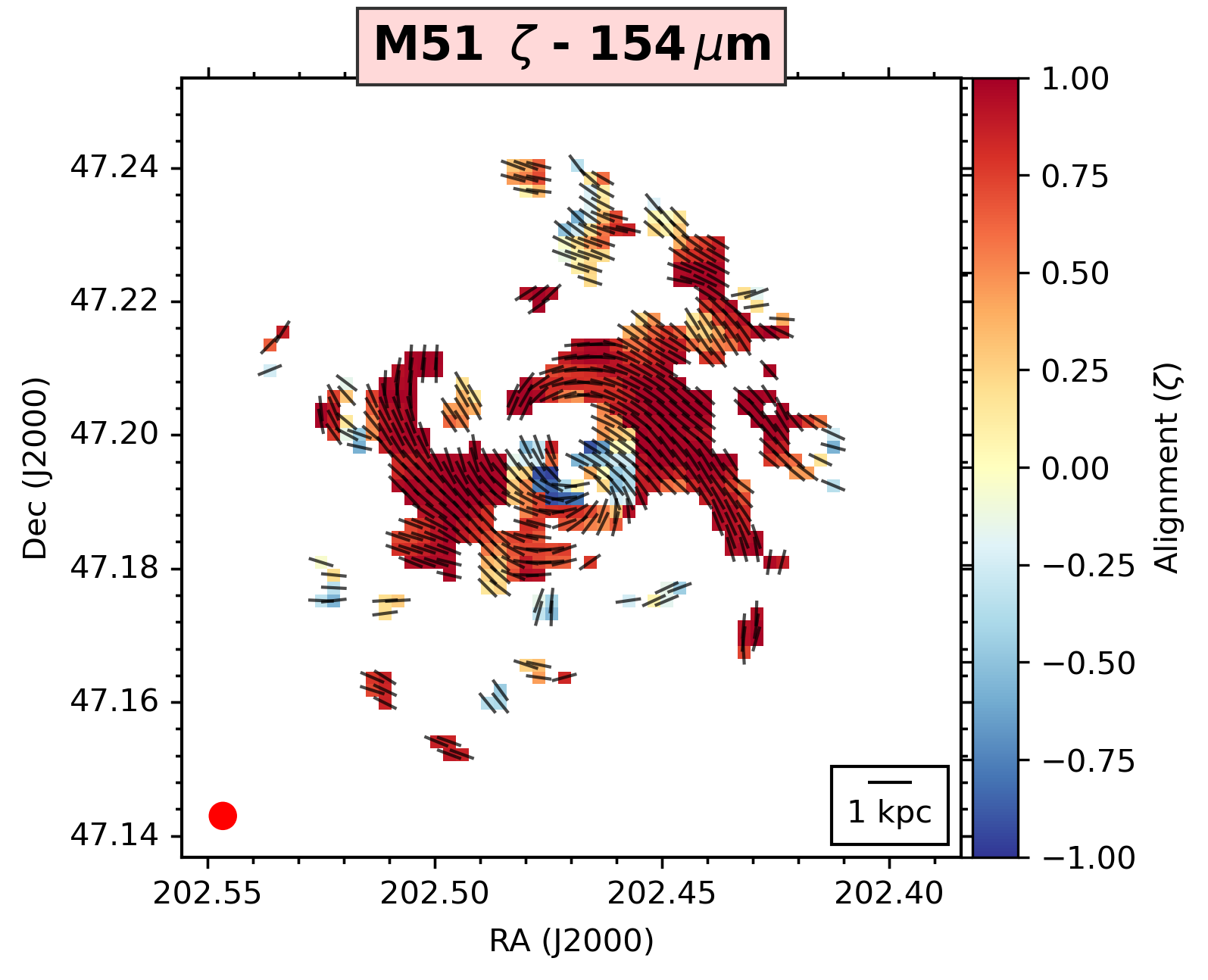}
\includegraphics[width=0.24\textwidth, trim=0 0 0 0]{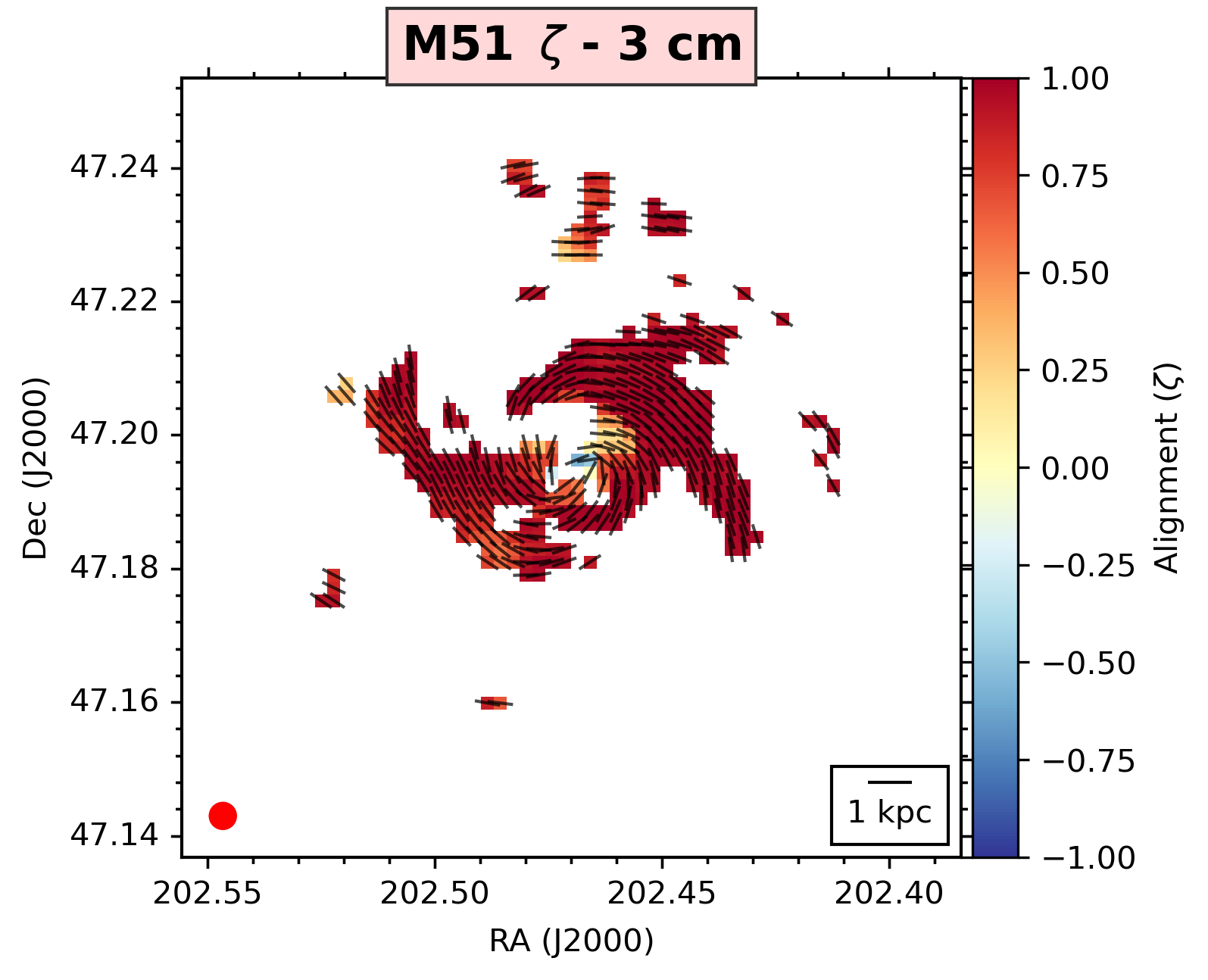}
\includegraphics[width=0.24\textwidth, trim=0 0 0 0]{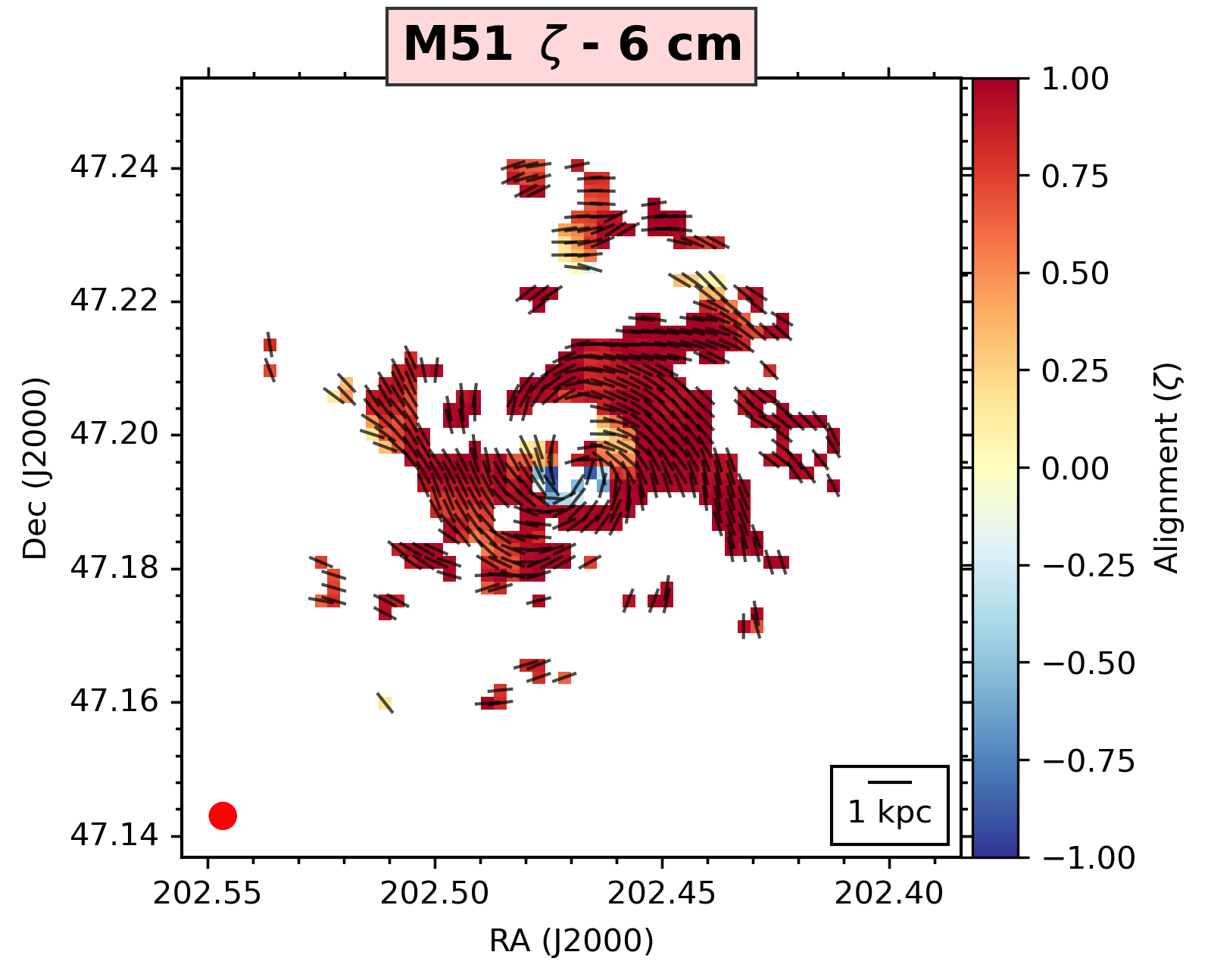}
\includegraphics[width=0.24\textwidth, trim=0 0 0 0]{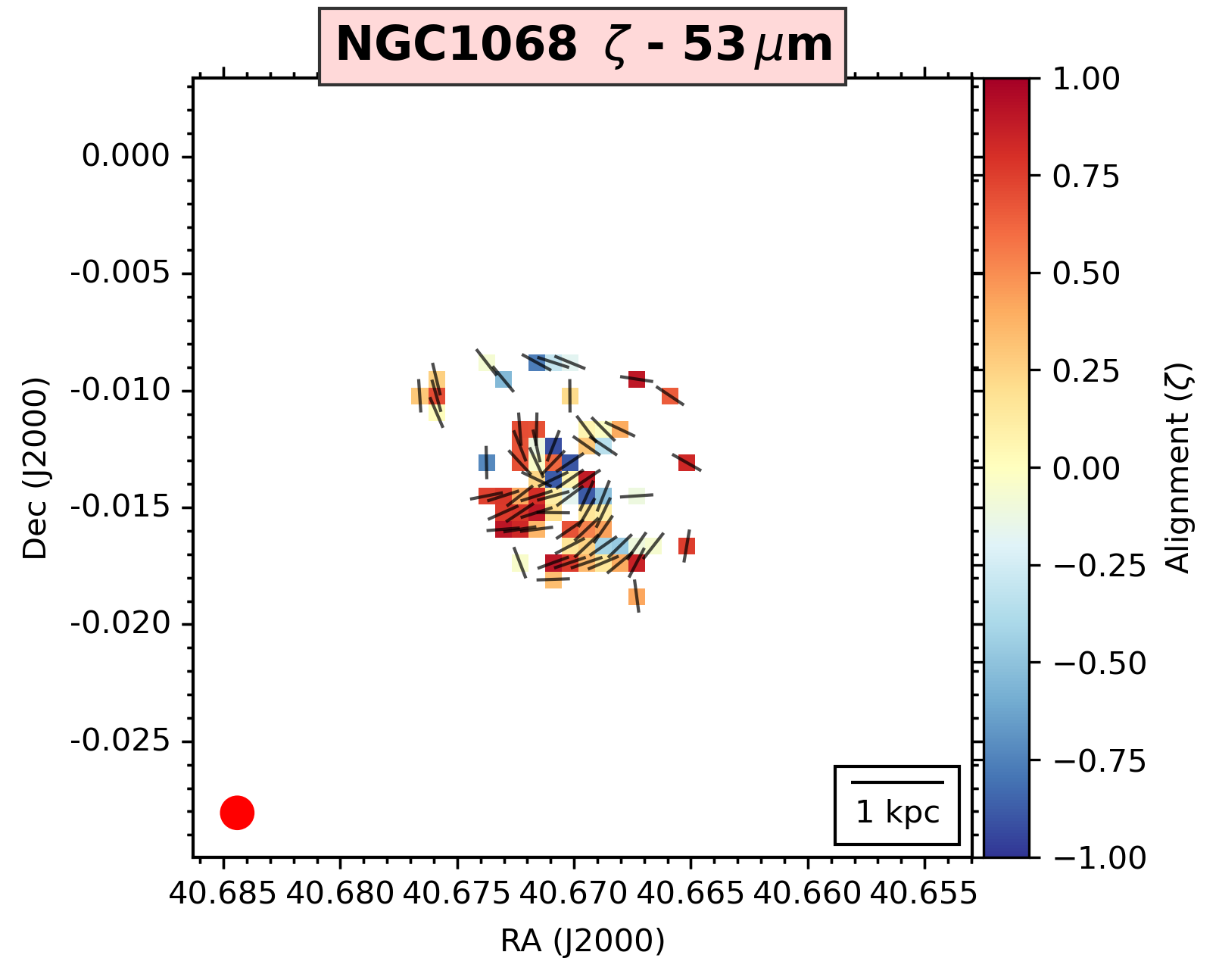}

\includegraphics[width=0.24\textwidth, trim=0 0 0 0]{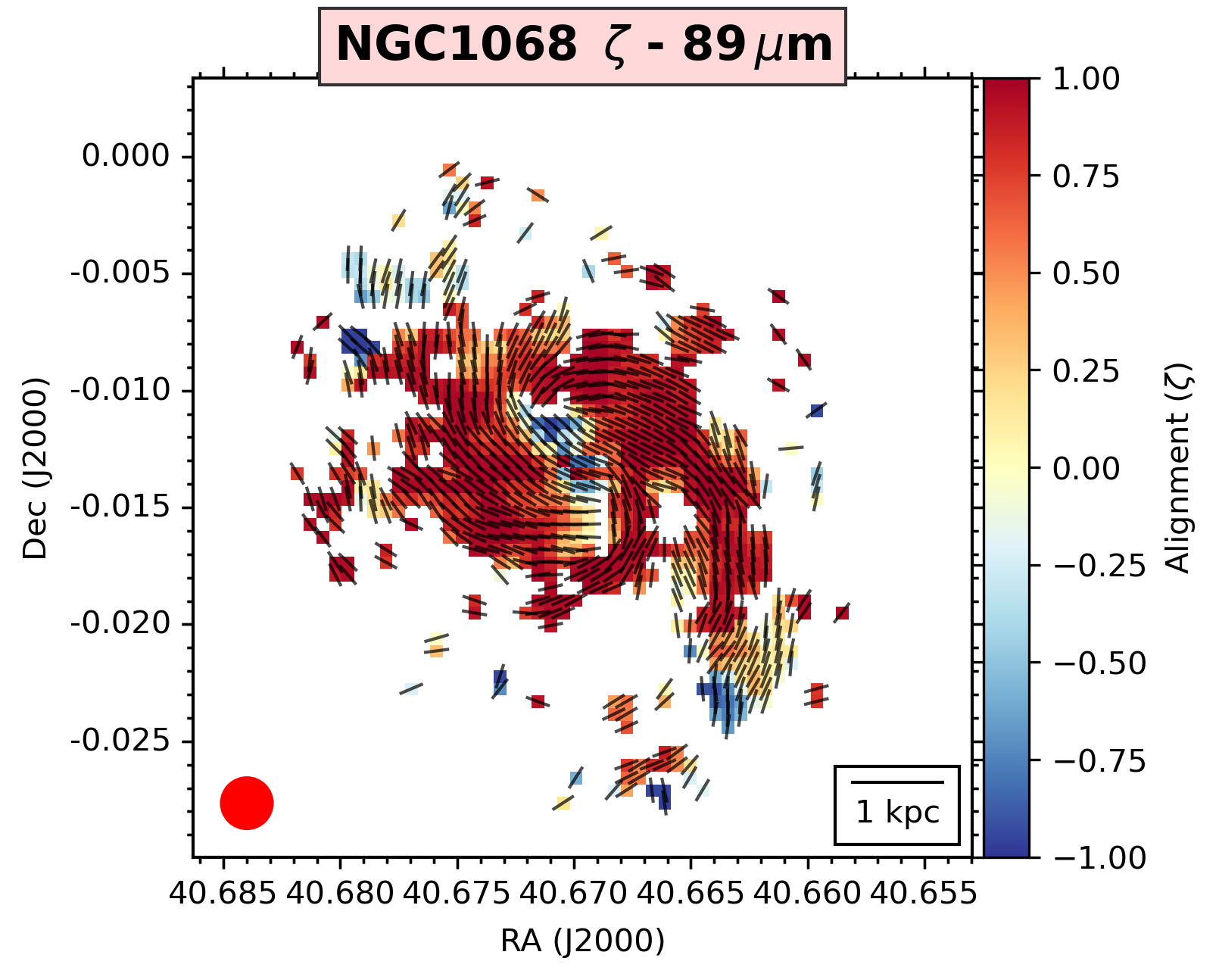}
\includegraphics[width=0.24\textwidth, trim=0 0 0 0]{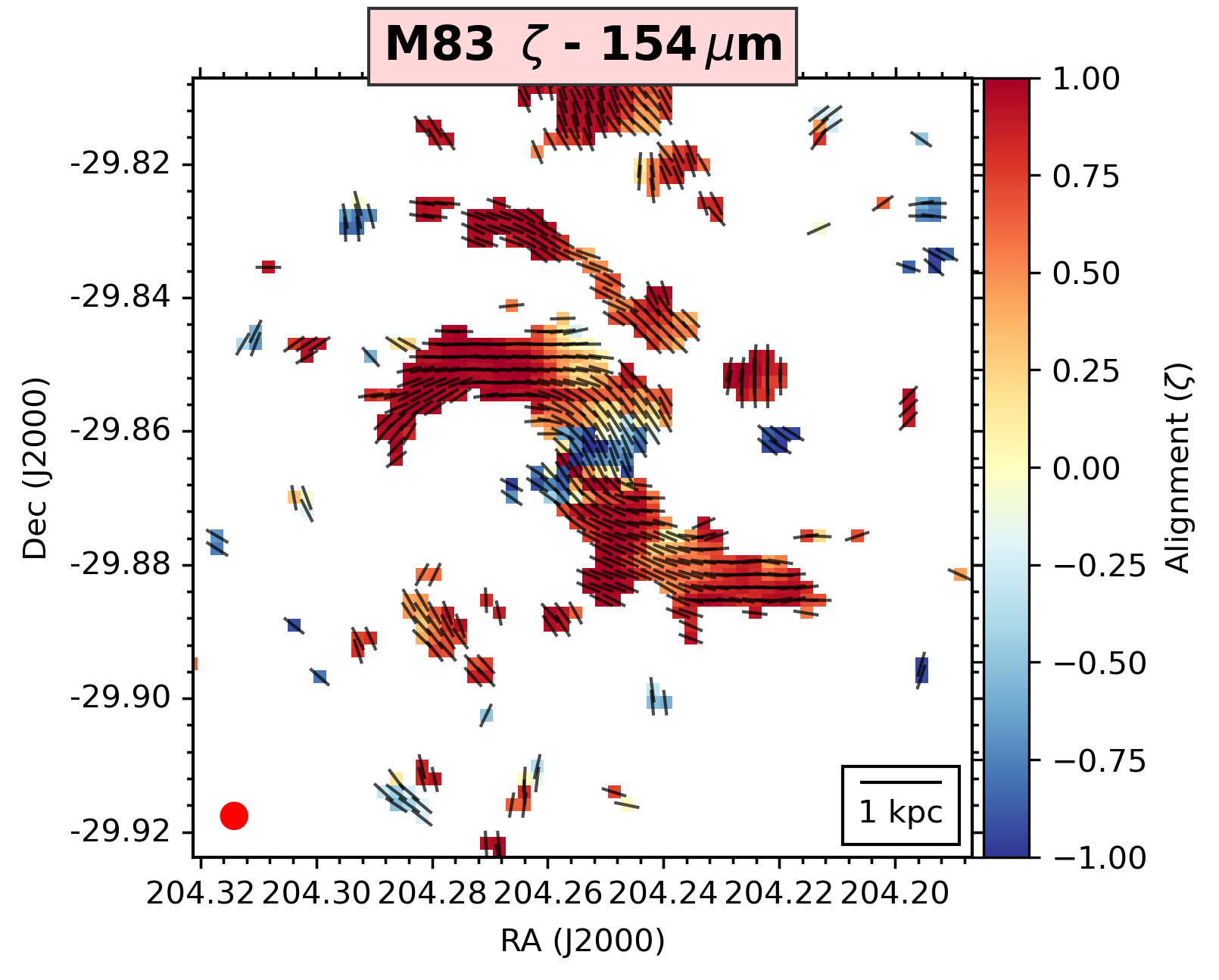}
\includegraphics[width=0.24\textwidth, trim=0 0 0 0]{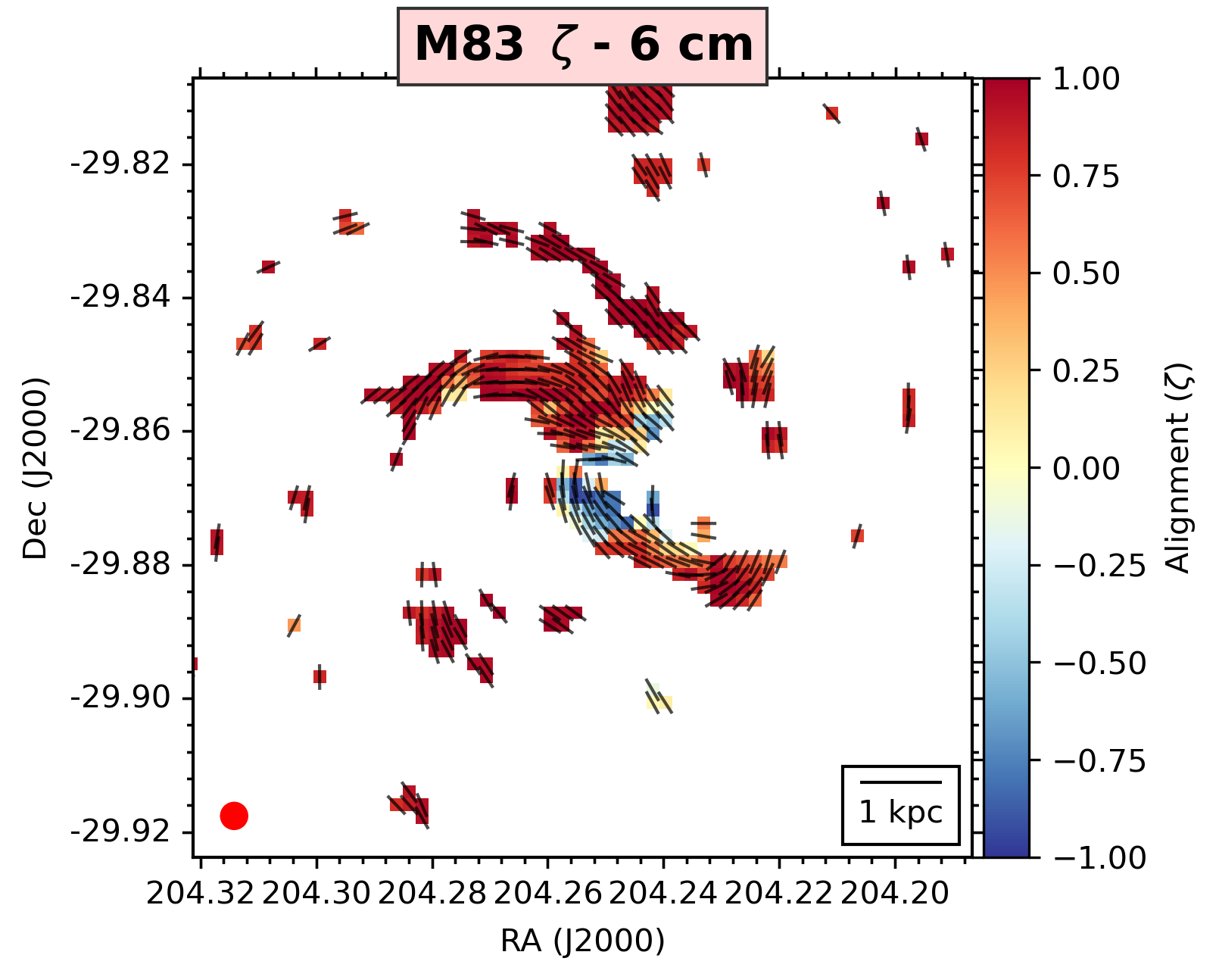}
\includegraphics[width=0.24\textwidth, trim=0 0 0 0]{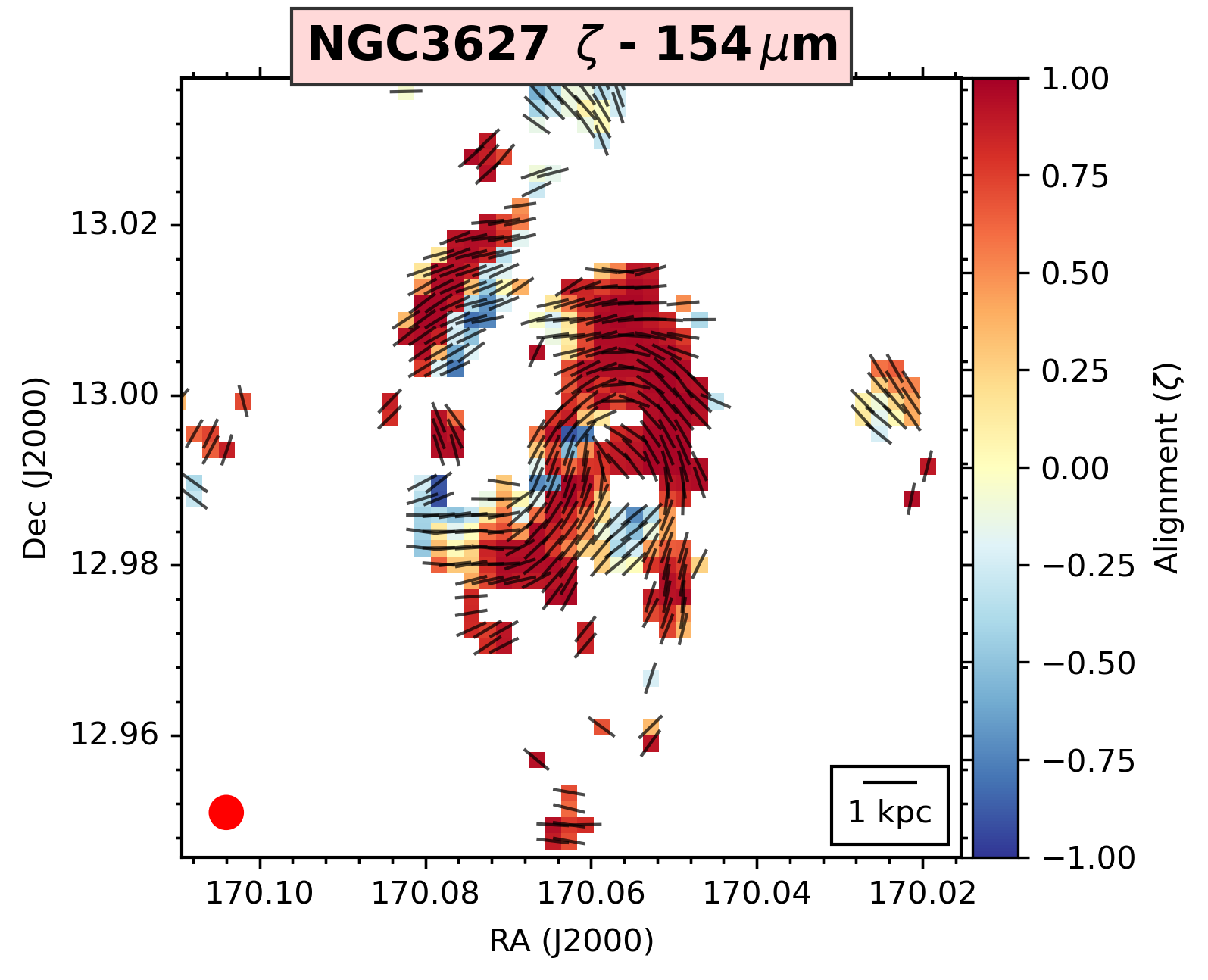}

\includegraphics[width=0.24\textwidth, trim=0 0 0 0]{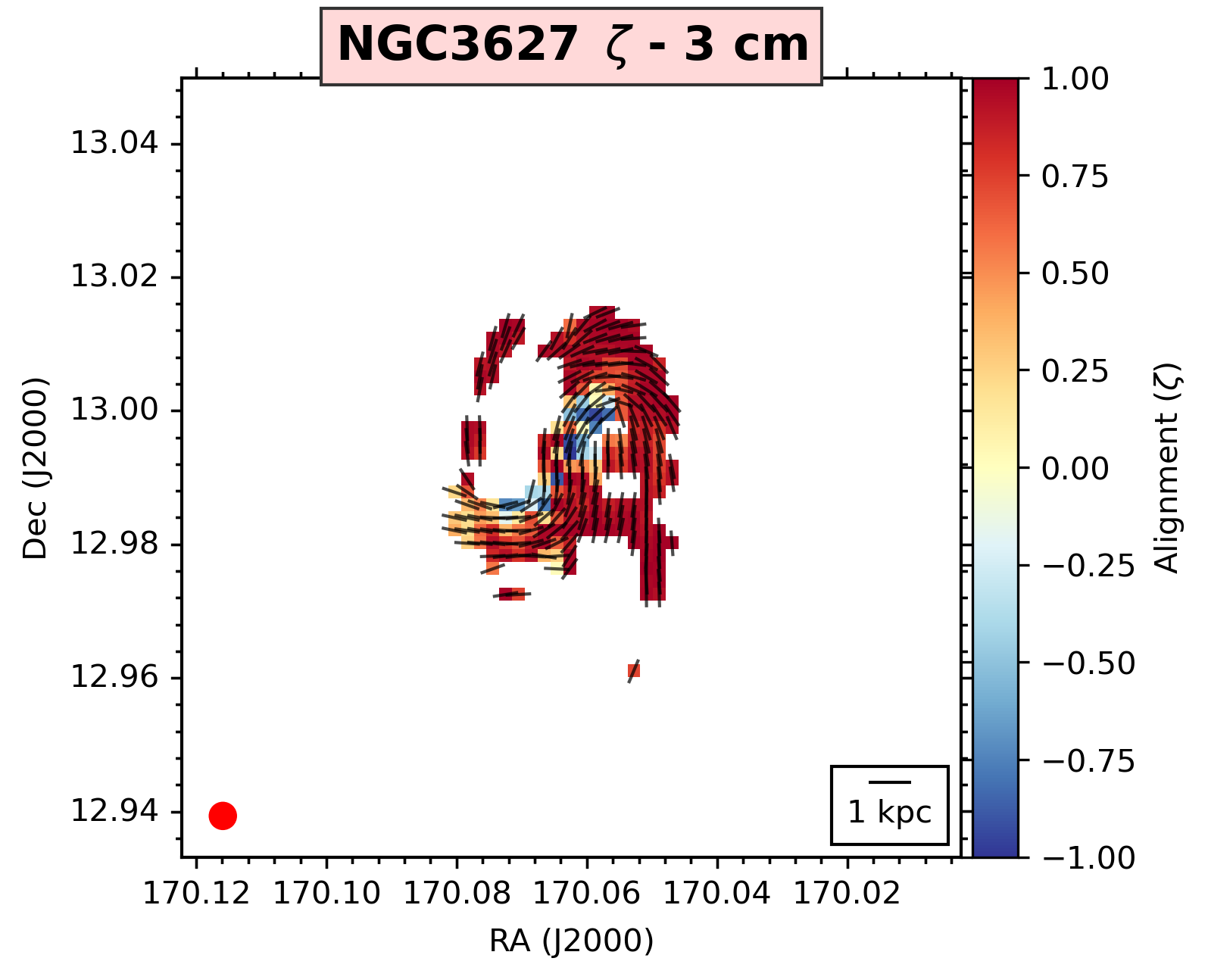}
\includegraphics[width=0.24\textwidth, trim=0 0 0 0]{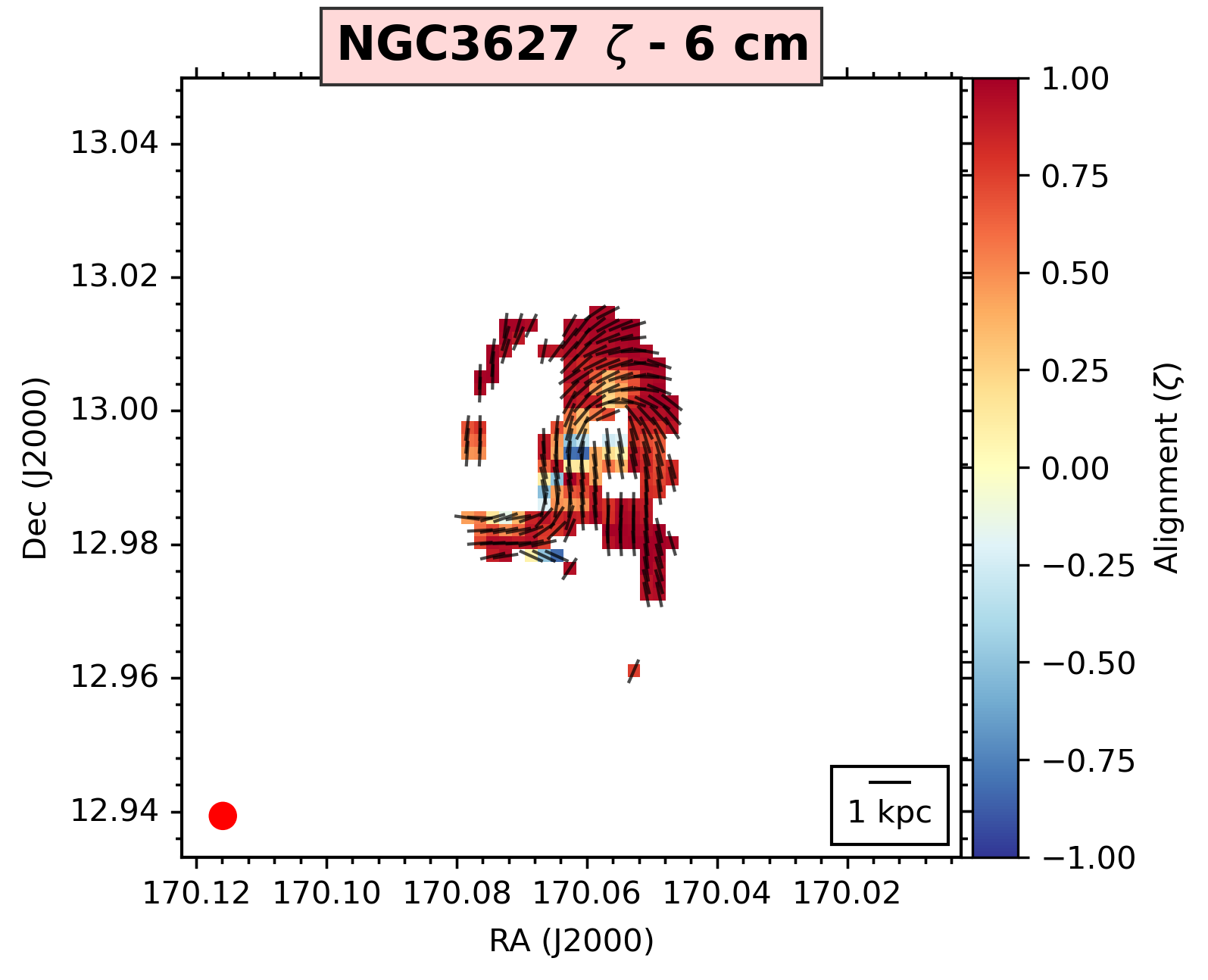}
\includegraphics[width=0.24\textwidth, trim=0 0 0 0]{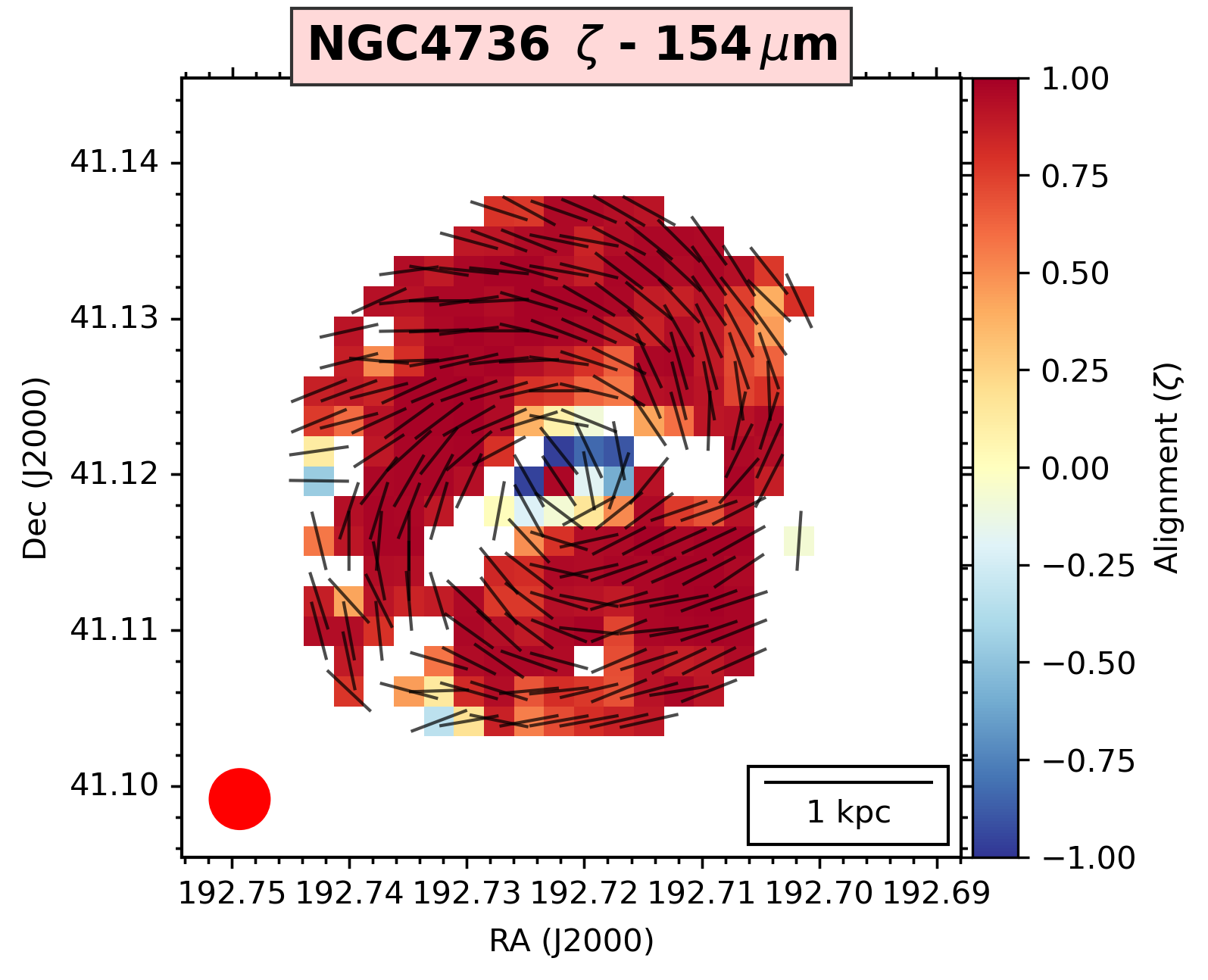}
\includegraphics[width=0.24\textwidth, trim=0 0 0 0]{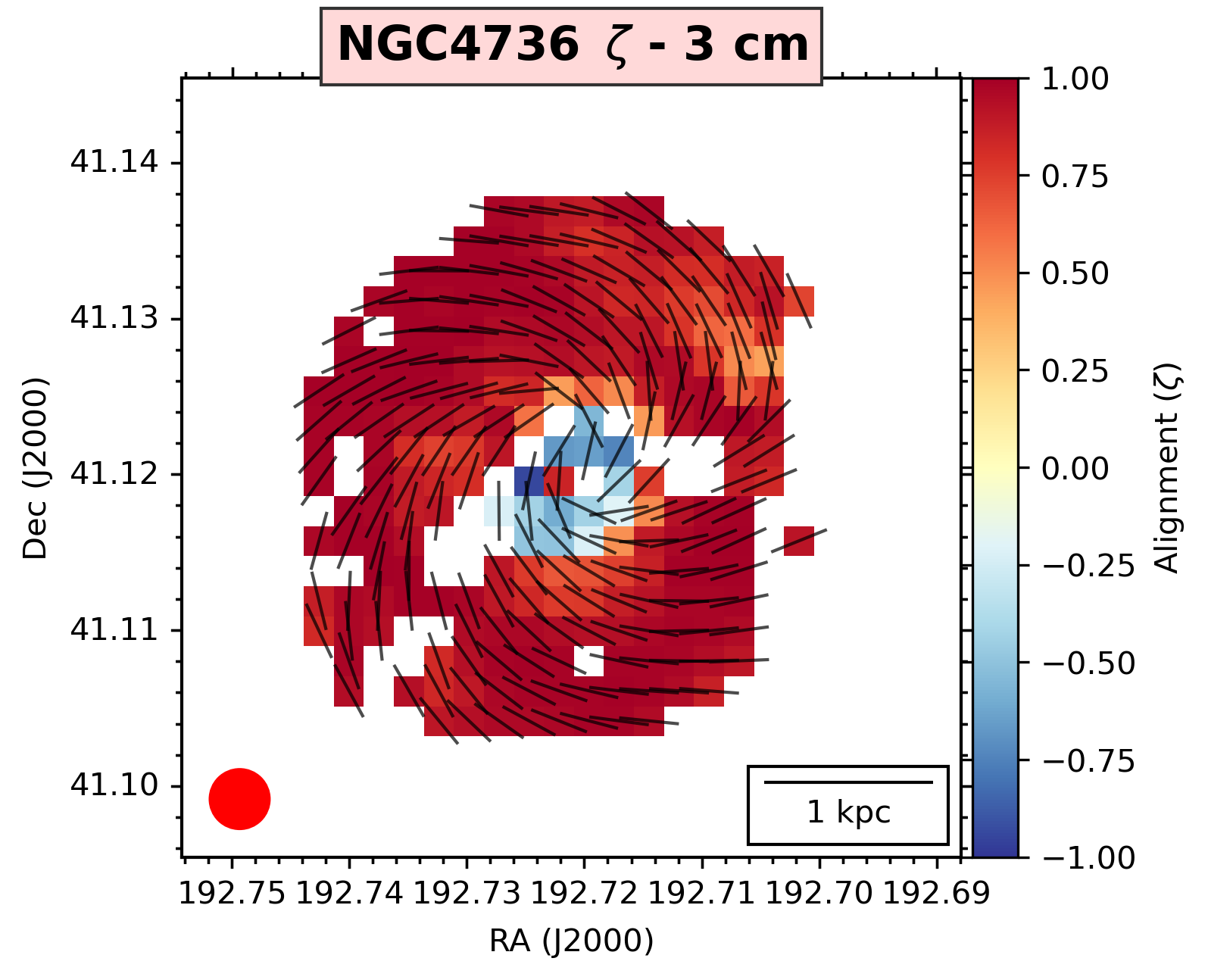}

\includegraphics[width=0.24\textwidth, trim=0 0 0 0]{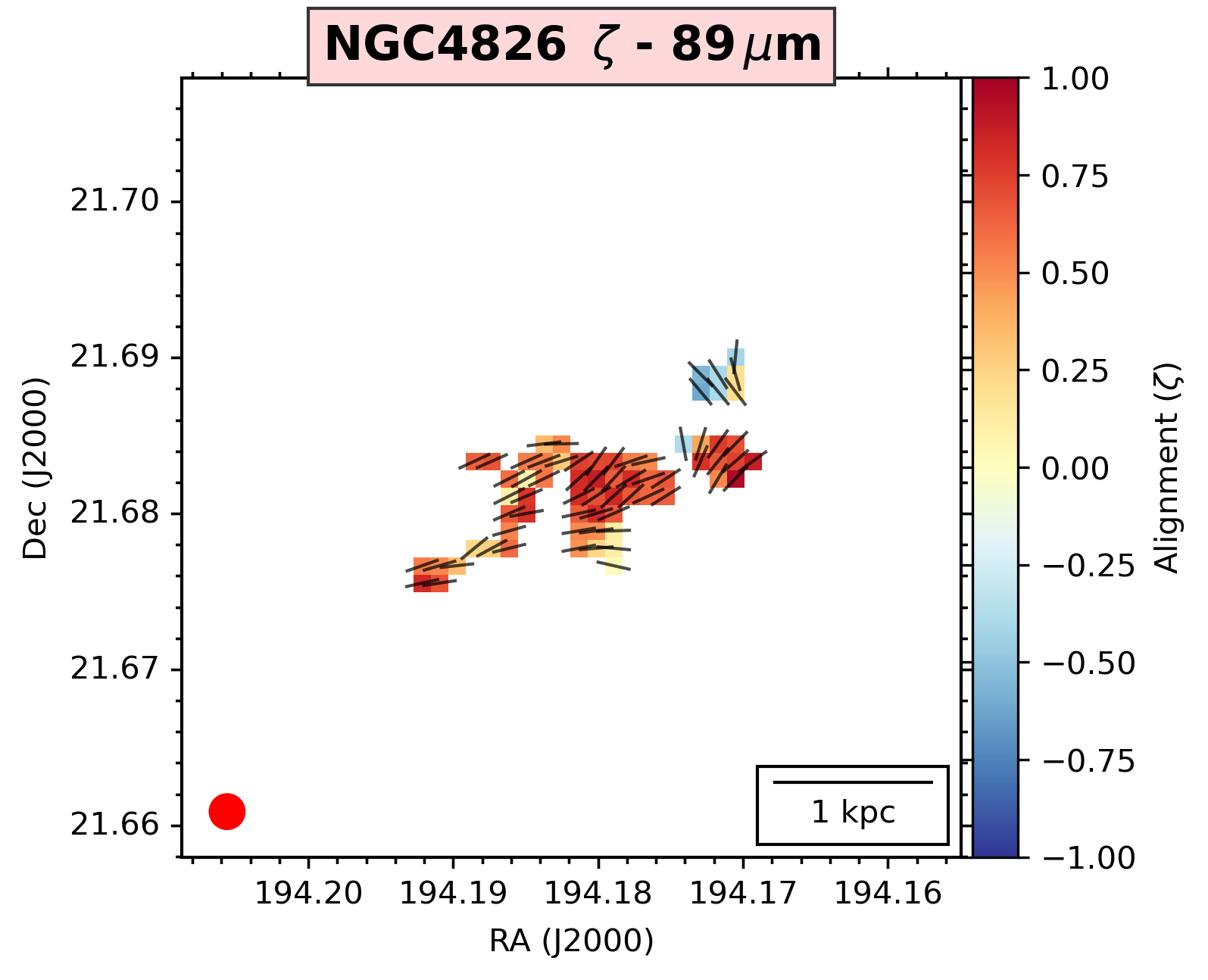}
\includegraphics[width=0.24\textwidth, trim=0 0 0 0]{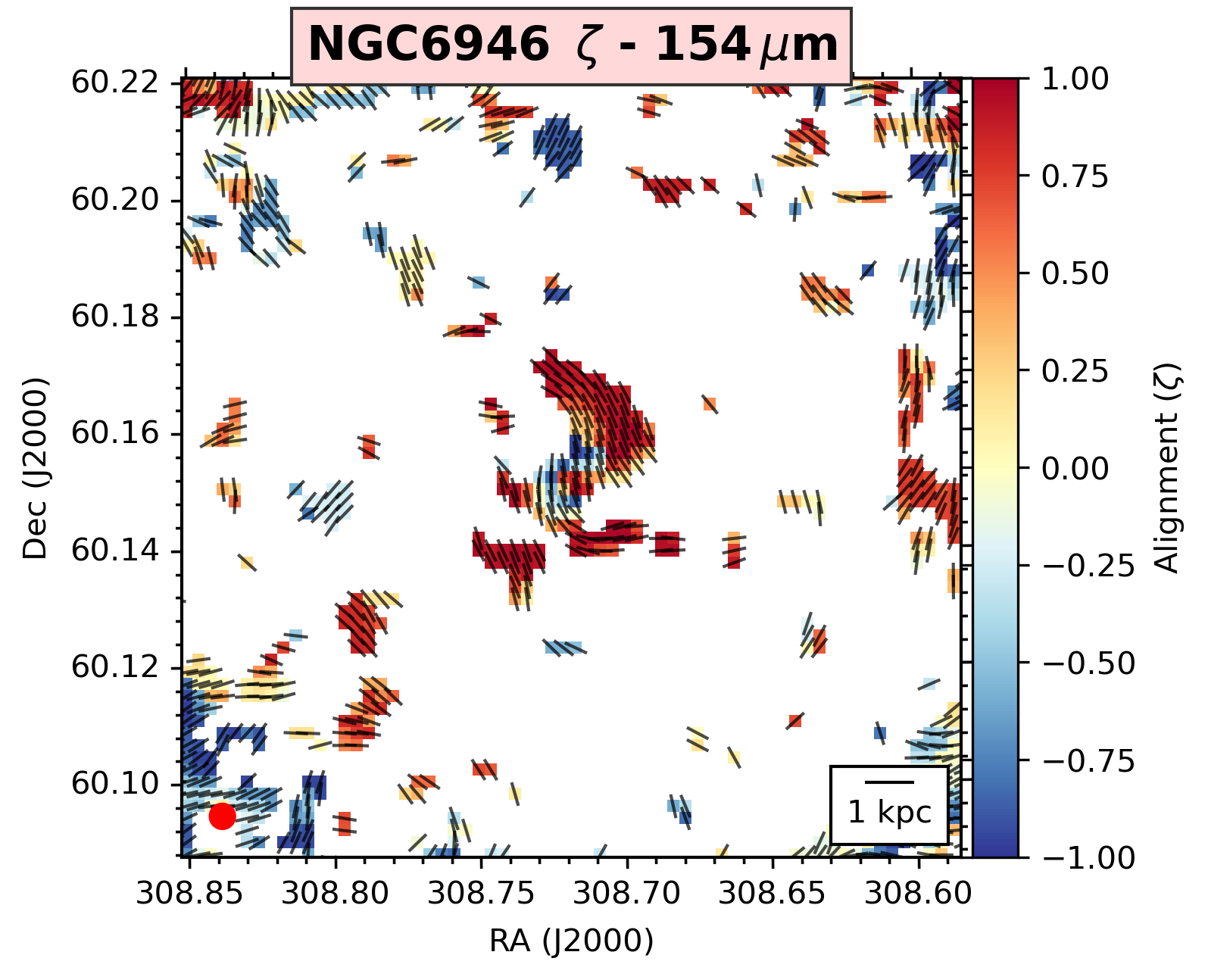}
\includegraphics[width=0.24\textwidth, trim=0 0 0 0]{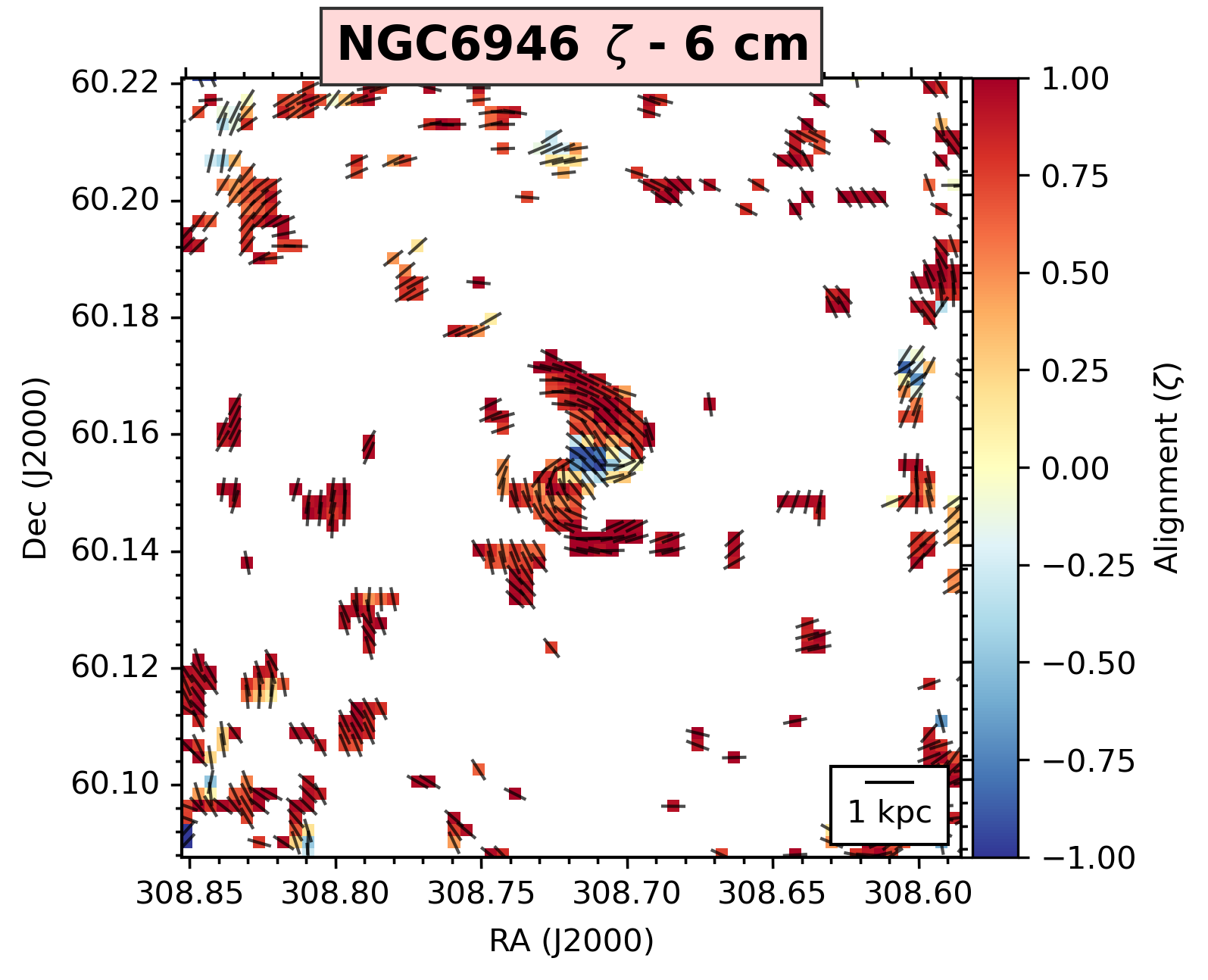}
\includegraphics[width=0.24\textwidth, trim=0 0 0 0]{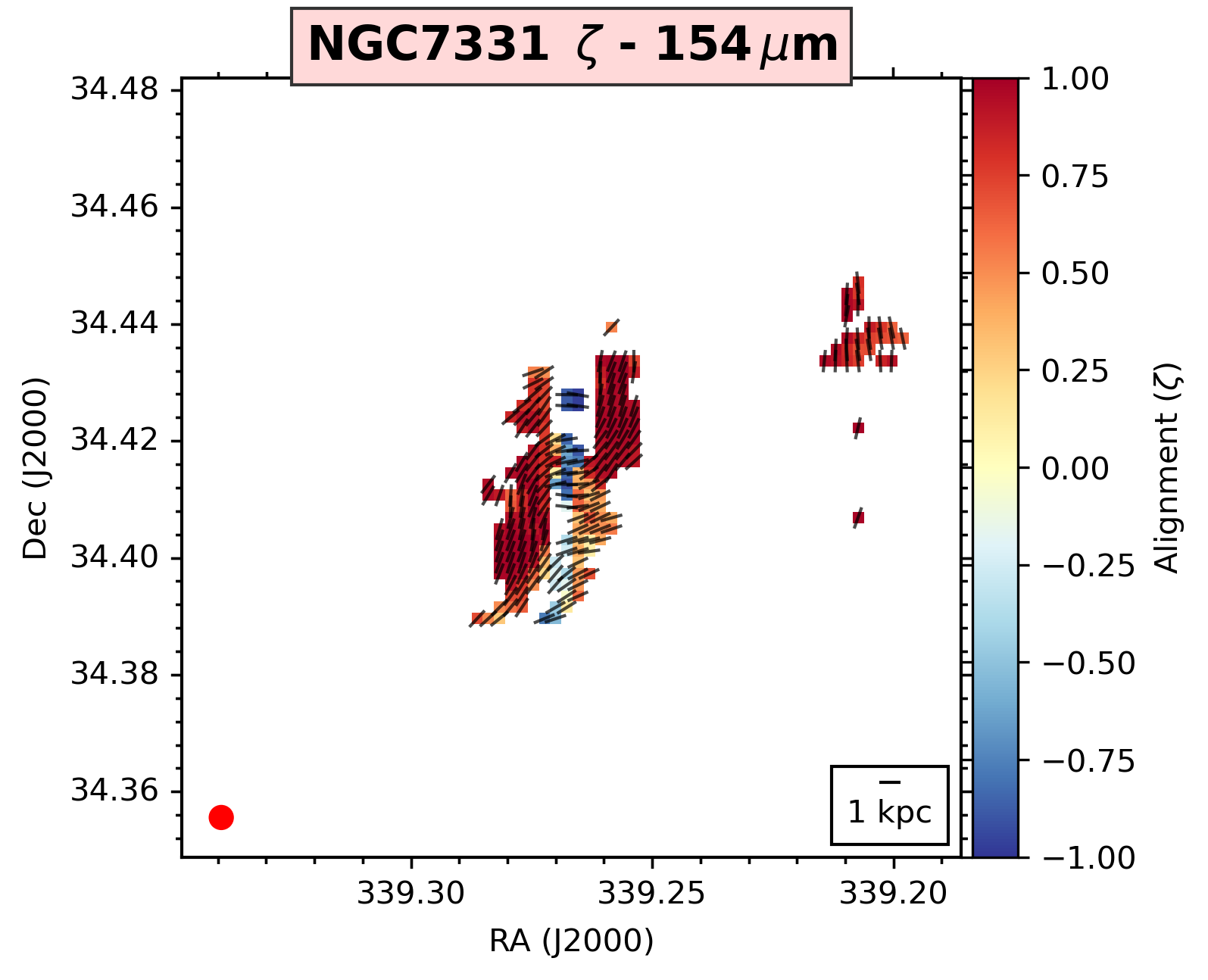}

\end{center}
\caption{Disordered-to-ordered $B$-field ratio (magnetic alignment; $\zeta$) maps at FIR and radio wavelengths of spiral galaxies. The measured (black dashes) $B$-field orientations are shown. Maps are shown with pixels at the Nyquist sampling (Table \ref{tab:HAWC}).
\label{fig:turbulence_maps}}
\end{figure*}

\begin{figure*}[ht!]
\centering
\begin{overpic}[width=\textwidth]{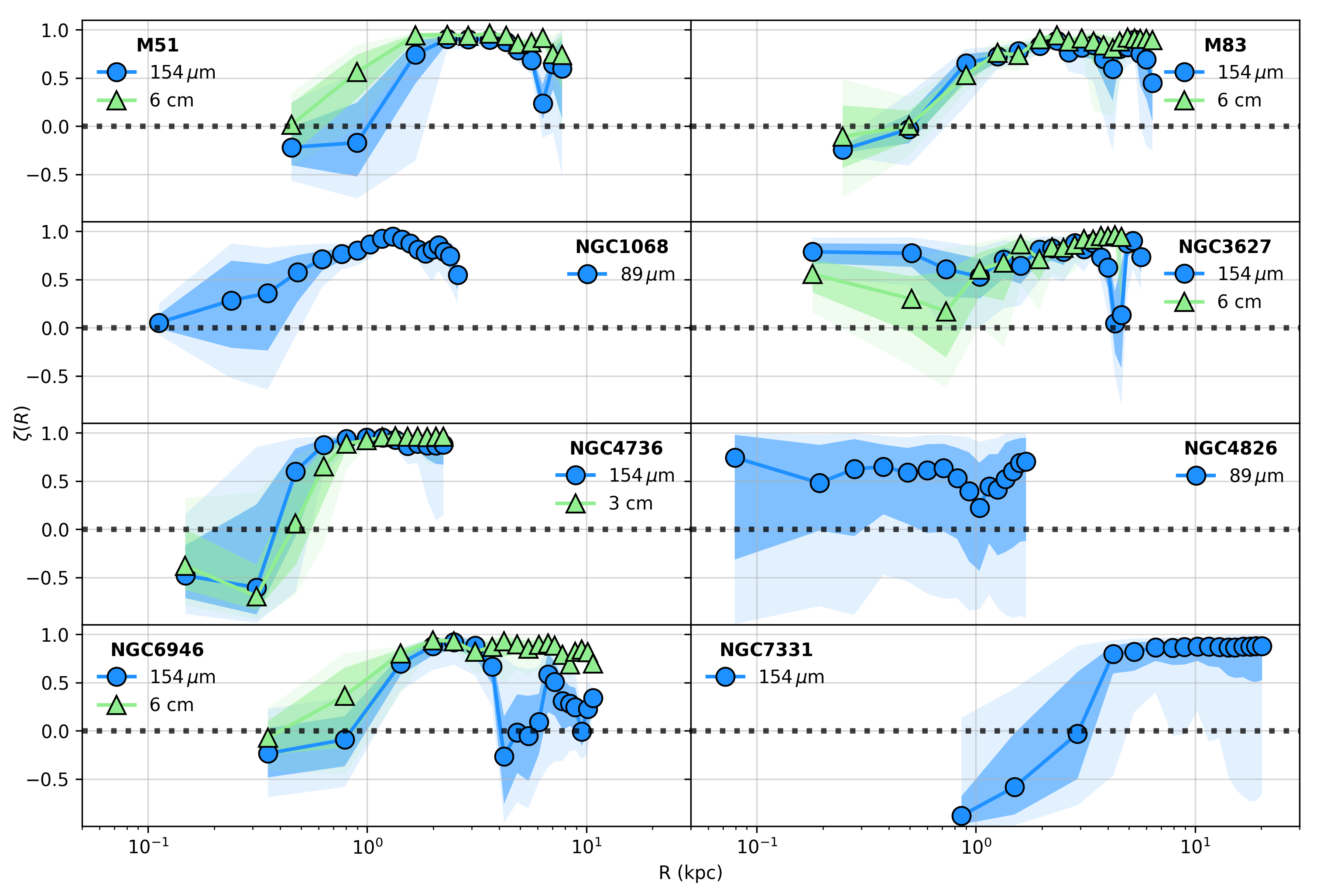}
\put(45,63.822){\textbf{\colorbox{white}{Perfect alignment}}}
\put(48,57.2){\textbf{\colorbox{white}{Random}}}
\put(46,51.885){\textbf{\colorbox{white}{Perpendicular}}}
\end{overpic}
\caption{Radial profiles of the magnetic alignment parameter $\zeta$, of spiral galaxies at FIR (blue) and radio (green) wavelengths. $\zeta=+1$ corresponds to a perfect axisymmetric spiral alignment. $\zeta=0$ corresponds to a completely random $B$-field. $\zeta=-1$ corresponds to a perpendicular alignment to the average $B$-field. The 68\% ($1\sigma$) and 95\% ($2\sigma$) statistical significance error contours are shown as shadowed areas.
\label{fig:zeta_angle_profile}}
\end{figure*}

\begin{figure*}[ht!]
\centering

\includegraphics[width=0.3\textwidth, trim=0 0 0 0]{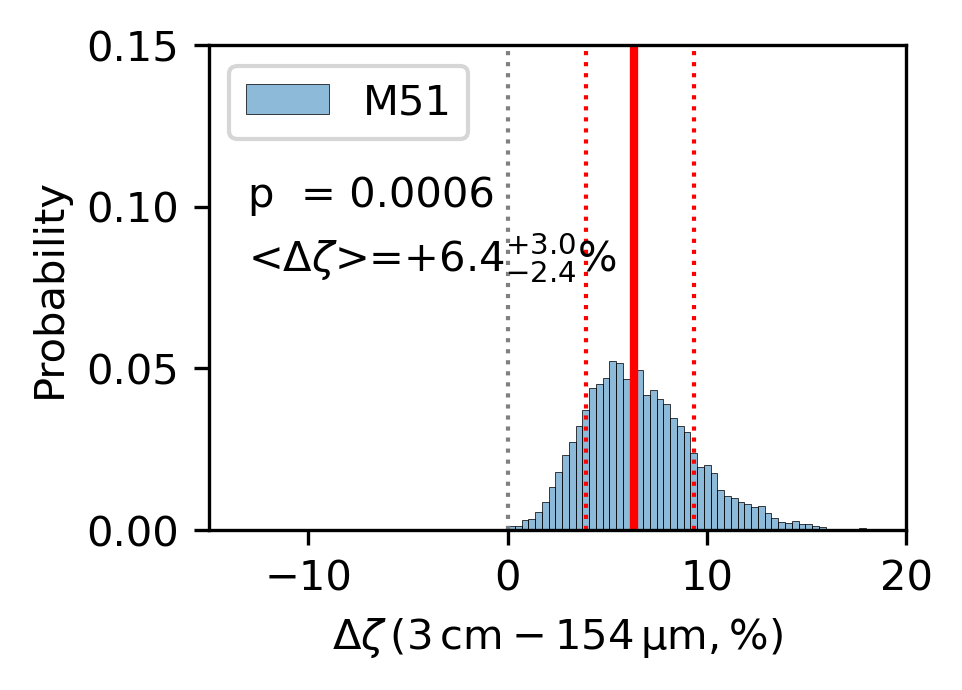}
\includegraphics[width=0.3\textwidth, trim=0 0 0 0]{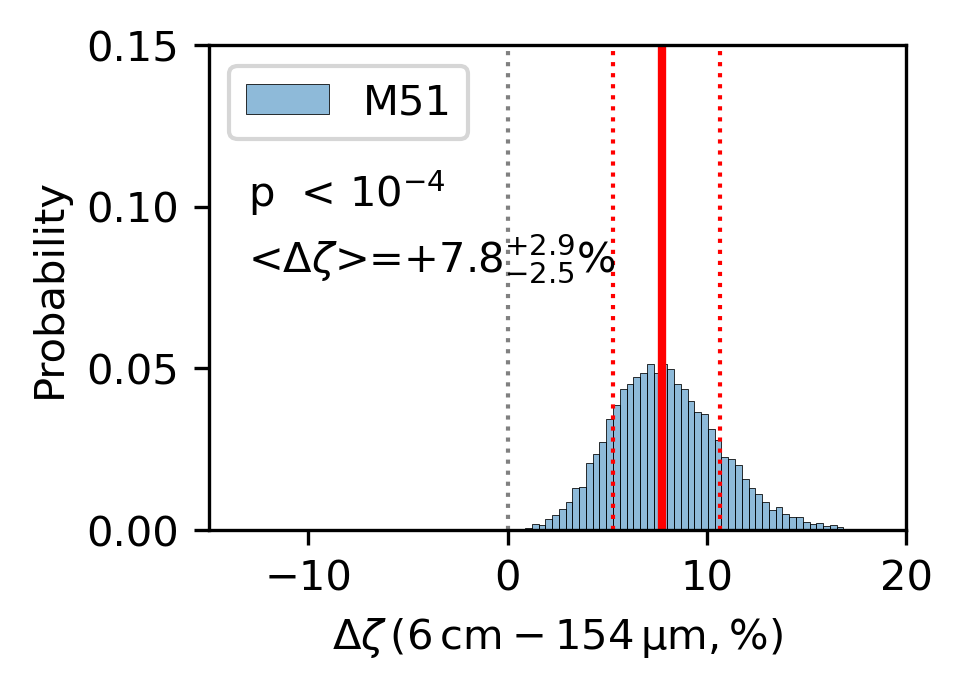}
\includegraphics[width=0.3\textwidth, trim=0 0 0 0]{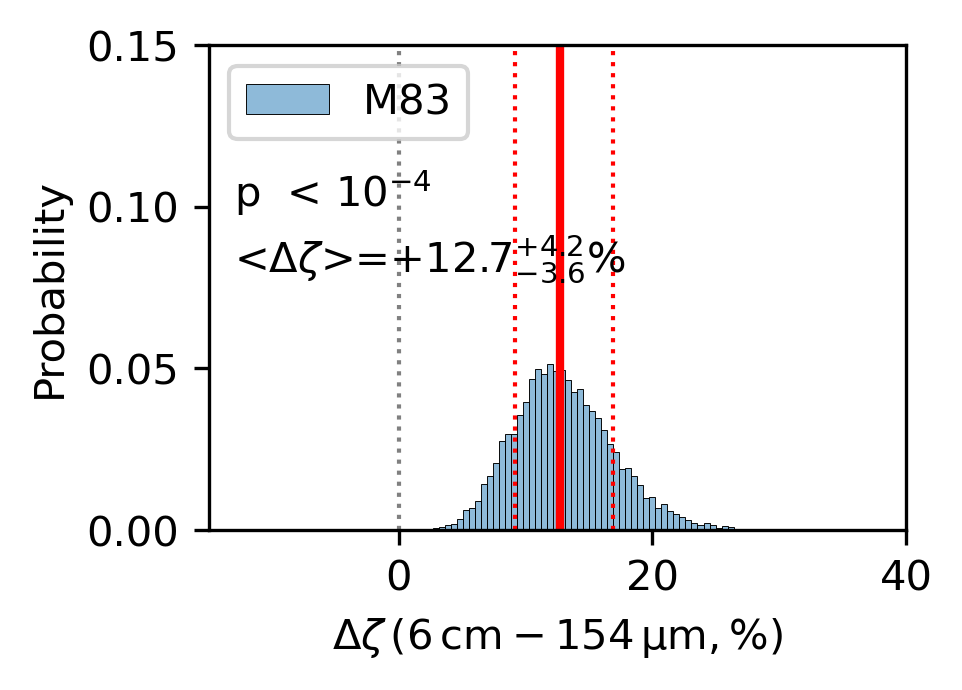}

\includegraphics[width=0.3\textwidth, trim=0 0 0 0]{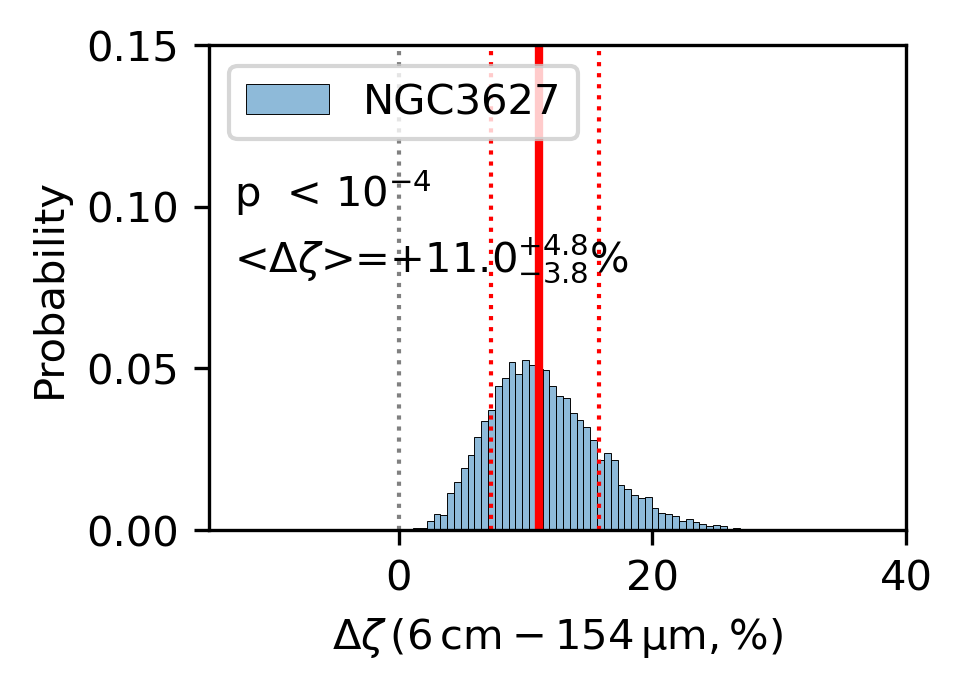}
\includegraphics[width=0.3\textwidth, trim=0 0 0 0]{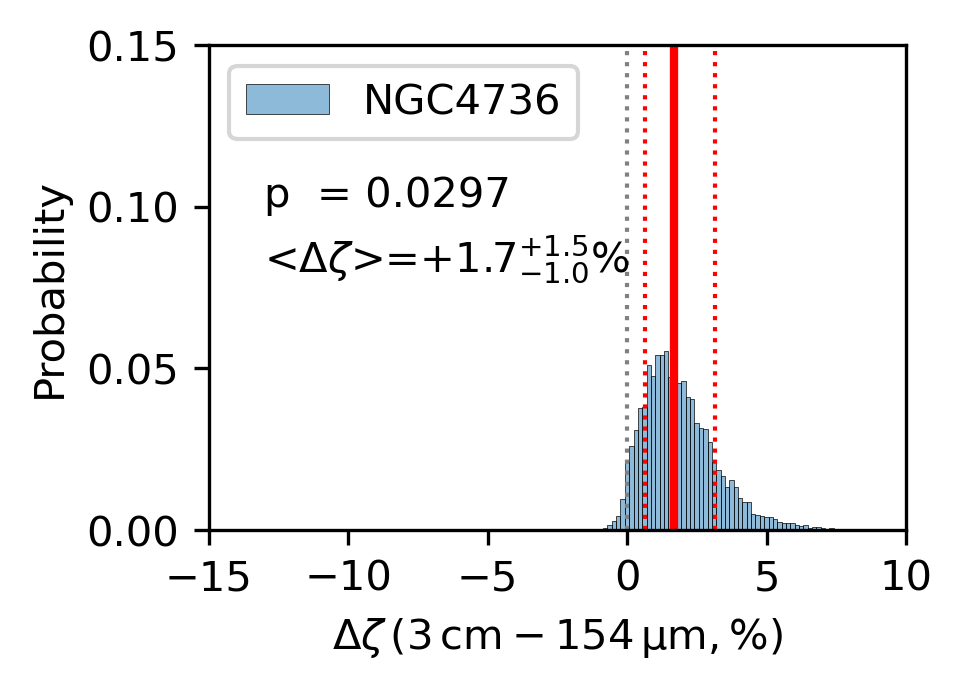}
\includegraphics[width=0.3\textwidth, trim=0 0 0 0]{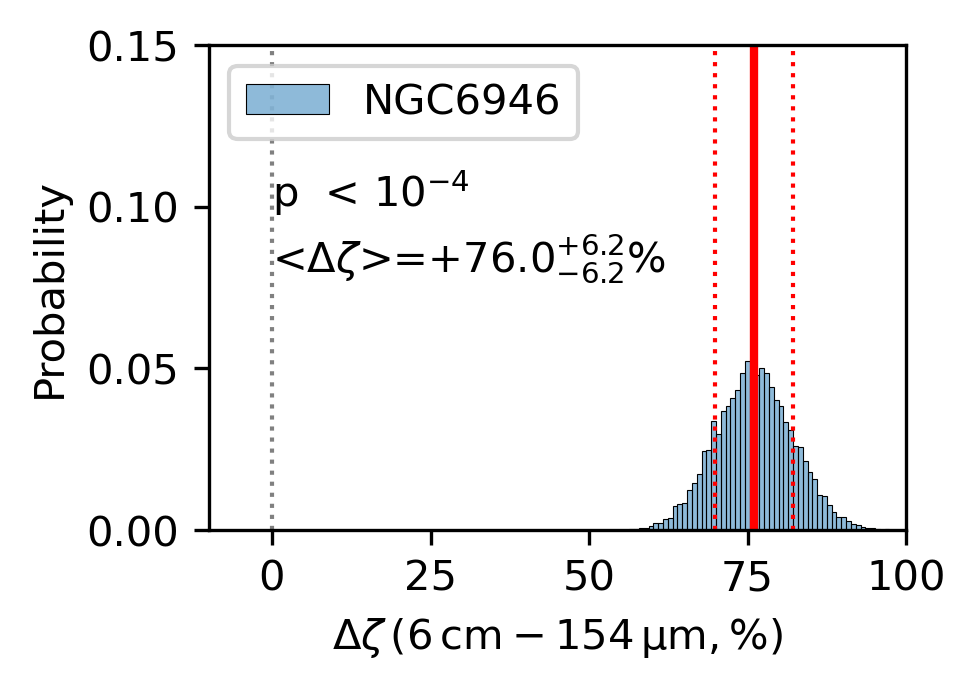}

\caption{Difference between radio and FIR spiral magnetic alignment parameter $\zeta$ of spiral galaxies. The median difference $\Delta\zeta$ (solid red line) and $1\sigma$ uncertainty (dashed red line) for each histogram are shown in the top right of each panel (Table \ref{tab:pB}). The statistical $p$-value for the null hypothesis that $\Delta\zeta<0$ is shown in the panel.} \label{fig:Deltazeta_hist}
\end{figure*}

Figure \ref{fig:turbulence_maps} shows the $\zeta$ maps of spiral galaxies on a pixel-by-pixel basis at FIR and radio wavelengths. Fig.\,\ref{fig:zeta_angle_profile} presents the $\zeta$ profiles of magnetic alignment as a function of galactocentric radius for all spiral galaxies in FIR and radio when available. To compare the magnetic alignment profiles as a function of wavelength (Fig.\,\ref{fig:Deltazeta_hist}), we show the histograms of the difference between radio and FIR, calculated as the difference between the $\zeta$ maps from Fig.\,\ref{fig:turbulence_maps}. The median and $1\sigma$ uncertainty of the histogram are shown in each plot and Table \ref{tab:pB}. 

We measure $\zeta > 0.8$ in the majority of the disks of the spiral galaxies of our sample beyond a certain galactocentric radius in both FIR and radio wavelengths (Fig.\,\ref{fig:zeta_angle_profile}), with exception of NGC\,6946 and NGC\,3627. This result indicates that a large-scale axisymmetric ordered $B$-field is the dominant $B$-field structure in the disk of galaxies at both FIR and radio wavelengths.  Note that the individual $\zeta$ measurements per LOS at the two wavelength regimes may differ.

The $\zeta$ profiles in Fig.\,\ref{fig:zeta_angle_profile} show that in the analyzed spiral galaxies with radio and FIR data, the magnetic alignment at FIR wavelengths tends to be systematically lower than at radio wavelengths at the same radii, with some exceptions in the inner regions of NGC\,3627 and NGC\,4736. We quantify this result by analyzing the difference between radio and FIR alignment maps ($\Delta\zeta$ is the \% increase of $\zeta_{\rm FIR}$ over $\zeta_{\rm Radio}$) in Fig.\,\ref{fig:Deltazeta_hist}. The average values of $\zeta$ are always statistically larger in radio polarization observations than FIR when compared in the same beams. The histograms show that M\,51, M\,83, NGC\,3627, NGC\,4736, and NGC\,6946 have a better alignment with the large-scale ordered $B$-field in radio than in FIR ($\Delta\zeta = [2-75]\%$, radio - FIR). We show the average differences $\Delta\zeta$ for each galaxy in Table \ref{tab:pB}, and we confirm that all of them are statistically significant ($p < 10^{-4}$). This suggests that the disordered component of the $B$-field within the beam (i.e, small-scale $B$-fields) of observations at FIR wavelengths may have a larger contribution than at radio wavelengths.



In summary:
\begin{itemize}

\item Spiral galaxies have a dominant, $\zeta>0.8$,  large-scale axisymmetric spiral $B$-field in both radio and FIR wavelengths.

\item Radio polarization show larger $\zeta$ (more ordered $B$-fields) than those from FIR polarization measurements (more disordered $B$-fields). Specifically, the FIR $B$-fields have $2-75$\% more disordered $B$-fields than those measured at radio wavelengths.

\end{itemize}

\section{Discussion}\label{sec:DIS}

One of the most striking results from this work is the differences on the patterns and angular dispersions between the FIR and radio $B$-fields shown in spiral and starburst galaxies. We will discuss these differences in terms of the intrinsic nature of the tracer and the $B$-field components, and separating the large- and small-scale $B$-fields.

Magnetic fields extending over kpc-scales have been detected in many spiral galaxies using IR and radio polarimetric observations \citep{beck+2004na48, beck2015aap24,SALSAIV}. These $B$-fields are a fundamental part of galaxy evolution, driving gas mass inflows into the galactic core \citep{kim+2012apj751}, regulating the star formation rate through the removal of angular momentum in contracting proto-stellar gas clouds \citep[][]{mouschovias+1999inproceedings_305,allen+2003apj599,dapp+2010aap521,PattlePPVII}, affecting the formation of molecular clouds \citep{Tahani2022A&A...660L...7T, Tahani2022A&A...660A..97T, Abe2021ApJ...916...83A}, and even potentially affecting the global morphology and gas kinematics in disk galaxies \citep{battaner+2007an328,ruizgranados+2010apj723, martinalvarez+2020mnras495_4475, vandevoort+2021mnras501_4888}. In spite of the substantial research efforts, the formation and structure of kpc-scale $B$-fields in galaxies, as well as their possible role in the evolution of their hosts, are still outstanding questions in modern astrophysics.

The magnetic fields in spiral galaxies are observed to generally follow spiral patterns resembling the morphological spiral arms \citep{Beck2019,lopezrodriguez+2020apj888_66}. This spatial correlation is in agreement with the current paradigm in which weak seed magnetic fields \citep{rees1987qjras28_197, subramanian2016RPP79_076901} are amplified by dynamo mechanisms such as small-scale ($<50$--$100$ pc) turbulence \citep{beck1996na1996_262,BNdynamoreview, haverkorn+2008apj680_362, bhat+2016mnras461_240,2022MNRAS.513.3326M}, and then reorganize them through differential galactic disk rotation and helical turbulence \citep[large-scale dynamos,][]{gressel+2008aap486,gressel+2008an329,gent2012thesis, bendre+2015an336_991,ENetal2020}.

While turbulent dynamos efficiently convert turbulent kinetic energy into magnetic energy \citep{widrow2002rmp74_775,kulsrud+2008RPP71_046901, federrath2016na82_535820601,rincon2019na85_205850401}, amplification takes place at the smallest scales and the resulting magnetic field follow the isotropic character of the driving turbulent field. The observations of kpc-scale magnetic field structures in spiral galaxies \citep{Beck2019,lopezrodriguez+2020apj888_66, Borlaff2021} thus suggest that small-scale dynamos do not operate alone. Theoretical and modeling studies suggest that the efficient small-scale dynamos provide strong seed $B$-fields for other less-efficient large-scale ordering mechanisms (i.e. differential rotation), which would subsequently align the field \citep{lou+1998apj493_102, arshakian+2009aap494_21, bhat+2016mnras461_240}. In addition to differential rotation, the interaction between small-scale dynamos and spiral density waves has been proposed as a mechanism for ordering the magnetic field in spiral galaxies \citep{beck+2013inbook_641}. Spiral density waves \citep{lin+1964apj140_646, bertin+1996book, sellwood2011mnras410, dobbs+2014pasa31, font+2019mnras482}  periodically compress and shear the interstellar medium (ISM), triggering star formation in their wake. Spiral density waves can also modulate parameters of the large scale dynamo and lead to enhanced non-axisymmteric fields (anti-) correlated with the spiral \citep{Mestel1991MNRAS.248..677M, Shukurov1998MNRAS.299L..21S, Chamandy2013MNRAS.428.3569C}.
As density waves travel through the ISM, magnetic field lines are sheared along the direction of the spiral arm, aligning the field with the morphological arms.

The mechanism described above to form kpc-scale ordered galactic magnetic fields scenario would work efficiently in grand-design spiral galaxies like M\,51 \citep{patrikeev+2006aap458_441} or M\,83 \citep{frick+2016aap585_21}, where star formation takes place preferentially on the spiral arms. Flocculent spirals (i.e., NGC\,253, NGC\,6946) would be less affected by the spiral density arm shear compression and as a consequence, they might present less ordered spiral magnetic fields, or lower alignment with their morphological counterparts (see NGC\,6946). Similar results should apply to starburst galaxies (e.g., M\,82), while AGN-dominated (e.g., Circinus), interacting, or recent merger galaxies \citep[like the Antennae galaxies or Centarus A,][]{chyzybeck2004, LopezRodriguez2021NatAs...5..604L,SALSAVI} may, in addition, present shock or AGN-induced aligned magnetic fields \citep{LopezRodriguez2021ApJ...914...24L}. 

\subsection{FIR vs. Radio magnetic pitch angles in spiral galaxies}\label{subsec:DIS_PA}

We found that the orientation of the large-scale spiral $B$-fields traced through FIR polarimetric observations is statistically different from those measured through radio-synchrotron polarization observations. Note that radio observations were convolved and rebinned to the angular resolution (13\farcs6) of the HAWC+ observations. These results confirm the conclusions from \citet[][SALSA I]{Borlaff2021} in M\,51 and extends the result to a larger and more diverse sample of galaxies. A recent independent study on the same datasets used a linear decomposition of the polarization fields that also resulted in the same conclusions \citep{Surgent2023arXiv230207278S}. We showed that a large-scale ordered spiral $B$-field is the dominant $B$-field pattern for both radio and FIR wavelengths (with the exception of NGC~6946). However, the magnetic pitch angle profiles traced with radio polarization observations present a more ordered distribution across the entire disk than those obtained with FIR (i.e., the outskirts of M\,51, M\,83, or the entire disk of NGC~6946). Specifically, the radio and FIR magnetic pitch angles show median angular dispersion of $8^{\circ}$ and $18^{\circ}$, respectively across the disk of spiral galaxies. In addition, the radio magnetic pitch angle is more open (i.e. larger pitch angles) with increasing galactrocentric radius, while the FIR pitch angle tends to be wrapped tighter (i.e. smaller pitch angles) in the outskirts of spiral galaxies (with exception of M\,83).

Among the potential causes for these $B$-field morphological differences we consider the intrinsic nature of the $B$-field tracer. FIR polarization observations trace the $B$-fields in the cold, $T_{\rm{d}} = [19,48]$ K, and dense, $\log_{10}(N_{\rm HI+H2} \rm{[cm^{-2}]}) = [19.9,22.9]$, ISM \citep[SALSA IV,][]{SALSAIV}. The measured FIR polarization is weighted by a combination of density, temperature, and grain alignment efficiency along the LOS and within the beam. Furthermore, the total FIR emission of galaxies has a typical vertical height of $0.1–0.6$ kpc \citep{Verstappen2013}. Thus, the FIR $B$-fields are sensitive to the morphology and physical conditions of the dense ISM in the central few hundred pc of the galactic disk.

Radio synchrotron polarization traces the $B$-field in a volume-filling medium of the warm and diffuse ISM \citep{Beck2019}. In spiral galaxies, the radio $B$-field morphology is observed to have a component parallel to the midplane of the disk and another component with an X-shaped structure extending several kpc above and below the disk \citep{Hummel1988,Krause2020}. Larger
radio halos containing $B$-fields and cosmic rays are found in galaxies with higher star formation rate surface densities \citep{Wiegert2015}. In addition, radio polarimetric observations suffer from Faraday rotation along the LOS. Radio $B$-fields are sensitive to the gradient of Faraday rotation \citep{Sokoloff1998} and the physical conditions of both the halo and the galactic disk, but at 3 and 6 cm this effect is expected to be very small (less than $\pm5^{\circ}$ rotation at 3 cm and less than $\pm20^{\circ}$ at 6 cm) \citep[see][Sec.\,2.4]{beck+2013inbook_641}. Beyond that limit, Faraday rotation systematic effects would greatly dominate the individual uncertainties of each beam, adding little information to our study. The main objective of the present work is to test observationally the structure of the magnetic field measured by different tracers (FIR vs. synchrotron). We attempt to do this by comparing datasets in the most compatible conditions (i.e., convolving the datasets of each galaxy to the largest beam size, applying quality cuts to both datasets, etc). While it is certainly interesting to explore a wider range in the polarization spectra at optical and near-IR wavelengths, these wavelengths are strongly affected by dust/electron scattering and have been shown to tell little about the B-fields in external galaxies \citep[e.g.,][]{Elvius1972,Elvius1978,Scarrott1991,Scarrott1993,Jones2000,Pavel2012} 

The large-scale dynamo (incorporating differential rotation) in galaxies can generate spiral $B$-fields over kpc-scales \citep{beck1996na1996_262,gressel+2008aap486,gressel+2008an329}. This mechanism generates B-fields with similar pitch angles to those of the morphological spiral arms \citep{beck+2013inbook_641}, although an offset between the $B$-fields and the spirals has been measured depending on the tracer \citep[e.g.,][]{Beck2019}. The large-scale dynamo could operate most efficiently in interarm regions because it may be suppressed or disturbed in the spiral arms \citep{Beck2015}. The $B$-field associated with the dense (e.g., molecular) gas may respond in relatively short time-scales to gravitational and kinematic effects like minor mergers, accretion of intergalactic material, and resonances. Diffuse atomic gas also responds collisionally, but its scale-height is significantly higher, and its structure is more affected by supernova explosions and galaxy-scale processes, such as tidal interactions, bar, or spiral formation. Observational effects compatible with these hypotheses have been measured in several galaxies: in the outskirts of M~51 \citep[SALSA~I,][]{Borlaff2021} due to the interaction with M~51b. Also in the central starburst of NGC~1097 \citep[SALSA~II,][]{SALSAII}, where the radio $B$-field has a spiral shape following the diffuse gas and the FIR $B$-field is cospatial with a galactic shock at the boundaries of the bar and the starburst ring. Finally, in the relic spiral arm of the Antennae galaxies \citep[SALSA~VI,][]{SALSAVI}, where the radio $B$-field is radial and cospatial with the HI gas dynamics, and the FIR $B$-field is spiral and cospatial with the dense and molecular ISM. 

We suggest that the differences of the radial magnetic pitch angles at radio and FIR wavelengths may be caused by the `averaging' of the several components of the $B$-fields along the LOS due to the different nature of the two tracers. Due to the relatively narrow extension of the molecular disk at above and below the galactic plane compared to more diffuse components of the ISM, the FIR magnetic pitch angle efficiently traces the $B$-field in the midplane of the galaxy (Fig.\,\ref{fig:pitch_angle_profiles}). Angular variations of the FIR magnetic pitch angle are driven by the dense ISM and star-forming regions (Sec.\,\ref{subsec:DIS_Tur}). However, the radio magnetic pitch angle may be more sensitive to the extraplanar $B$-field as the distance from the core increases. This may occur by the combination of a) an increase of the vertical height of the X-shaped $B$-field above and below the disk as galactocentric distance increases, b) an increase of the polarized flux from the X-shaped structure than from the disk, and/or c) depolarization in the arms due to an increase of random turbulent $B$-fields. 

We conclude that the FIR and radio magnetic pitch angles may trace similar $B$-fields in the midplane of the disk within the central few kpc of spiral galaxies. However, the radio pitch angle may be arising from different vertical heights as galactocentric radius increases. Specifically, the radio $B$-field may be extra-planar above the disk along the LOS, while the FIR is co-located with the midplane of the disk at large galactocentric radii.

\subsection{Disordered-to-ordered $B$-fields at FIR and radio}\label{subsec:DIS_Tur}

We have introduced the magnetic alignment parameter $\zeta$ as a proxy of the disordered-to-ordered $B$-field ratio on a pixel-to-pixel basis (see Sec.\, \ref{subsubsec:Dthetamethods}). The analysis of the $\zeta$ alignment maps allowed us to compare the disordered-to-ordered $B$-field ratio difference between FIR and radio polarimetric observations. Our results indicate that spiral galaxies have a dominant large-scale $B$-field with a $\zeta>0.8$ across their disk. We found that the FIR $B$-field is $2-75$\% more disordered than the radio $B$-field across the galaxy disk.

Individual measurements with $\zeta\neq1$ indicates that angular dispersion may be due to a) $B$-field variations below the physical scale of the beam of the observations, and/or b) $B$-field variations from some physical mechanisms larger than the physical scale of the beam of the observations. For the latter, systematic angular dispersion across several beams are expected, while random angular dispersion between beams can indicate the signature of small-scale $B$-fields. The azimuthally-averaged $\zeta$ radial profiles, $\zeta$($R$), allow us to identify the variation of the disordered-to-ordered magnetic field in the spiral galaxies of our sample. Variations of $\zeta(R)$ indicate deviations from the axisymmetric large-scale ordered $B$-field at scales larger than the beam of the observations. Angular variations larger than the beams of our observations can be driven by tangled random $B$-fields, $B$-field instabilities (Parker loops), and/or gas dynamics due to winds, outflows, supernovae, galaxy dynamics, or galaxy interactions. Angular variations smaller than the beam of our observations can also be driven by tangled random $B$-fields due to star-formation, supernovae, or the increase of gas turbulence due to galaxy interaction or gas dynamics.

\citet[][SALSA~IV]{SALSAIV} analyzed the trends of the FIR polarization fraction with the column density as a proxy to estimate the random $B$-fields in the ISM. SALSA~IV showed that the spiral galaxies have a relative increase of isotropic random FIR $B$-fields driven by star-forming regions in the spiral arms. Sec.\,\ref{subsec:DIS_PA} have shown that the FIR and radio polarized emission trace different components of the $B$-field in the disk of galaxies. In addition, $\Delta\zeta$ showed that the FIR $B$-field has is $2-75$\% more disordered than the radio $B$-field as measured by the independent measurements of $\zeta$. These results suggest that the $B$-field observations associated to dense, dusty, star-forming regions are less ordered than warmer, less-dense regions of the ISM. 

FIR observations of molecular clouds in the Milky Way showed that the $B$-field orientation can be drastically different as a function of the density of the ISM \citep{Pillai2020NatAs...4.1195P}. It is possible then that observationally, the properties of both the small- and large-scale magnetic fields differ as a function of the tracer and the phases of the ISM. This has been suggested by \citet{Seta2022MNRAS.514..957S} using MHD turbulent boxes with a multi-phase ISM. These authors measured that the `cold' (T$<10^{3}$ K) phase is more tangled that the `warm' (T$\ge10^{3}$ K) phase. 

Our results indicate that the FIR polarimetric observations can be used as an indicator of the small-scale $B$-fields associated to the activity of the star-forming regions and the morphology of the molecular clouds within a vertical height of few hundred pc in the disk of spiral galaxies. In addition, $\zeta$($R$) also shows that the FIR $B$-fields have larger fluctuation across the disk than radio $B$-fields. This result may indicate that FIR polarization is more sensitive to large-scale mechanisms (i.e. outflows, Parker loops, galaxy interaction) affecting the galaxy.


\subsection{Starburst galaxies}\label{subsec:DIS_SB}

The differences between the magnetic field traced by FIR and radio polarimetric observations are even more striking in our sub-sample of starburst galaxies (M\,82, NGC\,253, and NGC\,2146). We found that FIR polarimetric observations are able to trace the $B$-field by means of magnetically aligned dust grains along the galactic outflows within the $53-214$ \um~wavelength range.

For M\,82, we estimate a large change of the average magnetic field position angle (\PAhist\ and \PAint) with wavelength in both FIR (from $\sim150^{\circ}$ in 53 $\mu$m to $\sim60^{\circ}$ in 214 $\mu$m), and radio ($\sim90^{\circ}$ in 6~cm, see Fig.\,\ref{fig:PAhist} and Table \ref{tab:PPA}). The radio polarimetric observations trace the $B$-field by means of synchrotron emission in a magnetized bar mostly coincident with the galactic disk \citep{Adebahr2013,Adebahr2017}. A similar conclusion can be obtained from the observations of NGC~253 \citep{Heesen2011}. While the inner regions show $B$-field orientations parallel to the disk of the galaxy in all wavelengths, strong discrepancies are observed in the outskirts, which might be a signature of galactic outflows. Although some Faraday depolarization may be affecting the radio polarimetric observations, we suggest that the discrepancy may be produced by the limited lifetime of the cosmic rays (CR) along the galactic outflows. This may be explained by the weak total synchrotron emission in the outflow \citep[][Fig.\, 6]{Adebahr2013}. In addition, the steep spectral indexes may be an indication of rapid aging of the CR electrons. FIR total intensity observations with \textit{Herschel} have shown $\sim10$ kpc scale extensions of dust above and below the disk \citep{Roussel2010}. The galactic outflow may have expelled $\ge25$\% of the total dust mass of the galaxy \citep{Roussel2010}, making the cold phase of the ISM to be the bulk of the mass along the galactic outflow. Thus, the $B$-field along the galactic outflows of starburst galaxies are better traced by using FIR polarimetric observations.


Radio polarimetric observations have measured an X-shaped structure extending several kpc above the disks of edge-on spiral galaxies \citep{Krause2020}. This vertical $B$-field is different to that observed here in starburst galaxies. Our FIR polarimetric observations in edge-on galaxies do not show signs of a X-shaped $B$-field structure. This is expected as the vertical height of dust emission is $0.1-0.6$ kpc \citep{Verstappen2013}, while the X-shaped $B$-field observed in radio corresponds to synchrotron polarized emission from warm gas in the halo at several kpc scales.

FIR polarimetric observations also reveal large structural differences of the $B$-field with wavelength in the 53 -- 214 \um\ range, as observed in Circinus (Fig.\,\ref{fig:FIR_morphology_Circinus}) and NGC~2146 (Fig.\,\ref{fig:FIR_morphology_NGC2146}). For Circinus, the $53$ \um~and $89$ \um~$B$-fields may be dominated by the starburst ring and the inner-bar, while the $214$ \um~$B$-field may be tracing the $B$-field in the outer bar and spiral arms of the host galaxy \citep{Maiolino2000,Izumi2018}. Further modeling is required to quantify these results.

The most extreme case is NGC~2146. This galaxy shows a $B$-field with an orientation perpendicular to the galactic disk in the 53 -- 89 \um\ wavelength range associated with the galactic outflow. The 214 \um\ observations reveal a parallel $B$-field to the galactic disk associated to the galactic $B$-field. \citet[][SALSA~IV]{SALSAIV} computed the polarized spectral energy distribution (SED) of starburst galaxies showing a minimum within the $89-154$ \um~wavelength range. The falling $53-154$ \um~polarized spectrum was suggested to be produced by a decrease in the dust grain alignment efficiency due to a gradient of dust temperatures along the LOS. The rise spectrum at $154-214$ \um~was suggested to be produced by the dust becoming more optically thin at larger wavelengths. Thus, the measured $154$ \um~$B$-field orientation is expected to be the transition between the two regimes (see Sec.\,\ref{subsec:results_Bmaps_edgeon}). In addition, the extremely complex morphological structure of NGC~2146 suggests that it is the remnant of a recent interaction that triggered star-formation bursts \citep{Martini2003ApJS..146..353M,  Tarchi2004, Erwin2005MNRAS.364..283E}. While it is beyond the scope of this paper to explore this, the $B$-field orientation differences detected within the different SOFIA HAWC+ bands suggest that FIR polarized emission could be dominated by different populations of dust grains as a function of wavelength, revealing different physical regimes in terms of density, radiation, or turbulence.

\section{Conclusions}\label{sec:CON}

The HAWC+/FIR polarimetric observations of the SALSA Legacy Program of 14 nearby ($<20$ Mpc) galaxies analyzed in this paper, and presented in the previous publications from the project (SALSA I -- IV, VI) provide a previously unexplored view of a fundamental factor in galaxy evolution: the $B$-fields in the dense and cold ISM of galaxies. While extragalactic $B$-fields have been observed for decades in radio polarization observations \citep[e.g.,][]{Beck2019}, high-spatial resolution FIR polarimetric maps from molecular clouds in the Milky Way \citep[e.g.,][]{Pillai2020NatAs...4.1195P} have demonstrated that the structure of magnetic fields is a strong function of the density of the ISM at scales of hundreds of pc. 

Here, we provide quantitative evidence that the kpc-scale of galactic magnetic fields depends strongly of the tracer: radio polarization observations of the synchrotron emission at 3 and 6 cm, and FIR polarization observations of the thermal polarized emission from magnetically aligned dust grains. In addition, we introduce the alignment parameter $\zeta$ (Sec.\,\ref{subsec:Dtheta}), as a quantitative measurement of the dispersion between the measured B-field orientation and a large-scale axisymetric spiral magnetic field. $\zeta$ allows us to estimate how well-ordered a spiral magnetic field is, and to compare it as a function of different observational tracers.

Our results suggest that radio and FIR polarization observations do not necessarily trace the same magnetic field structures. For most of our galaxies, significant differences are found in terms of local and global orientation of the magnetic field (Sec.\,\ref{subsec:BmapsFIR_results}), the pitch angle of spiral galaxies (Sec.\,\ref{subsec:PitchB}), and the dispersion of the magnetic spiral structure (Sec.\,\ref{subsubsec:Dthetaresults}). We find evidence that even FIR polarization observations can trace different structures of the magnetic field as a function of wavelength as observed in NGC\,2146 (53\,\um\ -- 214\,\um, see Fig.\,\ref{fig:FIR_morphology_NGC2146}). The differences in the magnetic pitch angle profiles (Fig.\,\ref{fig:pitch_angle_profiles}), and the alignment $\zeta$ profiles (Fig.\,\ref{fig:zeta_angle_profile}) indicate that radio polarization observations tend to trace a more regular and stable structure of the magnetic field, with less variation of the magnetic pitch angle, and a higher level of alignment $\zeta$. The five spiral galaxies studied with available FIR and radio polarization observations presented a statistically higher magnetic field dispersion in FIR than in radio (see Fig.\,\ref{fig:Deltazeta_hist}). 

While our observations are limited by spatial resolution and thus not able to resolve individual molecular clouds, all results point towards the conclusion that FIR polarization observations trace a less-ordered structure of the magnetic field than radio polarization observations. We suggest that FIR polarization observations are dominated by the emission from magnetically aligned dust grains cospatial with the cold, dense ISM, while radio polarization observations trace more diffuse regions of the ISM, less dominated by kinematic turbulence. Each beam from both radio and FIR polarization maps would be tracing an average of the magnetic field inside them, but dominated by the the dense, dusty molecular clouds in FIR. Star-formation, outflows, and accretion, and the morphological complexity of the galactic filaments of dust would increase the dispersion of the magnetic field from the average kpc-scale magnetic field orientation, with a lower effect on less-dense phases of the ISM. 

High-resolution, FIR polarization observations of galaxies such as those provided by SOFIA/HAWC+ until its last flight in September 2022 are vital to understand the role of magnetic fields in the evolution of the Universe. While facilities like ALMA, South Pole Telescope, or James Clerk Maxwell Telescope provide polarimetric observations in sub-millimetric and/or FIR wavelength ranges, they lack the sensitivity and/or spatial resolution in the required wavelengths to achieve the same scientific requirements in nearby galaxies. In the absence of facilities such as SOFIA, unique in their ability to investigate the dusty component of galaxies as well as their magnetic fields, the astronomical community has no current alternatives to explore the magnetic fields inside the dust component of nearby galaxies through FIR polarimetry. Only the support of the astronomical community through proposed missions like PRIMA \citep[][potential launch no earlier than 2030]{Bradford2022AAS...24030407B} for the upcoming NASA far-infrared (FIR) Probe missions or future FIR polarimetric missions will ensure a continuation of this field in the next decades.


\newpage
\begin{acknowledgments}

Based on observations made with the NASA/DLR Stratospheric Observatory for Infrared Astronomy (SOFIA) under the 07\_0034, 08\_0012 Program. SOFIA is jointly operated by the Universities Space Research Association, Inc. (USRA), under NASA contract NNA17BF53C, and the Deutsches SOFIA Institut (DSI) under DLR contract 50 OK 0901 to the University of Stuttgart.

%

A.B. acknowledges the incredible support and tireless effort from Christine Martinez (NASA/Ames STA), that made this work possible. A.B. was supported by an appointment to the NASA Postdoctoral Program at the NASA Ames Research Center, administered by Oak Ridge Associated Universities under contract with NASA. A.B. is supported by a NASA Astrophysics Data Analysis grant (22-ADAP22-0118), program Hubble Archival Research project AR 17041, and Chandra Archival Research project ID \#24610329, provided by NASA through a grant from the Space Telescope Science Institute and the Center for Astrophysics Harvard \& Smithsonian, operated by the Association of Universities for Research in Astronomy, Inc., under NASA contract NAS 5-03127. K.T. has received funding from the European Research Council (ERC) under the European Unions Horizon 2020 research and innovation programme under grant agreement No. 771282 M.T. is supported by the Banting Fellowship (Natural Sciences and Engineering Research Council Canada) hosted at Stanford University and the Kavli Institute for Particle Astrophysics and Cosmology (KIPAC) Fellowship. I.M.C. acknowledges funding from U. La Laguna through the Margarita Salas Fellowship from the Spanish Ministry of Universities ref. UNI/551/2021-May 26, and under the EU Next Generation funds. This research was sponsored by the National Aeronautics and Space Administration (NASA) through a contract with ORAU. The views and conclusions contained in this document are those of the authors and should not be interpreted as representing the official policies, either expressed or implied, of the National Aeronautics and Space Administration (NASA) or the U.S. Government. The U.S.Government is authorized to reproduce and distribute reprints for Government purposes notwithstanding any copyright notation herein. 

The authors thank the careful revision and useful comments from the anonymous referee, who contributed to improve this manuscript greatly. The authors thank Dr. Yik Ki Ma from the Research School of Astronomy \& Astrophysics of The Australian National University in Canberra, Australia for his careful revision of the manuscript.

\end{acknowledgments}

\facilities{SOFIA (HAWC+), ALMA, \textit{Herschel}, VLA.}


\software{\textsc{astropy} \citep{astropy}, 
\textsc{APLpy} \citep{aplpy},
\textsc{matplotlib} \citep{hunter2007},
          }

\clearpage
\appendix 

\section{Measurements of the polarization angles at FIR and radio wavelengths}\label{App:PAradio}

\begin{figure*}[ht!]
\centering
\includegraphics[angle=0,width=\textwidth]{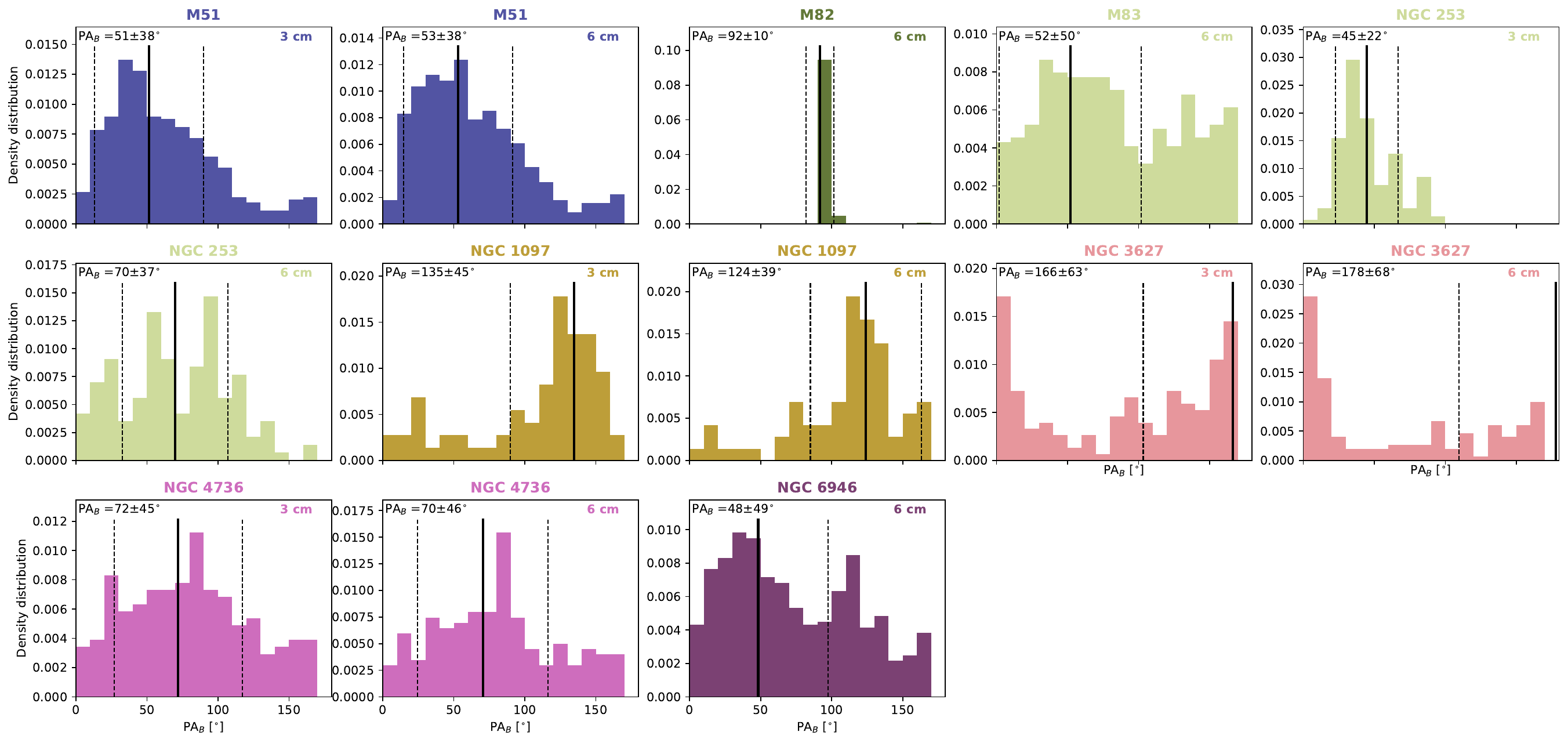}
\caption{Histograms of the $B$-field orientations for the galaxies with radio polarimetric observations (Table \ref{tab:FIRRadioObs}). The histograms are set to $10^{\circ}$ bins, each galaxy is shown in a unique color, and the median (black solid line) and $1\sigma$ uncertainty (black dashed line) for the radio observations are shown. Given the similarities of the $B$-field orientations between $3$ and $6$ cm wavelengths, only the histograms of the $B$-field orientations at $6$ cm are shown in Fig.\,\ref{fig:PAhist}.
\label{fig:figA1}}
\epsscale{2.}
\end{figure*}

This section shows the histograms of the $B$-field orientations at 3 and 6 cm, which are shown in Fig.\,\ref{fig:figA1}. Table \ref{tab:PPA} show the circular average and integrated polarization fraction and $B$-field orientation of the galaxies included in the SALSA sample. Table \ref{tab:quality_cuts_table} shows the noise levels obtained and the quality cuts applied to each dataset.

\begin{deluxetable*}{lcccccccccc}[h]
\tablecaption{Median and integrated polarization fraction and $B$-field orientation of galaxies. \emph{Columns, from left to right:} 
a) Galaxy name, 
b) central wavelength of the HAWC+ band, 
c)  median polarization angle based on individual measurements at FIR wavelengths
d) median polarization angle based on the integrated Stokes $IQU$ of the full galaxy at FIR wavelengths,
e) central wavelength of the band at radio wavelengths,
f) median polarization angle based on individual measurements at radio wavelengths,
g) median polarization angle based on the integrated Stokes $IQU$ of the full galaxy at radio wavelengths.
\label{tab:PPA} 
}
\tablecolumns{7}
\tablewidth{0pt}
\tablehead{\colhead{Galaxy} & 	\colhead{Band}  & \colhead{$\langle \rm{PA}_{\rm{B},\rm{FIR}}^{\rm{hist}} \rangle$}  & \colhead{$\langle \rm{PA}_{\rm{B,FIR}}^{\rm{int}} \rangle$} & 	\colhead{Band}  & \colhead{$\langle \rm{PA}_{\rm{B,Radio}}^{\rm{hist}} \rangle$} & \colhead{$\langle \rm{PA}_{\rm{B,Radio}}^{\rm{int}} \rangle$}  \\ 
  &	\colhead{(\um)}	& \colhead{($^{\circ}$)}  &	\colhead{($^{\circ}$)} & 	\colhead{(cm)}  & \colhead{($^{\circ}$)} & \colhead{($^{\circ}$)}\\
\colhead{(a)} & \colhead{(b)} & \colhead{(c)} & \colhead{(d)} & \colhead{(e)} & \colhead{(f)}  & \colhead{(g)}} 
\startdata
Centaurus A	&	89	&	$	124	\pm	28	$	&	$	126	\pm	9	$	&	-	&	-	&	-	\\
Circinus		&	53	&	$	52	\pm	23	$	&	$	59	\pm	13	$	&	-	&	-	&	-	\\
			&	89	&	$	28	\pm	37	$	&	$	30	\pm	12	$	&	-	&	-	&	-	\\
			&	214	&	$	1	\pm	77	$	&	$	179	\pm	2	$	&	-	&	-	&	-	\\
M~51			&	154	&	$	51	\pm	36	$	&	$	56	\pm	12	$	&	3	&	$51\pm38$	&	$47\pm1$ \\
			&		&						&						&	6	&	$53\pm38$	&	$53\pm1$\\
M~82			&	53	&	$	156	\pm	52	$	&	$	155	\pm	4	$	&	6	&	$92\pm10$	&	-$^{\star}$ \\
			&	89	&	$	154	\pm	55	$	&	$	153	\pm	5	$	&	-	&	-	&	-	\\
			&	154	&	$	83	\pm	43	$	&	$	86	\pm	7	$	&	-	&	-	&	-	\\
			&	214	&	$	58	\pm	41	$	&	$	53	\pm	3	$	&	-	&	-	&	-	\\
M~83			&	154	&	$	59	\pm	40	$	&	$	50	\pm	10	$	&	6	&	$52\pm50$	&	$44\pm1$ \\
NGC~253		&	89	&	$	50	\pm	41	$	&	$	39	\pm	19	$	&	3	&	$45\pm22$	&	$44\pm2$ \\
			&	154	&	$	62	\pm	23	$	&	$	164	\pm	56	$	&	6	&	$70\pm26$	&	$70\pm3$ \\
NGC~1068	&	53	&	$	124	\pm	49	$	&	$	39	\pm	84	$	&	-	&	-	&	-	 \\
			&	89	&	$	139	\pm	48	$	&	$	135	\pm	10	$	&	-	&	-	&	-	\\
NGC~1097	&	89	&	$	131	\pm	32	$	&	$	132	\pm	5	$	&	3	&	$135\pm45$	&	$134\pm1$ \\
			&	154	&	$	126	\pm	35	$	&	$	134	\pm	16	$	&	6	&	$124\pm39$	&	$128\pm1$ \\
NGC~2146	&	53	&	$	64	\pm	16	$	&	$	65	\pm	10	$	&	-	&	-	&	- \\
			&	89	&	$	74	\pm	29	$	&	$	80	\pm	9	$	&	-	&	-	&	-	\\
			&	154	&	$	165	\pm	61	$	&	$	160	\pm	9	$	&	-	&	-	&	- \\
			&	214	&	$	120	\pm	20	$	&	$	119	\pm	4	$	&	-	&	-	&	-\\
NGC~3627	&	154	&	$	177	\pm	60	$	&	$	17	\pm	19	$	&	3	&	$166\pm63$	&	$174\pm1$ \\
			&		&						&						&	6	&	$178\pm68$	&	$0\pm1$		\\
NGC~4736	&	154	&	$	85	\pm	42	$	&	$	88	\pm	7	$	&	3	&	$72\pm45$	&	$72\pm1$\\
			&		&						&						&	6	&	$70\pm46$	&	$71\pm1$		\\
NGC~4826	&	89	&	$	106	\pm	31	$	&	$	104	\pm	10	$	&	-	&	-	&	-\\
NGC~6946	&	154	&	$	151	\pm	55	$	&	$	161	\pm	11	$	&	6	&	$48\pm49$	&	$47\pm1$	 \\
NGC~7331	&	154	&	$	127	\pm	30	$	&	$	121	\pm	6	$	&	-	&	-	&	-\\
\enddata
\tablenotetext{{\star}}{Stokes $QU$ were not available.}
\end{deluxetable*}

\begin{deluxetable*}{lcccccccccc}[h]
\tablecaption{Noise level in total intensity (I) and polarized intensity (PI) and quality cuts of the FIR and radio polarization observations of the galaxies in the sample. \emph{Columns, from left to right:} 
a) Galaxy name, 
b) central wavelength of the HAWC+ band, 
c) total intensity noise level in FIR,
d) polarized intensity noise level in FIR,
e) central wavelength of the band at radio wavelengths,
f) total intensity noise level in radio,
g) polarized intensity noise level in radio, h) limiting SNR in polarization intensity, i) limiting polarization fraction, and j) limiting SNR in total intensity. The polarization fraction limit of $P = 20\%$ is given
by the maximum polarization fraction ($p_{\rm max} = 19.8\%$) of dust measured by Planck observations \citep{Planck2015}. Note that more stringent quality cuts were applied to some specific galaxies, to avoid systematic effects. For more specific information about the quality cuts chosen for each galaxy we refer to \citep{LopezRodriguez2021ApJ...914...24L}.  
\label{tab:quality_cuts_table} 
}
\tablecolumns{10}
\tablewidth{0pt}
\tablehead{\colhead{Galaxy} & 	\colhead{Band}  & \colhead{$\sigma_{\rm{I}}^{\rm{FIR}}$}  & \colhead{$\sigma_{\rm{PI}}^{\rm{FIR}}$} & 	\colhead{Band}  & \colhead{$\sigma_{\rm{I}}^{\rm{Radio}}$} & \colhead{$\sigma_{\rm{PI}}^{\rm{Radio}}$} & \colhead{SNR$_{\rm{FIR, PI}}$} & \colhead{$p_{\rm{lim}}$} & \colhead{SNR$_{\rm{FIR, I}}$}  \\ 
  &	\colhead{(\um)}	& \colhead{({Jy arcsec$^{-2}$})}   &	\colhead{({Jy arcsec$^{-2}$})} & 	\colhead{(cm)}  & \colhead{({Jy arcsec$^{-2}$})}  & \colhead{({Jy arcsec$^{-2}$})} & & \colhead{(\%)} &  \\
\colhead{(a)} & \colhead{(b)} & \colhead{(c)} & \colhead{(d)} & \colhead{(e)} & \colhead{(f)}  & \colhead{(g)} & \colhead{(h)}  & \colhead{(i)} & \colhead{(j)} } 
\startdata
Centaurus A	&	89	& $1.39\cdot 10^{-4}$ 	& $3.39\cdot 10^{-3}$ &	-	&	-	&	-	& 2.5 & 20 & 10 \\
Circinus		& 53 &	$9.17\cdot 10^{-4}$	&	$10^{-2}$	&	-	&	-	&	-	& 2.3 & 	20 & 	10	\\
			&	89	& $1.22\cdot 10^{-4}$ &	$2.96\cdot 10^{-3}$	&	-	&	-	&	-	& 3.0	& 20	& 10 \\
			&	214	&	$3.72\cdot 10^{-5}$ &	$4.74\cdot 10^{-3}$	&	-	&	-	&	-	& 2.5	& 20	& 10\\
M~51			&	154	&	$2.23\cdot 10^{-5}$	&	$1.52\cdot 10^{-3}$	&	3	&	$1.75\cdot 10^{-7}$	&	$2.32\cdot 10^{-7}$ & 3.0& 	20 &	10	\\
			&		&						&						&	6	&	$2.16\cdot 10^{-7}$ &	$2.28\cdot 10^{-7}$ & & \\
M~82			&	53	& $6.27\cdot 10^{-4}$ & $6.84\cdot 10^{-3}$ &	6	& * & * & 3.0	& 20	& 80\\
			&	89	& $1.65\cdot 10^{-4}$ &	$4.02\cdot 10^{-3}$ &	-	&	-	&	-	& 3.0&20&80\\
			&	154	&	$3.57\cdot 10^{-5}$	& $2.46\cdot 10^{-3}$ &	-	&	-	&	-	& 3.0&20&20\\
			&	214	& $3.96\cdot 10^{-5}$ &	$5.0\cdot 10^{-3}$	&	- &	 -	&	-	& 3.0&20&20\\
M~83			&	154	& $2.57\cdot 10^{-5}$ &$1.78\cdot 10^{-3}$&	6	& $2.2\cdot 10^{-7}$ & $2.5\cdot 10^{-7}$ & 2.5&20&10\\
NGC~253		&	89	& $2.80\cdot 10^{-4}$ & $6.52\cdot 10^{-3}$ &	3	& $4.10\cdot 10^{-7}$ & $4.63\cdot 10^{-7}$ & 2.5&10&20\\
			&	154	&  $3.30\cdot 10^{-4}$ & $2.12\cdot 10^{-2}$ &	6	& $7.88\cdot 10^{-7}$ &  $6.13\cdot 10^{-7}$ & 2.0&10&20\\
NGC~1068	&	53	& $6.32\cdot 10^{-4}$ & $7.00\cdot 10^{-3}$ &	-	&	-	&	-	& 2.0&10&10 \\
			&	89	& $2.60\cdot 10^{-4}$ & $1.18\cdot 10^{-3}$ &	-	&	-	&	-	& 2.0&10&10\\
NGC~1097	&	89	& $2.16\cdot 10^{-4}$ & $4.98\cdot 10^{-3}$ &	3	&  $4.65\cdot 10^{-8}$ & $1.47\cdot 10^{-7}$ & 3.0&10&10	\\
			&	154	& $1.47\cdot 10^{-4}$ & $9.88\cdot 10^{-3}$ &	6	& $5.41\cdot 10^{-8}$ & $2.00\cdot 10^{-7}$ & 2.0&10&10\\
NGC~2146	&	53	&  $4.03\cdot 10^{-4}$ &  $4.39\cdot 10^{-3}$ &	-	&	-	&	- & 3.0&20&30	\\
			&	89	& $1.39\cdot 10^{-4}$ & $3.38\cdot 10^{-3}$ &	-	&	-	&	-	& 3.0&20&20 \\
			&	154	& $3.27\cdot 10^{-5}$ & $2.25\cdot 10^{-3}$ &	-	&	-	&	- & 3.0&20&20	\\
			&	214	& $2.24\cdot 10^{-5}$ & $2.85\cdot 10^{-3}$ &	-	&	-	&	-	& 3.0&20&20 \\
NGC~3627	&	154	& $2.59\cdot 10^{-5}$ & $1.79\cdot 10^{-3}$ &	3	& $4.33\cdot 10^{-7}$ & $5.10\cdot 10^{-8}$ & 	2.0&20&40 \\
			&		&						&						&	6	& $1.13\cdot 10^{-6}$ & $8.51\cdot 10^{-8}$ & &\\
NGC~4736	&	154	&  $4.74\cdot 10^{-5}$ &  $3.22\cdot 10^{-3}$ &	3	& $6.32\cdot 10^{-8}$ & $6.94\cdot 10^{-8}$ & 2.0&20&68	\\
			&		&						&						&	6	& $2.05\cdot 10^{-8}$ &	$4.76\cdot 10^{-8}$ & &		\\
NGC~4826	&	89	& $1.82\cdot 10^{-4}$ & $4.42\cdot 10^{-3}$	&	-	&	-	&	-	& 2.3&20&20 \\
NGC~6946	&	154	& $3.10\cdot 10^{-5}$ &  $2.14\cdot 10^{-3}$ &	6	& $9.02\cdot 10^{-9}$ & $5.94\cdot 10^{-8}$ & 2.3&20&20 \\
NGC~7331	&	154	& $2.19\cdot 10^{-5}$ &  $1.50\cdot 10^{-3}$ &	-	&	-	&	- & 2.8&20&45 \\
\enddata
\tablenotetext{{\star}}{Stokes $QU$ were not available.}
\end{deluxetable*}

\section{Measuring the pitch angle of the FIR and radio polarization fields}\label{App:PitchBmethods}

We analyze the magnetic pitch angles of the $B$-field maps following the methodology (\Mohawc) presented by \citet{Borlaff2021} on M~51. Here, we summarize the steps to estimate the magnetic pitch angle, $\Psi_{B}$, as a function of the galactrocentric radius of spiral galaxies. \Mohawc\ uses the Stokes $IQU$ maps to derive the $\Psi_{B}$ profiles on a pixel-per-pixel basis, assuming that all the $B$-field orientations are on the galactic plane viewed at a certain inclination, $i$, with respect to the observer, and position angle, $\theta$, on the plane of the sky. The uncertainties of the Stokes $IQU$, position angle, and inclination are taken into account through Monte Carlo simulations. Our magnetic pitch angle estimation method entails processing the data on a pixel-by-pixel basis, allowing the user to separate different regions of the galaxy by using masks (if desired). The steps followed by \Mohawc\ can be summarized as follows: 

\begin{enumerate}
    \item The debiased polarization degree and its associated uncertainty are computed using the Stokes $IQU$ parameters and their uncertainties $\delta I, \delta Q, \delta U$:
\begin{equation}
\label{eq:polarization_level}
P_{\rm debias} = \sqrt{P^2 - \delta P^2}
\end{equation}

following \citet{Serkowski1974MExP...12..361S}, where:

\begin{equation}
\label{eq:polarization_level2}
P = \sqrt{\left(\frac{Q}{I}\right)^2 + \left(\frac{U}{I}\right)^2}
\end{equation}

and:

\begin{equation}
\label{eq:polarization_level_uncert}
    \delta P = \frac{1}{I} \sqrt{\frac{(Q \cdot \delta Q)^2 + (U \cdot \delta U)^2}{Q^2 + U^2} + \delta I^2\frac{Q^2 + U^2}{I^2}}
\end{equation}

\item The $B$-field orientations are then re-projected to a new reference frame where the galaxy is observed face-on, using coordinates of the galactic center ($\alpha$, $\delta$), the galactic disk inclination, $i$, and tilt angle, $\theta$. The morphological parameters for each galaxy are summarized in Table \ref{tab:GalaxySample}.

\item An azimuthal angular mask is created. The azimuthal mask is generated such that the deprojected $B$-field orientation at each pixel location is perpendicular to the radial direction. We will refer to this idealized field as the \emph{zero pitch angle} field or $\Xi$.


\item The magnetic pitch angle at each pixel, $\Psi_{B}(x,y)$, is calculated as the difference between the measured PAs of the $B$-field orientation and the $\Xi$ vector field. 

\item The magnetic pitch angle profile $\Psi_{B}(x,y)$ is then averaged at each radius from the core. The radial bins are linearly spaced, and their number is optimized as a compromise between SNR and spatial resolution. The angular average as a function of the galactrocentric radius, $R$, is performed as follows: 

\begin{equation}
\label{eq:vector_averaging}
    \mPsi(R) = {\rm atan2} \left( \frac{ \langle \cos\,\Psi_{B}(x,y) \rangle}{\langle\sin\, \Psi_{B}(x,y) \rangle} \right)
\end{equation}
\noindent
where the $\langle \rangle$ operator indicates a robust median value (based on randomized Monte Carlo simulations) and $\mPsi$(R)\ is the averaged magnetic pitch angle value for a certain radial bin. 
\end{enumerate}

For each galaxy, the process detailed above is repeated 10,000 times, using Monte Carlo simulations to include the uncertainties of the tilt angle, inclination, and the Stokes parameters. We assume that each parameter follows a Gaussian probability distribution, with a standard deviation $\sigma$ equal to their uncertainties. We use the probability distribution for the pitch angle of each pixel to calculate the median magnetic pitch angle, \mPsi\, radial profiles, and their uncertainties (68\%, 95\%, equivalent to the 1$\sigma$, 2$\sigma$). For all the analyses, we consider a critical level of at least $p=0.05$ (95\%) to declare statistical significance. 

Note that one of the limitations of \Mohawc\ resides in the estimation of the pitch angle within the central 2-3 beams at the core of the galaxies. This limitation is due to the small number of measurements per annulus to obtain statistically significant measurements of the pitch angles \citep[see Fig.\,22 by][]{Borlaff2021}. Our analysis only takes into account the polarization measurements in the galaxy disks after $2$ beams ($4$ pixels diameter) from the core. Systematic errors are expected due to the lack of statistics in regions too close to the center, or affected by internal core phenomena, like rings, bars, or AGNs. NGC\,1068, M\,83, or M\,51 are examples of extreme values on their innermost bins, due to statistics, presenting more acceptable values in the second bin.\\

\section{Line Integral Convolution (LIC) RGB background components}\label{App:RGBComponents}

In this appendix we summarize the components used as the RGB backgrounds in Fig.\,\ref{fig:SALSA_poster}. Note that this figure is only for illustrative purposes, and no analysis were derived from these datasets:

\begin{enumerate}
\item \textbf{M51:} \emph{Hubble}/ACS, F658N (H$\alpha$) and F814W (red), F555W (green), and F435W (blue) - NASA/ESA S.\,Beckwith (STScI) and the \emph{Hubble} Heritage Team (STScI/AURA). 

\item \textbf{M82:} The image combines visible starlight (gray) and H-alpha emission (red) from the Kitt Peak 2.1 m (\emph{Spitzer} Infrared Nearby Galaxy Survey Legacy project), and near-infrared and mid-infrared starlight plus dust from \emph{Spitzer} IRAC and SOFIA (3.6 and 53 $\mu$m; yellow, credit: J.\,Moustakas et al.). 

\item \textbf{M83:} \emph{Spitzer}/IRAC 3.6 $\mu$m (blue), 4.5 $\mu$m (green), 8 $\mu$m (red) microns (NASA/JPL-Caltech). 

\item \textbf{NGC\,253:} MPG/ESO 2.2-metre telescope Wide Field Imager (WFI, $R$, $V$, H$\alpha$ and O\,III composition) - La Silla Observatory (Credit: ESO). 

\item \textbf{NGC\,1068:} SDSS RGB image, combined with hard X-ray (magenta, NASA/\emph{NuSTAR}). Credit: NASA/JPL-Caltech/Roma Tre Univ. 

\item \textbf{NGC\,1097:} VLT/NACO $J$(blue), $H$ (green), and $Ks$-band (red). Credit: ESO/Prieto et al. 

\item \textbf{NGC\,2146:} \emph{Hubble}/ACS: F814W (blue), F658N (red), F814W + F658N (green). Credit: ESA/\emph{Hubble} \& NASA. 

\item \textbf{NGC\,3627:} \emph{Spitzer}/IRAC 3.6 $\mu$m (blue), 4.5 $\mu$m (green), 8 $\mu$m (red) microns. Credit: NASA/JPL-Caltech/R. Kennicutt (University of Arizona) and the SINGS Team.

\item \textbf{NGC\,4736:} \emph{Hubble}/WFPC2: F450W (blue), F555W (green), F814W (orange), F336W (red). Credit: ESA/\emph{Hubble} \& NASA. 

\item \textbf{NGC\,4826:} \emph{Hubble}/WFC3: F336W (purple), F438W (blue), F555W (green), F814W (orange), F275W (red). Credit: ESA/\emph{Hubble} \& NASA, J. Lee and the PHANGS-HST Team. Acknowledgement: Judy Schmidt.

\textbf{NGC\,6946:}  \emph{Spitzer}/IRAC 3.6 $\mu$m (blue), 4.5 $\mu$m (green), 8 $\mu$m (red) microns (NASA/JPL-Caltech). Credit: NASA/JPL-Caltech L.Proudfit. 

\textbf{NGC\,7331:} LRGB color composite CCD image taken with the 32-inch Schulman Telescope at Mt. Lemmon. Credit: Adam Block/Mount Lemmon SkyCenter/University of Arizona.

\textbf{Antennae:} \emph{Hubble}/ACS and WFC3: F336W (blue), F435W (blue), F550W (cyan), F555W (cyan), F625W (green), F814W (orange), F658N (red). Credit:
ESA/\emph{Hubble} \& NASA.

\textbf{Centaurus A:} MPG/ESO 2.2-metre telescope
WFI ($B$ - blue, $V$ - green), Atacama Pathfinder Experiment
LABOCA (870 $\mu$m - orange), \emph{Chandra} X-ray Observatory (blue). Credit:
ESO/WFI (Optical); MPIfR/ESO/APEX/A.Weiss et al. (Submillimetre); NASA/CXC/CfA/R.Kraft et al. (X-ray).
\end{enumerate}

\bibliography{references}



\end{document}